\documentclass[fleqn,usenatbib]{mnras}

\usepackage{newtxtext,newtxmath}

\usepackage[T1]{fontenc}

\usepackage[normalem]{ulem} 

\DeclareRobustCommand{\VAN}[3]{#2}
\let\VANthebibliography\thebibliography
\def\thebibliography{\DeclareRobustCommand{\VAN}[3]{##3}\VANthebibliography}

\usepackage{graphicx}	
\usepackage{amsmath}	
\usepackage{amssymb}	
\usepackage{color}
\usepackage{booktabs}
\usepackage{breakcites}
\usepackage[table]{xcolor}
\usepackage{empheq}

\mathchardef\mhyphen="2D

\newcommand{\di}{\mathrm{d}}
\newcommand{\bfx}{\mathbf{x}}

\newcommand{\degree}{\ensuremath{^\circ}}

\newcommand{\bfv}{\mathbf{v}}
\newcommand{\pc}{\,{\rm pc}}
\newcommand{\kpc}{\,{\rm kpc}}

\newcommand{\Gyr}{\,{\rm Gyr}}
\newcommand{\kms}{\,{\rm km\, s^{-1}}}

\newcommand{\pa}{\partial}
\newcommand{\Msun}{\, \rm M_\odot}

\title[Jeans modelling of the NSD]{Jeans modelling of the Milky Way's nuclear stellar disc}

\author[Sormani et al.]{Mattia C. Sormani,$^{1}$\thanks{E-mail: mattia.sormani@uni-heidelberg.de}
John Magorrian,$^{2}$
Francisco Nogueras-Lara,$^{3}$ \newauthor
Nadine Neumayer,$^{3}$
Ralph Sch\"{o}nrich,$^{4}$
Ralf S. Klessen,$^{1,5}$ \newauthor
and Alessandra Mastrobuono-Battisti$^{6}$
\\
$^1$ Universit\"{a}t Heidelberg, Zentrum f\"{u}r Astronomie, Institut f\"{u}r theoretische Astrophysik, Albert-Ueberle-Str. 2, 69120 Heidelberg, Germany \\
$^2$ Rudolf Peierls Centre for Theoretical Physics, Clarendon Laboratory, Parks Road, Oxford OX1 3PU \\
$^3$ Max Planck Institute for Astronomy. K\"{o}nigstuhl 17. D-69 117 Heidelberg, Germany \\
$^4$ Mullard Space Science Laboratory, University College London, Holm-bury St. Mary, Dorking, Surrey, RH5 6NT, UK \\
$^5$Universit\"{a}t Heidelberg, Interdiszipli\"ares Zentrum f\"{u}r Wissenschaftliches Rechnen, Im Neuenheimer Feld 205, 69120 Heidelberg, Germany \\
$^6$ Department of Astronomy and Theoretical Physics, Lund Observatory, Box 43, SE-221 00, Lund, Sweden
}

\date{Accepted XXX. Received YYY; in original form ZZZ}
\pubyear{2015}

\begin{document}
\label{firstpage}
\pagerange{\pageref{firstpage}--\pageref{lastpage}}
\maketitle

\begin{abstract}
The nuclear stellar disc (NSD) is a flattened stellar structure that dominates the gravitational potential of the Milky Way at Galactocentric radii $30 \lesssim R \lesssim 300\pc$. In this paper, we construct axisymmetric Jeans dynamical models of the NSD based on previous photometric studies and we fit them to line-of-sight kinematic data of APOGEE and SiO maser stars. We find that (i) the NSD mass is lower but consistent with the mass independently determined from photometry by Launhardt et al. (2002). Our fiducial model has a mass contained within spherical radius $r=100\pc$ of $M(r<100\pc) = 3.9 \pm 1 \times 10^8 \Msun$ and a total mass of $M_{\rm NSD} = 6.9 \pm 2 \times 10^8  \Msun$. (ii) The NSD might be the first example of a vertically biased disc, i.e. with ratio between the vertical and radial velocity dispersion $\sigma_z/\sigma_R>1$. Observations and theoretical models of the star-forming molecular gas in the central molecular zone suggest that large vertical oscillations may be already imprinted at stellar birth. However, the finding $\sigma_z/\sigma_R > 1$ depends on a drop in the velocity dispersion in the innermost few tens of parsecs, on our assumption that the NSD is axisymmetric, and that the available (extinction corrected) stellar samples broadly trace the underlying light and mass distributions, all of which need to be established by future observations and/or modelling. (iii) We provide the most accurate rotation curve to date for the innermost $500\pc$ of our Galaxy.
\end{abstract}



\section{Introduction} \label{sec:introduction}

The nuclear stellar disc (NSD) is a flattened stellar structure that dominates the gravitational potential of the Milky Way at Galactocentric radii $30 \lesssim R \lesssim 300\pc$ (see for example Figure 14 in \citealt{Launhardt+2002}). Current observational constraints are consistent with the NSD being an axisymmetric structure \citep{GerhardMartinezValpuesta2012}, although it cannot be ruled out that it actually consists of a secondary nuclear bar \citep{Alard2001,RFC2008}. The radius and exponential scale-height determined from near-infrared photometry and star counts are $R\simeq 100\mhyphen200\pc$ and $H\simeq 45\pc$ respectively \citep{Catchpole+1990,Launhardt+2002,Nishiyama+2013,GallegoCano+2020}. 

The NSD is co-spatial with the central molecular zone (CMZ), a ring-like accumulation of molecular gas at $R \lesssim 200 \pc$, which is the Milky Way's counterpart of the star-forming nuclear rings that are commonly found at the centre of barred galaxies \citep[][]{Molinari+2011,Henshaw+2016,Tress+2020}. This co-spatiality is presumably not a coincidence, and suggests that the NSD is made of stars born in the dense CMZ gas \citep{BabaKawata2020,Sormani+2020}. This picture is consistent with kinematic observations that show that the NSD is rotating with velocities similar to those of the molecular gas in the CMZ  \citep{Schoenrich+2015}. The rotation of the NSD has been detected in APOGEE data by \citet{Schoenrich+2015}, in OH/IR and SiO maser stars by \citet{Lindqvist+1992} and \citet{Habing+2006}, in ISAAC (VLT) near-infrared integral-field spectroscopy by \cite{Feldmeier+2014} and in classical cepheids by \cite{Matsunaga+2015}.

Since the CMZ gas currently flows in the gravitational potential created by the NSD, having an accurate representation of the NSD mass and density distribution is crucial to understand gas flows in the CMZ. Hydrodynamical simulations confirm this by showing that macroscopic properties such as the size of the CMZ strongly depend on the mass and density profile of the NSD \citep[e.g.][]{Sormani+2018a,Tress+2020,Li+2020}. However,\\ \citet{BlandHawthornGerhard2016} note that the kinematic data from \citet{Schoenrich+2015} suggest a mass which is on the lower side of that determined from near-infrared photometry by \citet{Launhardt+2002}: the former report a rotation velocity of $v\simeq120\kms$ at $R\simeq 100\pc$, which naively suggests (ignoring asymmetric drift) a mass of $M_{\rm NSD} \simeq R v^2 / G \simeq 3 \times 10^8 \Msun$, while the latter report a mass of $M_{\rm NSD} = 6\pm 2 \times 10^8 \Msun$ at the same radius. It is thus important to constrain the NSD mass more precisely.

The mass and structure of the NSD can be constrained by constructing stellar dynamical models of the NSD and comparing them with the available kinematic/star counts data. The only attempt available in the literature is a very simple spherical Jeans modelling from \citet{Lindqvist+1992} based on a sample of 148 OH/IR maser stars. However, this model neglects that the stellar density distribution of the NSD is strongly flattened \citep{Launhardt+2002,Nishiyama+2013,GallegoCano+2020} and is based on a limited number of stars.

Dynamical modelling of the NSD is also interesting from a general theoretical perspective. Nuclear stellar discs are common in the centre of spiral galaxies \citep{Pizzella+2002,Cole+2014,Gadotti+2019,Gadotti+2020}. The radii of nuclear rings in the sample of \cite{Gadotti+2019,Gadotti+2020} range from $R\sim100\pc$ to $R\sim1000\pc$, so the size of the MW's NSD is consistent with but on the lower side of their distribution (see Figure 5 and Table 2 in \citealt{Gadotti+2020}). Since nuclear stellar discs have different formation and evolution history than more well-studied disc systems such as galactic discs, they may be expected to have qualitatively different structural and kinematic properties. 

In this paper, we aim to construct Jeans-type dynamical models of the NSD which are consistent with previous photometric/star counts studies and to compare them with line-of-sight kinematic data. This will provide constraints on the mass and structure of the NSD.

The paper is structured as follows. In Section \ref{sec:data} we describe the observational data. In Section \ref{sec:jeans} we describe the Jeans modelling methodology. In Section  \ref{sec:results} we present our results and in Section \ref{sec:discussion} we discuss them. We sum up in Section \ref{sec:conclusion}.

\section{Observational data} \label{sec:data}

\subsection{APOGEE data} \label{sec:apogee}

We use data from the SDSS-{\sc IV}/APOGEE survey \citep{Majewski+2017} data release 16 (DR16, \citealt{Ahumada+2019}), which is publicly available at \url{https://www.sdss.org/dr16/irspec/}. APOGEE is the first large-scale spectroscopic survey of the Milky Way in the near infrared ($H$-band, $1.51\mhyphen1.70\, \micron$). Most of the stars that we will use for the modelling in this paper are part of the ``GALCEN'' field, which is a special additional target not part of the main survey targets (see Section 8 in \citealt{Zasowski+2013}). Since observations of the Galactic centre is hampered by the extreme crowding and the high extinction and differential reddening \citep{Nishiyama+2008,Schoedel+2010,NoguerasLara+2018a,NoguerasLara+2019a,NoguerasLara+2020b}, the majority of stars that we can observe using APOGEE are bright giants \citep{Bovy+2014,Bovy+2016a}.

\begin{figure}

	\includegraphics[width=\columnwidth]{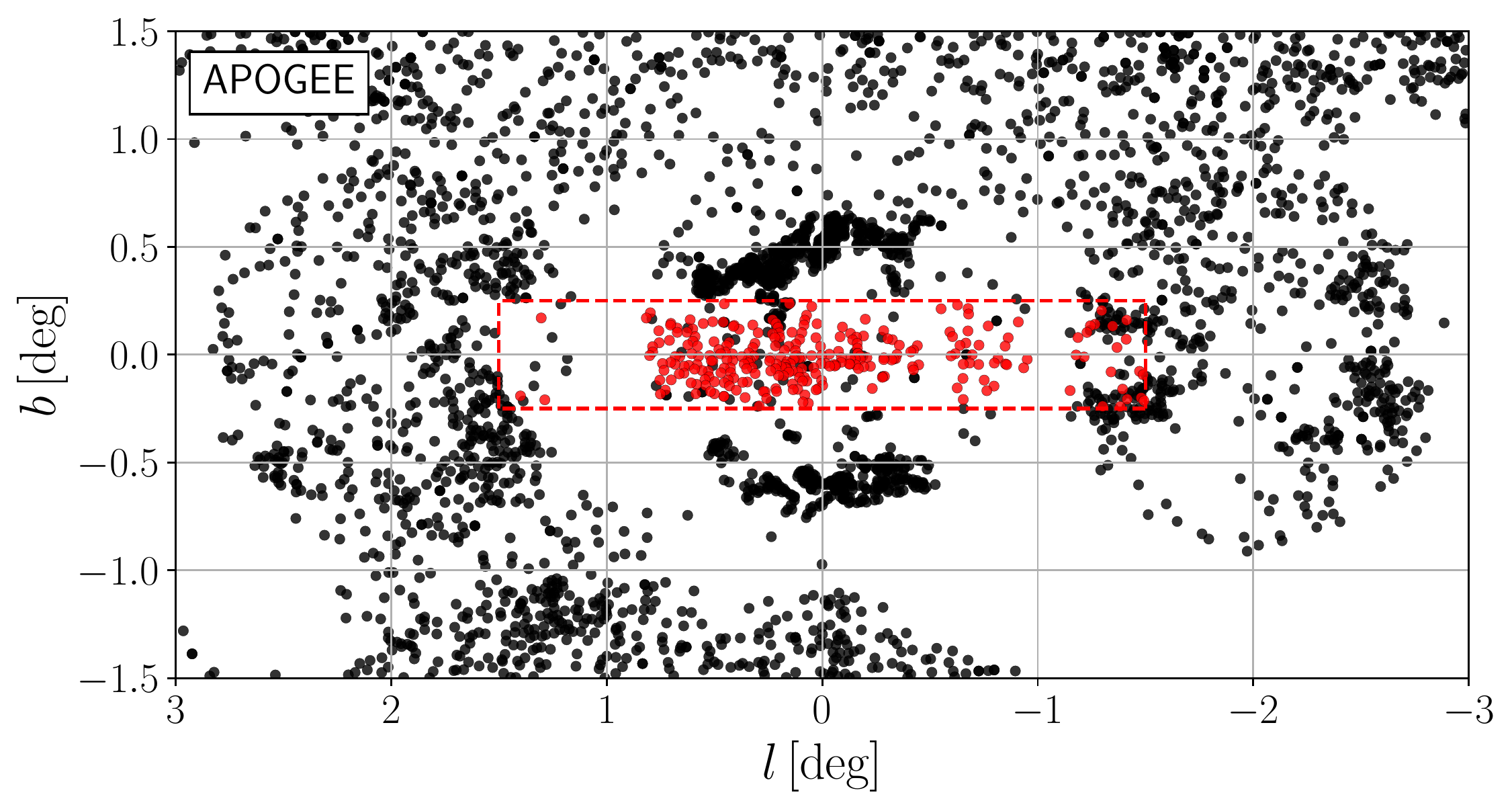}
	\includegraphics[width=\columnwidth]{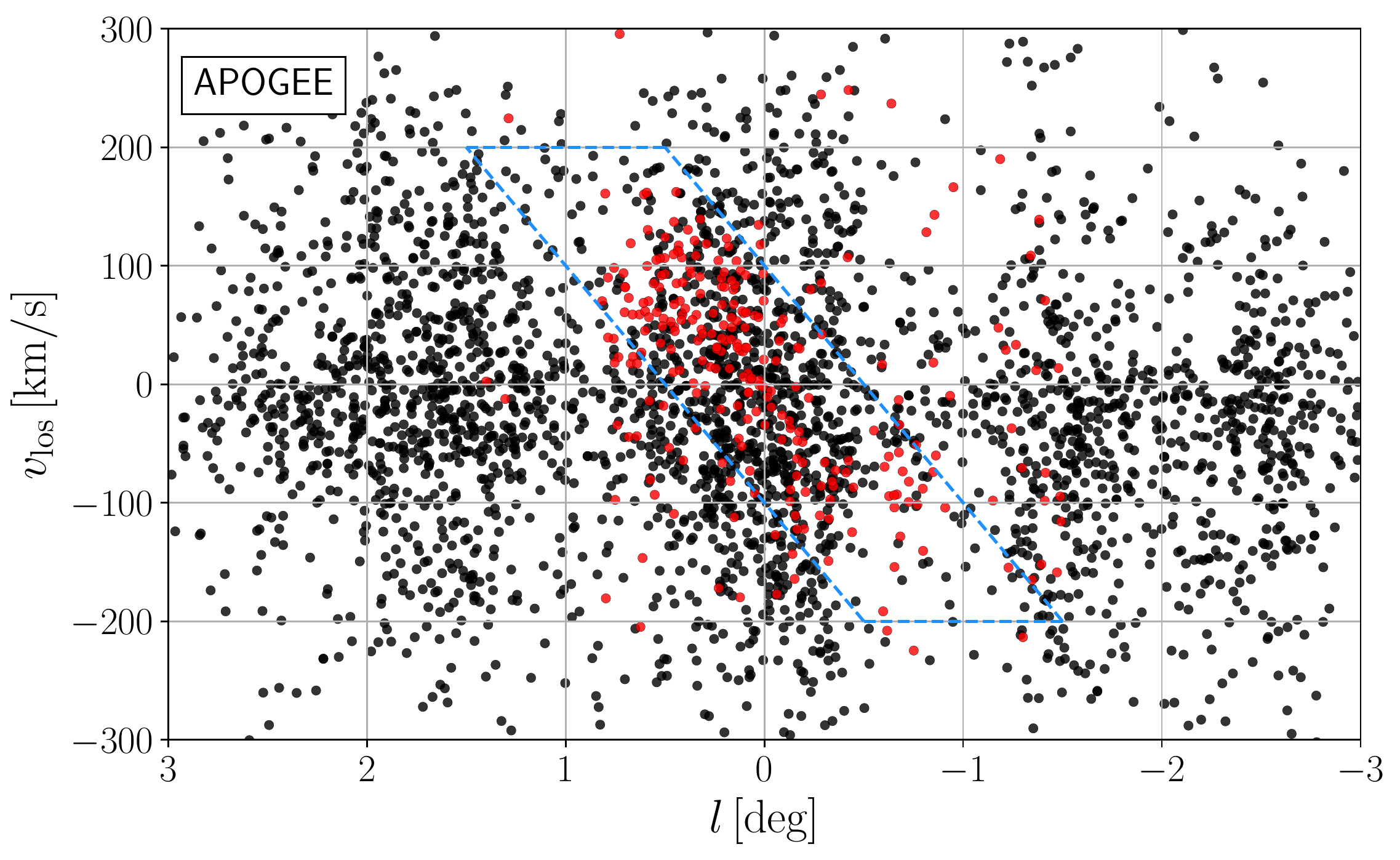}
	\includegraphics[width=\columnwidth]{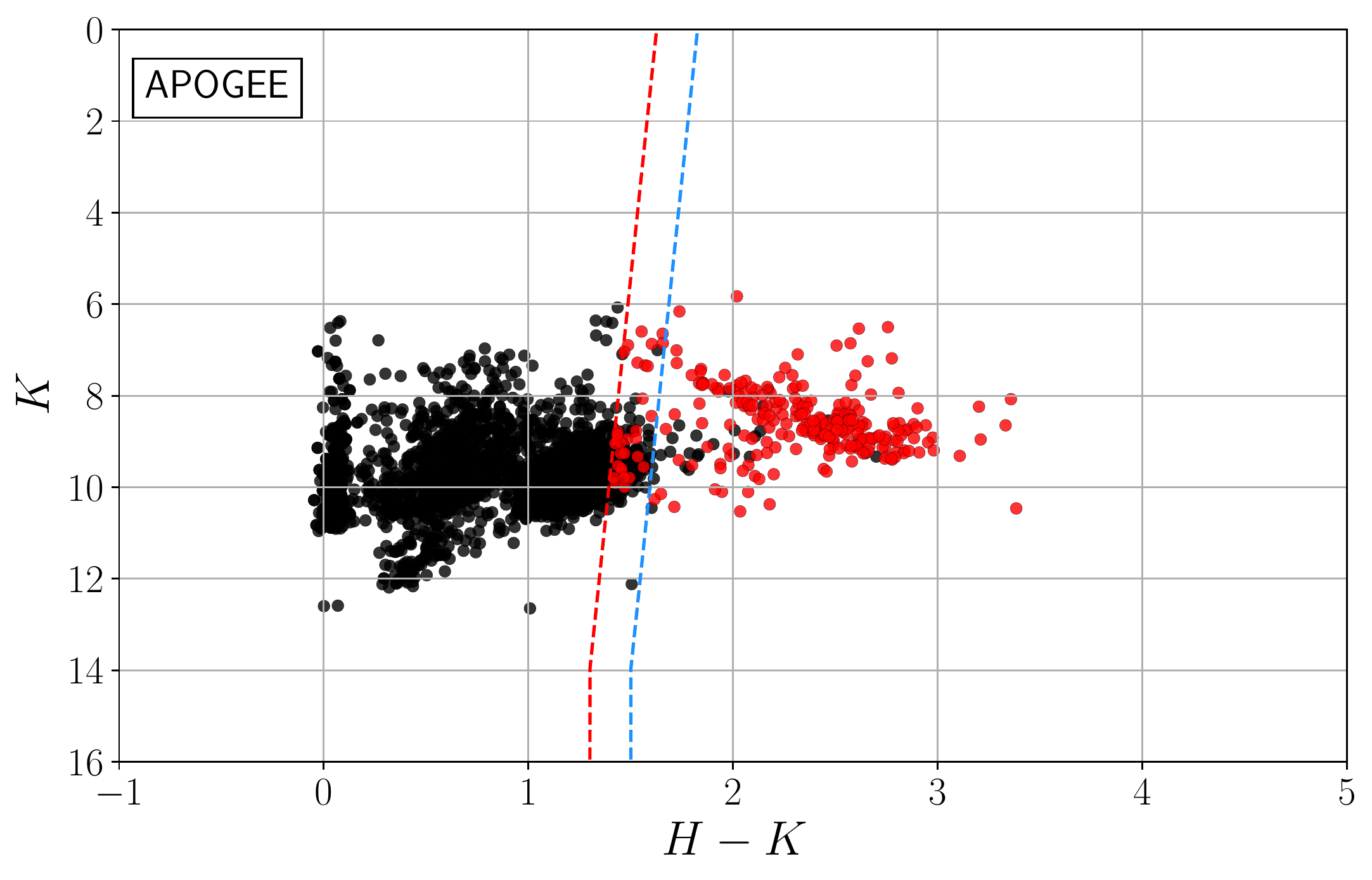}
	\includegraphics[width=\columnwidth]{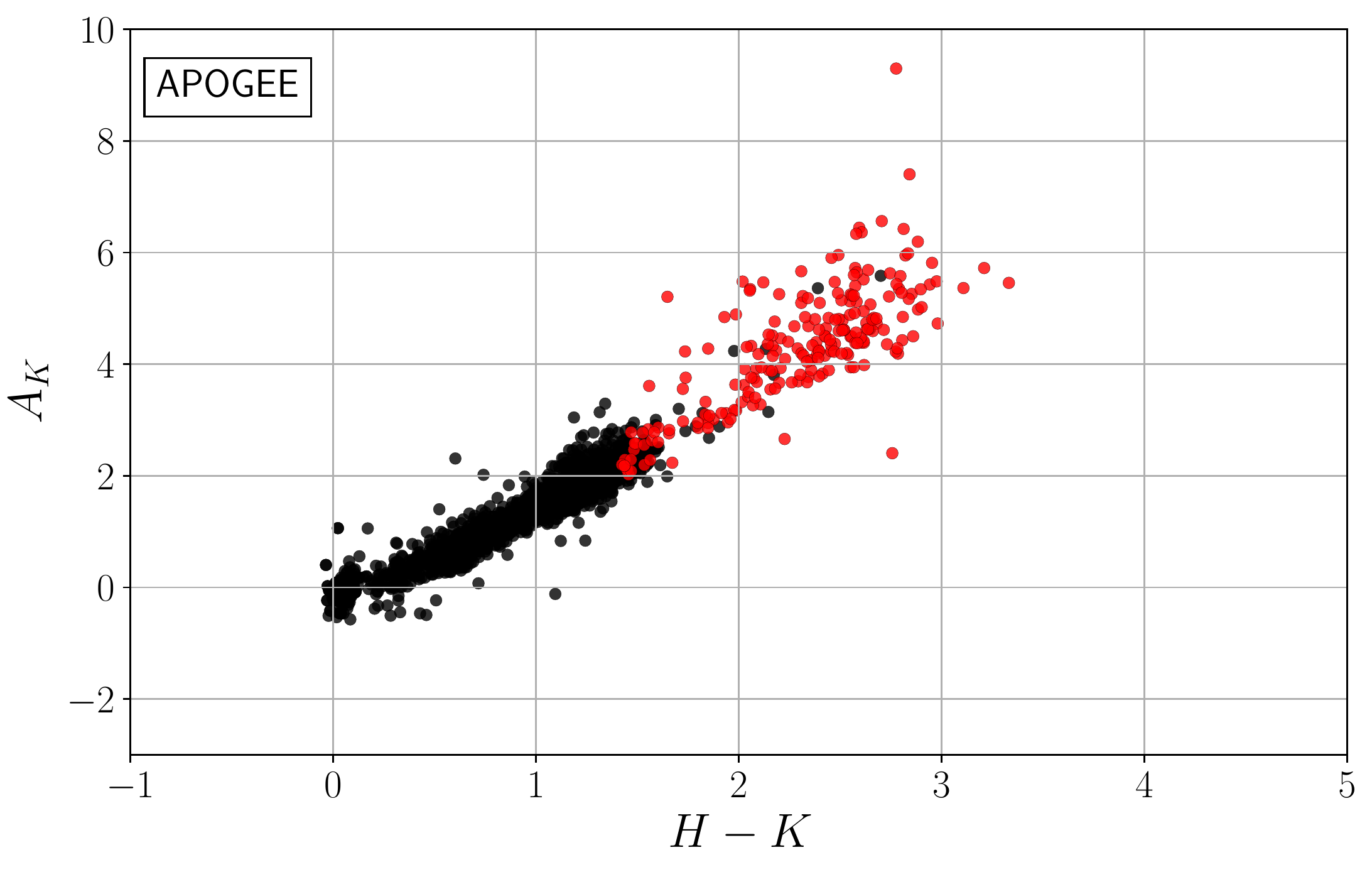}
	
    \caption{All APOGEE stars contained in the region $|b|<0.3\degree$, $|l|<3\degree$. Each point represents an individual star. Red stars are those that satisfy all the selection criteria defined in Section \ref{sec:apogee}, while black stars are those excluded by the various cuts. The red dashed lines indicate these cuts: stars outside of the box in the top panel, or to the left of the red-dashed line in the third panel, are excluded. The blue dashed lines indicate additional cuts that we use to check the robustness of our results against variations in the selection criteria (see Section \ref{sec:results}). $v_{\rm los}$ is the line-of-sight velocity, $H$ and $K$ are the 2MASS $H$-band and $K$-band magnitudes and $A_K$ is the $K$-band extinction from the WISE survey \citep{Wright+2010}.}
    \label{fig:apogee_01}
\end{figure}

In order to diminish foreground contamination, we apply a series of cuts to the data. Our ``standard'' filter is constructed as follows. First, we exclude all stars outside the range $|l|<1.5\degree$ and $|b|<0.25\degree$ (see red dashed box in the top panel of Figure \ref{fig:apogee_01}). Assuming a Sun-Galactic centre distance of 8.2 kpc \citep[e.g.][]{Gravity2019}, this correspond to projected radial and vertical distances of $|R|<215\pc$ and $|z|<36\pc$, roughly the NSD radius and scale-height (see Section \ref{sec:introduction}). In this region, the surface density of the NSD is higher than that of the Galactic disc and therefore the percentage of contaminating stars is expected to be relatively low (see for example Table 5 in \citealt{Catchpole+1990}). A total of 405 APOGEE stars are contained in this region. We then apply a parallax cut by excluding stars that according to the APOGEE datafile have $p - 3\delta_p>1/d_{\rm min}$, where $p$ is the Gaia DR2 parallax, $\delta_p$ is the Gaia DR2 parallax uncertainty and $d_{\min}=7 \kpc$ (so we remove stars that are closer than this distance). Only a small subset of stars has Gaia parallax defined, so this cut only removes 2 stars from the 405, leaving 403. Then we apply a proper motion cut by excluding stars that have $\mu_{\alpha}-3 \delta \mu_{\alpha}>\mu_{\rm max}$ or $\mu_{\delta}-3 \delta \mu_{\delta}>\mu_{\rm max}$ where $\mu_{\alpha}$ and $\mu_{\delta}$ are the Gaia DR2 proper motions in RA and DE directions, $\delta \mu_{\alpha}$ and $\delta \mu_{\delta}$ are the associated uncertainties and $\mu_{\rm max}=400/(4.74\times 7000)\times 1000 \rm \, mas \,yr^{-1}$ corresponds to a proper motion velocity of 400 km/s at 7kpc (i.e., we exclude stars that at distance of $d>7\kpc$ move faster than $400\kms$). After applying this cut, we are left with 366 stars. Finally, we apply a colour-magnitude cut. We follow the methodology explained in \citet{NoguerasLara+2020} and consider only stars with $H-K > \operatorname{max}(-0.0233 K + 1.63 ,1.3)$, see red dashed line in the third panel of Figure \ref{fig:apogee_01}. Due to the high extinction that characterises the Galactic centre ($A_K \sim 2.5 \rm \, mag$, e.g.\ \citealt{Nishiyama+2008,NoguerasLara+2018a,NoguerasLara+2019a,NoguerasLara+2020}), this colour cut effectively excludes the foreground stellar population belonging to the Galactic disc, whose absolute extinction is significantly lower, and also the majority of stars from the inner bulge ($A_K \sim 1.2 \rm \, mag$, corresponding to $(H~-~K)~\sim~1$, \citealt{NoguerasLara+2018b}). The final set of stars, which consists of 273 stars, is shown in red in Figure \ref{fig:apogee_01}.

\begin{figure}
	\includegraphics[width=\columnwidth]{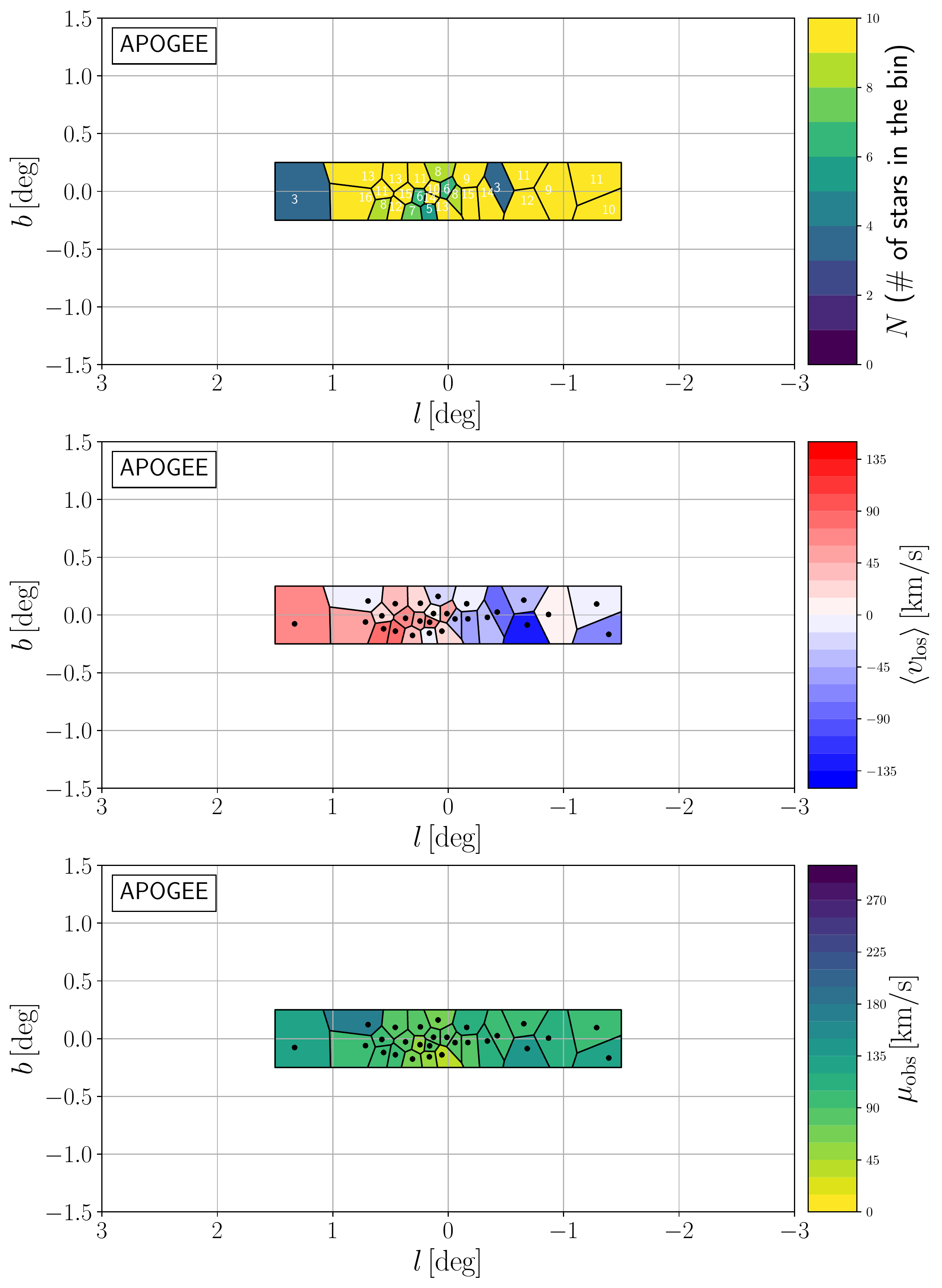}
        \caption{Voronoi binning of the APOGEE stars that satisfy all the selection criteria defined in Section \ref{sec:apogee} (the red stars in Figure \ref{fig:apogee_01}). \emph{Top}: the number of stars in the bin. \emph{Middle}: average line-of-sight velocity (Equation \ref{eq:vlosmean}). \emph{Bottom}: root mean square velocity (Equation \ref{eq:vlos2mean}). The latter is the quantity that we fit in our Jeans modelling.}
    \label{fig:apogee_02}
\end{figure}

In order to compare the data with Jeans models, we bin the final set of stars using the {\sc vorbin} package from \url{https://pypi.org/user/micappe/}. This is an implementation of the two-dimensional adaptive spatial binning method of \citet{CappellariCopin2003}, which uses a Voronoi tessellation to bin data with given minimum signal-to-noise ratio. Here, we only use this as a convenient method to define a Voronoi tessellation which has approximately the same number of stars in each bin. The signal-to-noise parameter essentially controls the average number of stars in each bin: a higher (lower) value results in less (more) bins with a higher (lower) number of stars in each of them. We assign constant $\rm signal=1$, $\rm noise=1$ and use a target signal-to-noise ratio of $3.2$, which gives an average of $\simeq 10$ stars per bin. The result is shown in Figure \ref{fig:apogee_02}. 

\subsection{SiO maser data} \label{sec:SiO}

We use the 86 Ghz SiO maser survey of the inner Galaxy from \citet{Messineo+2002,Messineo+2004,Messineo+2005}. The SiO maser stars targeted in this survey are stars in the Asymptotic Giant Branch (AGB) phase with estimated ages in the range $0.2\mhyphen2\Gyr$ \cite[e.g.][and references therein]{Habing+2006}.

\begin{figure}

	\includegraphics[width=\columnwidth]{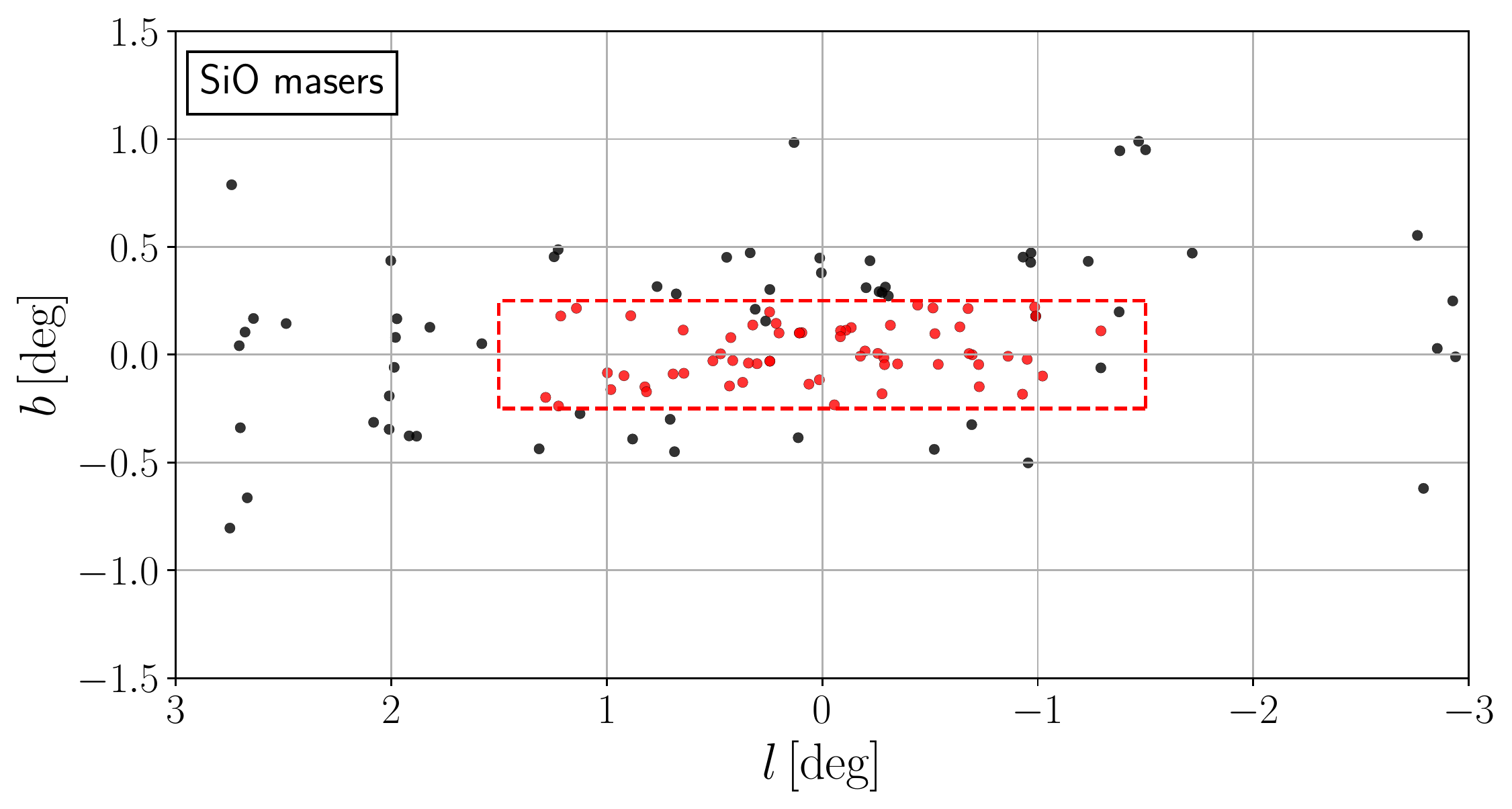}
	\includegraphics[width=\columnwidth]{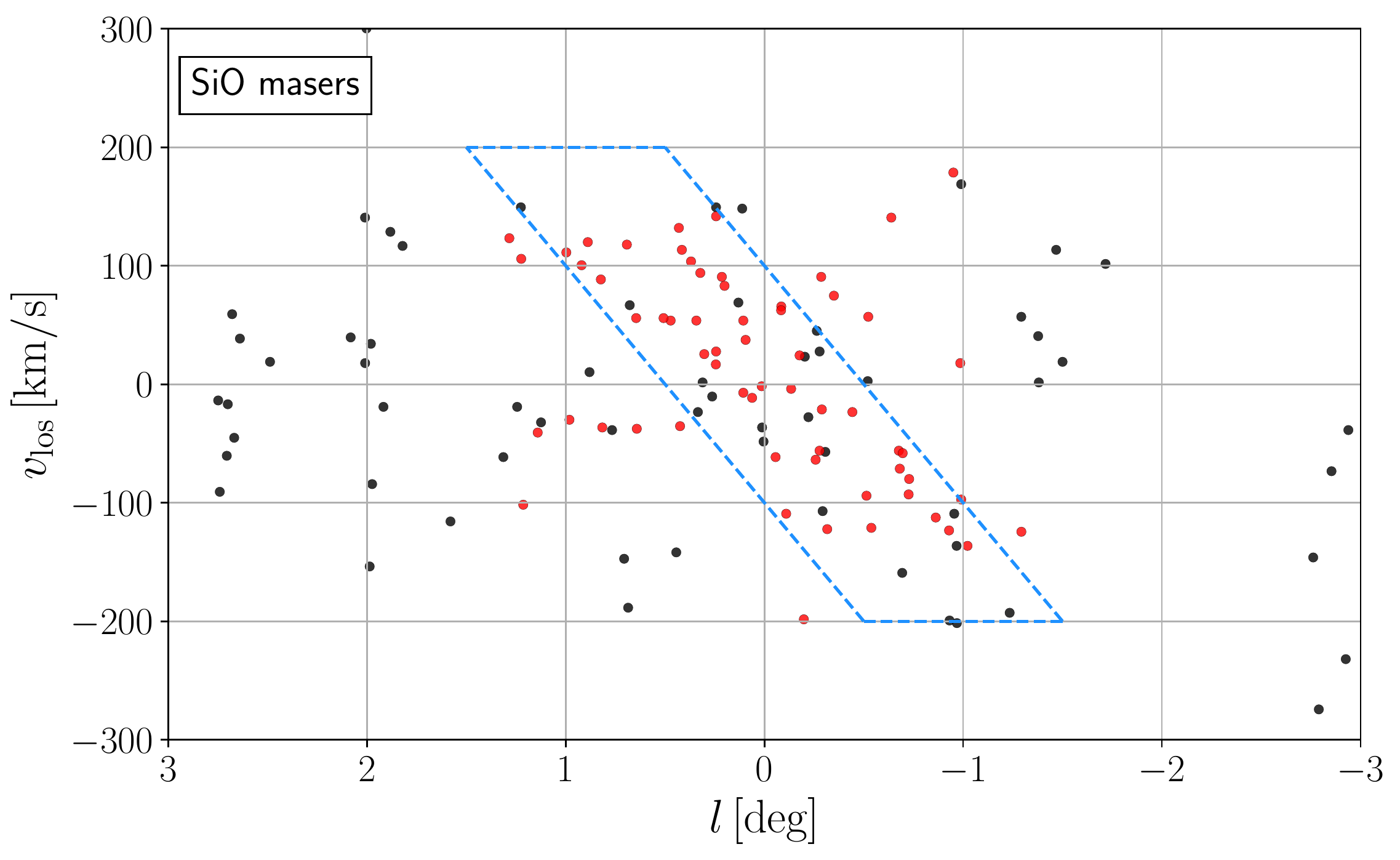}
	\includegraphics[width=\columnwidth]{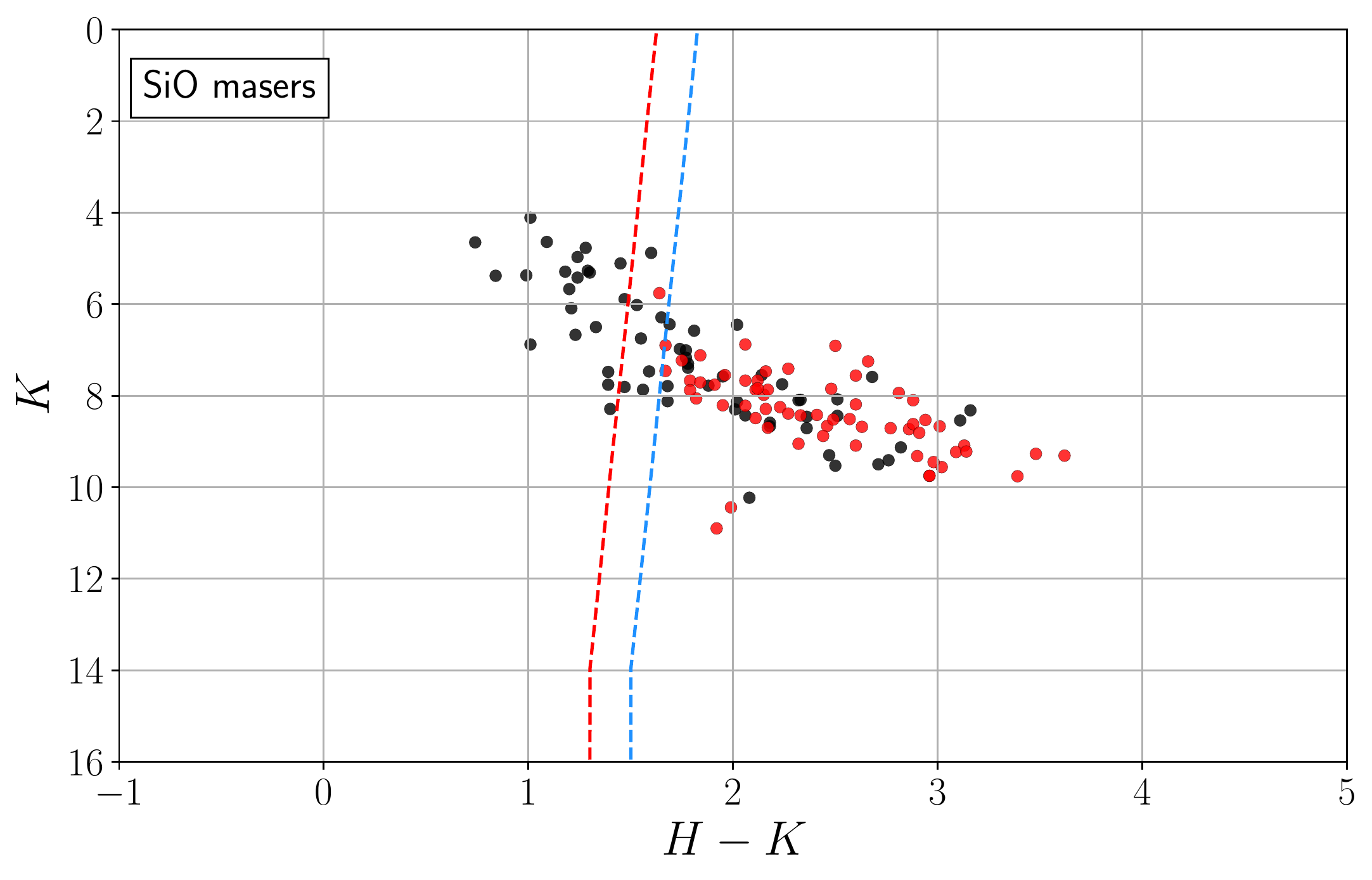}
	\includegraphics[width=\columnwidth]{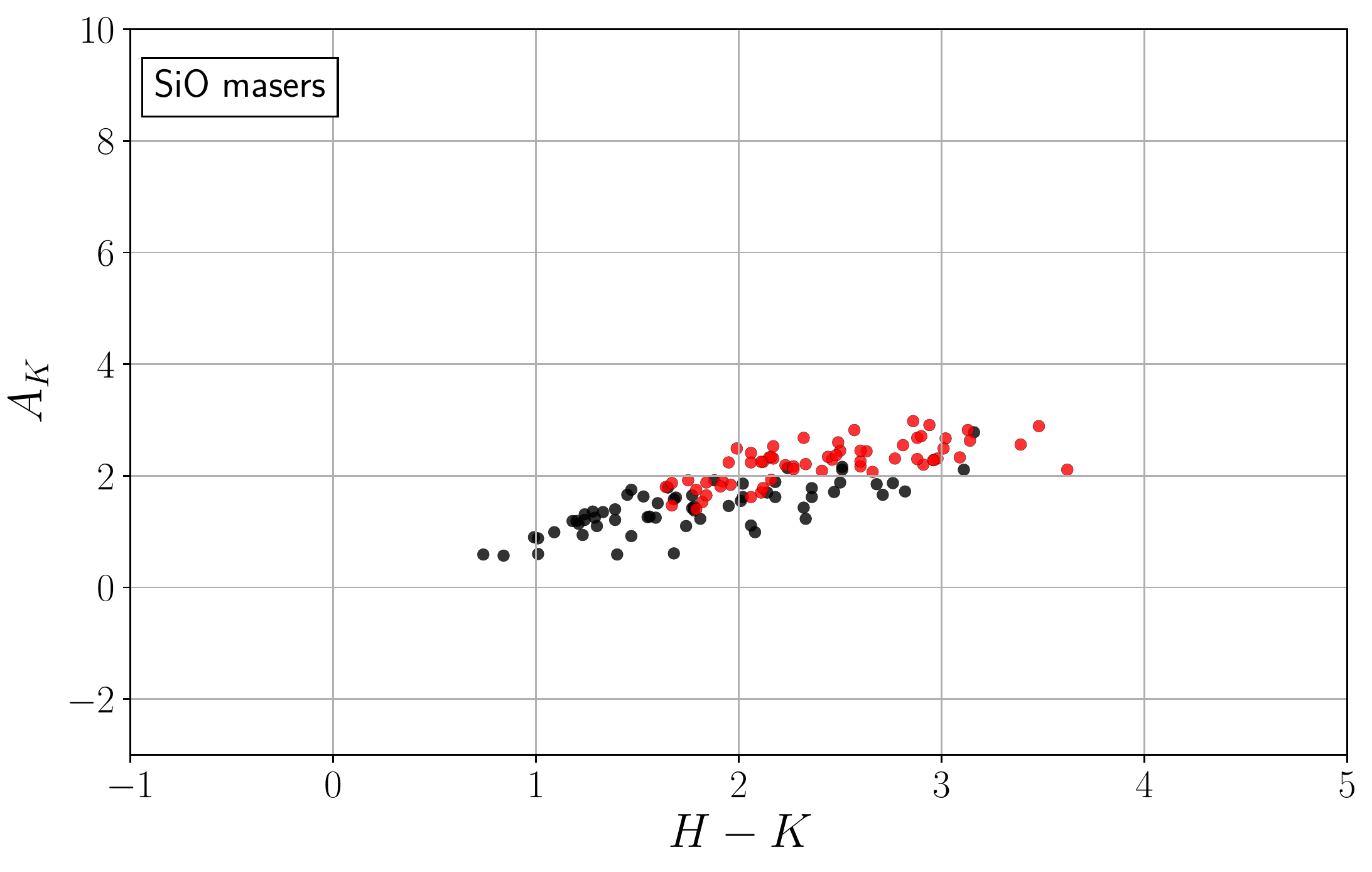}
	
    \caption{Same as Figure \ref{fig:apogee_01}, but for the SiO maser stars from \citet{Messineo+2002,Messineo+2004,Messineo+2005} and using the selection criteria defined in Section \ref{sec:SiO}.}
    \label{fig:SiO_01}
\end{figure}

Similarly to what we have done in Section \ref{sec:apogee} for the APOGEE data, in order to reduce foreground contamination we apply a series of cuts to the maser data. There are initially 67 maser stars in the region $|l|<1.5\degree$ and $|b|<0.25\degree$, 4 of which are flagged as foreground contamination by \citet{Messineo+2005}. After excluding these four stars, we apply the same color-magnitude cut defined in Section \ref{sec:apogee} using $H$ and $K$ determined from the 2MASS survey \citep{2MASS}, see red dashed line in the third panel of Figure \ref{fig:SiO_01}. This excludes only one more star. The final set therefore consists of 62 stars, which are shown in red in Figure \ref{fig:SiO_01}.

\begin{figure}
\includegraphics[width=\columnwidth]{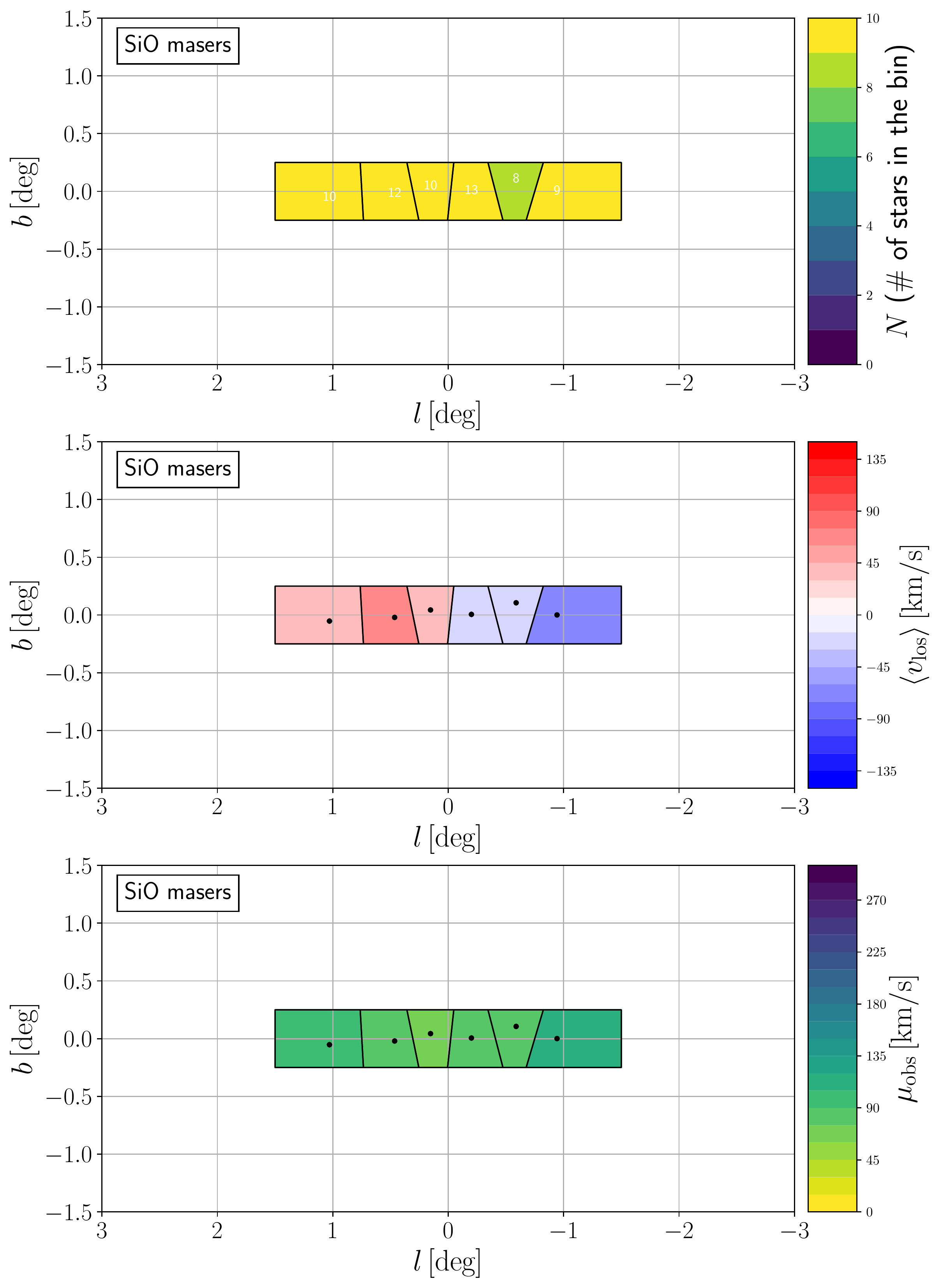}
    \caption{Same as Figure \ref{fig:apogee_02}, but for SiO maser stars shown in red in Figure~\ref{fig:SiO_01}.}
    \label{fig:SiO_02}
\end{figure}

As for the APOGEE data, we bin the final set of stars using the {\sc vorbin} package. Again we assign constant $\rm signal=1$, $\rm noise=1$ and use a target signal-to-noise ratio of $3.2$, which gives an average of $\simeq 10$ stars per bin. The result is shown in Figure \ref{fig:SiO_02}.

\section{Jeans modelling} \label{sec:jeans}

We model the line-of-sight stellar kinematics using an anisotropic axisymmetric Jeans formalism \citep{Cappellari2008}. Section \ref{sec:basic} reviews the basic equations of this formalism. Section \ref{sec:observables} describes how we compute the observables from the model, and Section \ref{sec:fitting} describes our fitting procedure. Section \ref{sec:mass} describes the mass distribution and gravitational potential models that we employ. 

\subsection{Review of Jeans equations} \label{sec:basic}

The dynamics of a collisionless stellar system is described by the collisionless Boltzmann equation, which in cylindrical coordinates $(R,\phi,z)$ for an axisymmetric system reads (see Equation 4-17 of \citealt{BT1987}):
\begin{align} \label{eq:boltzmann}
\frac{\pa f}{\pa t}  	& + v_R \frac{\pa f}{\pa R} + \frac{v_\phi}{R} \frac{\pa f}{\pa \phi} + v_z \frac{\pa f}{\pa z} + \left(\frac{v_\phi^2}{R} - \frac{\pa \Phi}{\pa R} \right) \frac{\pa f}{\pa v_R} \nonumber \\
			  	& - \frac{1}{R} \left( v_R v_\phi + \frac{\pa \Phi}{\pa\phi} \right)\frac{\pa f}{\pa v_\phi} - \frac{\pa \Phi}{\pa z}\frac{\pa f}{\pa v_z} = 0,
\end{align}
where $f(\bfx,\bfv,t)$ is the distribution function (DF), $f(\bfx,\bfv,t) \, \di^3 \bfx \, \di^3 \bfv$ is the number of stars in the small volume $\di^3 \bfx = R \, \di R \,\di \phi \,\di z$ centred on $\bfx$ and with velocities in the small range $\di^3 \bfv = \di v_R \, \di v_\phi \, \di v_z$ centred on $\bfv$, and $\Phi(\bfx,t)$ is the gravitational potential. Note that, for Equation \eqref{eq:boltzmann} to be valid, it is not necessary that $\Phi$ is the potential self-consistently generated by the (tracer) density distribution calculated from $f$ (see Equation \ref{eq:rho} below). For example, $f$ might describe a sub-population of stars which only partially contributes to the overall gravitational potential $\Phi$.

The spatial density of tracer stars $\rho(\bfx)$, the mean velocities $\bar{v}_i(\bfx)$, and the velocity ellipsoid $\sigma_{ij}(\bfx)$ are defined as:
\begin{align}
\rho 		   		& = \int f \di^3 \bfv , \label{eq:rho} \\ 
\overline{v}_i 	   	& = \frac{1}{\rho} \int f v_i \di^3 \bfv\,, \label{eq:vmean} \\
\overline{v_i v_j} 	& = \frac{1}{\rho} \int f v_i v_j \di^3 \bfv \,, \\
\sigma_{ij}^2             & = \overline{ (v_i - \bar{v}_i)(v_j - \bar{v}_j)} = \overline{v_i v_j} - \bar{v}_i \bar{v}_j \,,
\end{align}
where $i,j=R,\phi$ or $z$. Multiplying Equation \eqref{eq:boltzmann} by $v_R$, $v_\phi$, or $v_z$ respectively, assuming axisymmetry ($\pa_\phi=0$), and integrating over all velocities, we obtain the following Jeans equations\footnote{The steps involve integrating some terms by parts and assuming that $f\to0$ for $|\bfv|\to\infty$.} (see Equations 4-29a-4-29c in \citealt{BT1987}):

\begin{align} 
& \frac{\pa \left(\rho \bar{v}_R\right) }{\pa t}   + \frac{\pa \left( \rho \overline{v_R^2} \right)}{\pa R} + \frac{\pa  \left( \rho \overline{ v_R v_z } \right)}{\pa z} + \rho \left( \frac{\overline{v_R^2} - \overline{v_\phi^2}}{R} +  \frac{\pa \Phi}{\pa R}	\right)  = 0 \,,\label{eq:JR} \\
& \frac{\pa \left(\rho \bar{v}_\phi\right) }{\pa t}   + \frac{\pa \left( \rho \overline{v_R v_\phi} \right)}{\pa R} + \frac{\pa  \left( \rho \overline{ v_\phi v_z } \right)}{\pa z} + \frac{2 \rho}{R} \overline{v_\phi v_R}  = 0 \,, \label{eq:Jphi} \\
&\frac{\pa \left(\rho \bar{v}_z\right) }{\pa t}   + \frac{\pa \left( \rho \overline{v_R v_z} \right)}{\pa R} + \frac{\pa  \left( \rho \overline{ v_z^2 } \right)}{\pa z} + \frac{\rho}{R} \overline{v_R v_z} + \rho \frac{\pa \Phi}{\pa z}  = 0 \,. \label{eq:Jz}
\end{align}
The typical situation in Jeans modelling is one in which given $\rho$ and $\Phi$, and under the assumption of steady state ($\pa_t=0$), we want to use Equations \eqref{eq:JR}-\eqref{eq:Jz} to generate predictions for the six velocity moments ($\overline{v_R^2}$, $\overline{v_\phi^2}$, $\overline{v_z^2}$, $\overline{v_R v_\phi}$, $\overline{v_R v_z}$, and $\overline{v_\phi v_z}$) that can be compared with kinematic observations. However, for a steady state system ($\pa_t=0$), Equations \eqref{eq:JR}--\eqref{eq:Jz} provide only three constraints among these six moments. Therefore, in order to proceed, one has to make some assumptions that reduce the number of unknowns to match the number of equations. Following \citet{Cappellari2008} we assume that:
\begin{enumerate} 
\item $\bar{v}_R=\bar{v}_z=0$, i.e. any mean-streaming motion within the disc is purely tangential.
\item $\overline{v_Rv_\phi}=\overline{v_zv_\phi}=0$.
\item $\sigma_{Rz}=\overline{v_Rv_z}=0$, i.e., the principal axes of the velocity ellipsoid $\sigma_{ij}$ (which can always be diagonalised since it is a symmetric tensor) are parallel to the $R$ and $z$ axes.
\item $\overline{v_R^2}=b\overline{v_z^2}$, where the anisotropy $b$ is a constant.
\end{enumerate}
Assumptions (i) and (ii) are in the spirit of our assumption that the disc is axisymmetric. Assumption (iii) is stronger than (ii), and does not have such a natural justification.  It is reasonable to assume that $\overline{v_Rv_z}=0$ in the $z=0$ plane, since this follows if we assume reflection symmetry with respect to the plane $z=0$.  In the solar neighbourhood, as one moves away from the $z=0$ plane the velocity ellipsoid ``tilts'' in the sense that it is more closely aligned with the $r$ and $\theta$ axes of a spherical polar coordinate system \citep{Siebert+2008,Binney+2014,Everall+2019}.  Nevertheless even if the velocity ellipsoid does tilt like this then (using the standard rules for the transformation of tensors under rotations) we would have $\overline{v_R v_z}=(\sigma_r^2-\sigma_\theta^2)\sin\theta\cos\theta$, which is much smaller than the other terms in equations~\eqref{eq:JR} and \eqref{eq:Jz} when one is close to the plane ($\theta=\pi/2$). We have tested that assuming that the principal axes are aligned on spherical rather than cylindrical coordinates does not affect the conclusions of the paper (see Section \ref{sec:discussion_b} for more details). Assumption (iv) is harder to justify a priori and is mainly motivated by simplicity. We will see in Section \ref{sec:results} that it gives an adequate representation of the available data.

Under these assumptions, Equation \eqref{eq:Jphi} is identically zero, while Equations \eqref{eq:JR} and \eqref{eq:Jz} become
\begin{align} 
& \frac{\pa \left( \rho b \overline{v_z^2} \right)}{\pa R}  + \rho \left( \frac{ b\overline{v_z^2} - \overline{v_\phi^2}}{R} +  \frac{\pa \Phi}{\pa R}	\right)  = 0 \,, \label{eq:JR2} \\
& \frac{\pa  \left( \rho \overline{ v_z^2 } \right)}{\pa z} + \rho \frac{\pa \Phi}{\pa z}  = 0 \label{eq:Jz2}\,.
\end{align}
These two equations can be solved in the two unknowns $\overline{v_z^2}$ and $\overline{v_R^2}$. Integrating Equation~\eqref{eq:Jz2} using the boundary condition $\rho v_z^2\to 0$ as $z\to\infty$ and then substituting in Equation~\eqref{eq:JR2} we obtain:
\begin{align} 
&  \rho \overline{ v_z^2}(R,z)  = \int_z^\infty  \rho \frac{\pa \Phi}{\pa z} \di z   \,, \label{eq:Jz3} \\
&  \rho \overline{ v_\phi^2}(R,z) = b \left[ R \frac{\pa \left( \rho  \overline{v_z^2} \right)}{\pa R}  + \rho \overline{v_z^2} \right]  + R \rho  \frac{\pa \Phi}{\pa R} \,. \label{eq:JR3}
\end{align}
These equations allow one to generate predictions for $\overline{v_z^2}(R,z)$ and $\overline{v_\phi^2}(R,z)$ given $\rho(R,z)$, $\Phi(R,z)$ and the parameter $b$. In Section \ref{sec:observables} we show that it is straightforward to project these quantities along lines of sight and to compare the results against the (density-weighted) projected second moment constructed from the observed stellar samples. We stress however that Equations \eqref{eq:Jz3} and \eqref{eq:JR3} rely on the simplifying and somewhat arbitrary assumptions (iii) and (iv) above. One of the biggest shortcomings of Jeans modelling is that even if a tracer density model $\rho$ and a gravitational potential $\Phi$ are found such that the moments calculated using Equations \eqref{eq:Jz3} and \eqref{eq:JR3} project to give a good representation of the data, this does not guarantee that this model is physical: it may not exist a well-defined DF ($f>0$) in the potential~$\Phi$ that corresponds to the density $\rho$ and that satisfies all the assumptions made in this section (steady state, axisymmetry, i-iv; see for example Section 4.4.1 in \citealt{BT2008}).

\subsection{Calculation of observables} \label{sec:observables}

Equations \eqref{eq:Jz3} and \eqref{eq:JR3} allow one to calculate predictions for $\overline{v_z^2}(R,z)$ and $\overline{v_\phi^2}(R,z)$. However, we do not have direct observations of these two quantities for the NSD. In order to calculate the observables that can be compared to our data, we need to integrate them along the line of sight.

\begin{figure}
	\includegraphics[width=\columnwidth]{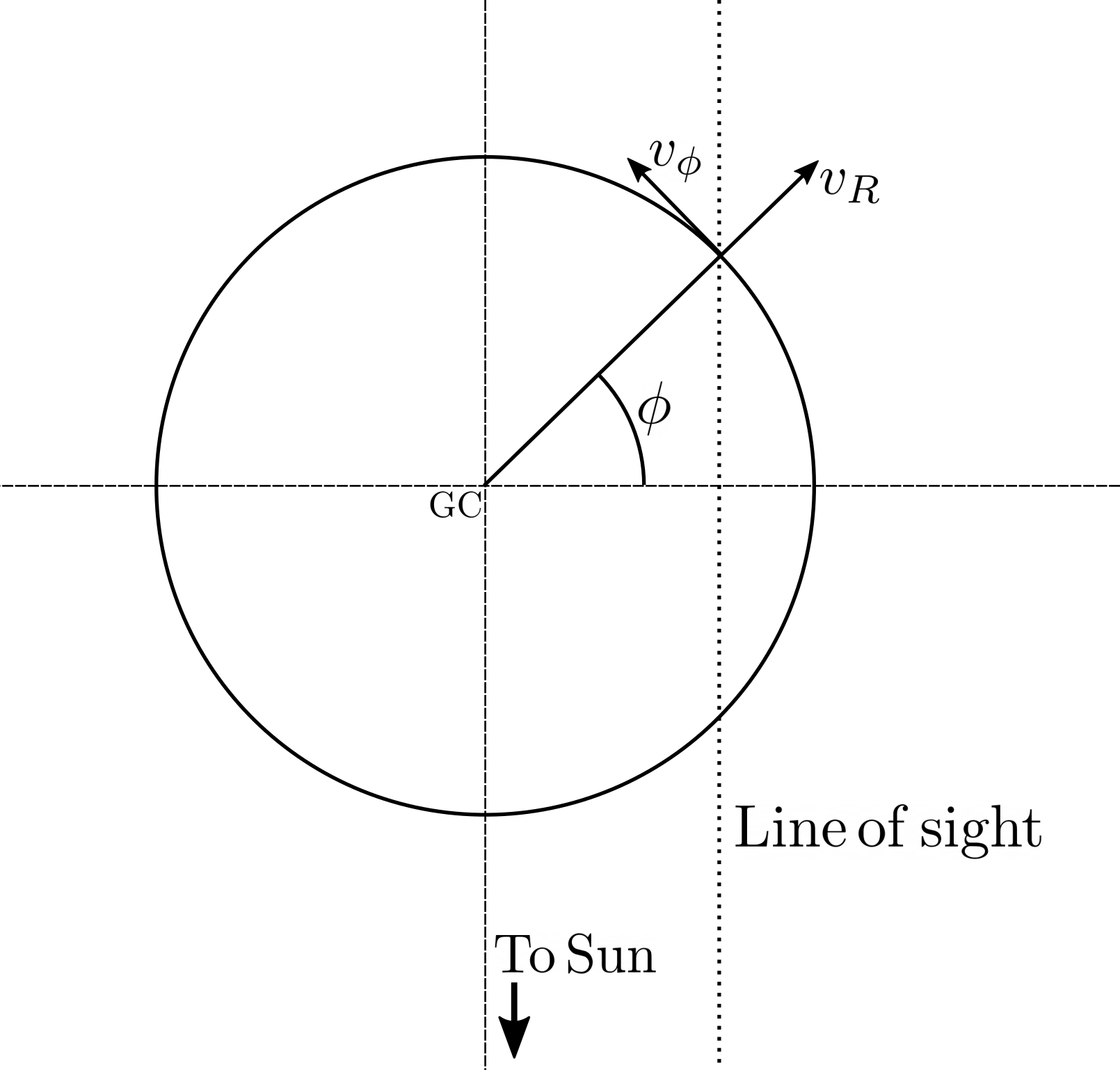}
    \caption{Geometry of the line-of-sight integration (see Section \ref{sec:observables}).}
    \label{fig:los}
\end{figure}

We assume that the NSD is exactly edge-on and that all lines of sight can be considered parallel at the distance of the Galactic centre (GC). Under these assumptions the line-of-sight velocity is given by (see Figure \ref{fig:los}):
\begin{equation} \label{eq:vlos}
v_{\rm los}(R,z) = v_\phi(R,z) \cos \phi + v_R(R,z) \sin \phi \,.
\end{equation}
Taking the square of this equation and then averaging\footnote{The average of a generic quantity $G(\bfx,\bfv)$ are defined here as $\bar{G}(R,z) = [\int f G \di^3 \bfv]/\rho$, where $\rho=\int f \di^3 \bfv$.}  we find (see for example Appendix A of \citealt{EvansDeZeeuw1994}):
\begin{equation} \label{eq:vlos2}
\overline{v_{\rm los}^2}(R,z) = \overline{v_\phi^2}(R,z) \cos^2 \phi + \overline{v_R^2}(R,z) \sin^2 \phi \,,
\end{equation}
where we have used that $\overline{v_\phi v_R} = 0$ (see Section \ref{sec:basic}).
The second moment of the line-of-sight velocity is obtained by integrating \eqref{eq:vlos2} along the line of sight weighting by density:\footnote{Note that we use the same Greek letter $\Sigma$ to denote both surface density and the summation symbol. The two can be distinguished since the latter is always accompanied by an index of summation ($i$ or $j$), while the former never is.}
\begin{equation} \label{eq:vlos3}
\Sigma \mu_{\rm los}^2 = \int_{-\infty}^{\infty} \rho \left( \overline{v_\phi^2} \cos^2 \phi + \overline{v_R^2} \sin^2 \phi \right) \di s \,,
\end{equation}
where $s$ indicates the distance along the line of sight and the surface density is defined as:
\begin{equation} \label{eq:sigma}
\Sigma = \int_{-\infty}^{\infty} \rho \di s \,.
\end{equation}

\subsection{Fitting procedure} \label{sec:fitting}

We calculate the likelihood of a model as follows:
\begin{enumerate}
\item For each bin $j$ (see Figures \ref{fig:apogee_02} and \ref{fig:SiO_02}), we calculate the mean line-of-sight velocity and the mean square line-of-sight velocity from the observed sample as:
\begin{align}
\langle {v}_{\rm los,obs} \rangle_j  & = \frac{1}{N_j} \sum_{i=0}^{N_j} v_{ {\rm los,obs},i} \,, \label{eq:vlosmean}\\
\mu_{ {\rm obs},j}^2 \equiv \langle {v}_{\rm los,obs}^2 \rangle_j & =  \frac{1}{N_j} \sum_{i=0}^{N_j} v_{ {\rm los,obs},i}^2  \label{eq:vlos2mean} 
\end{align}
where the sum is extended over all stars contained in the bin, the index $j$ labels the bins, the index $i$ labels individual stars, $v_{ {\rm los,obs},i}$ is the observed line-of-sight velocity of the star $i$ and $N_j$ is the number of stars in the bin. We use the notation $\langle \cdot \rangle_j = \sum_i \cdot / N_j $ to denote averages over the bin $j$, while we reserve the overline symbol $\bar{\cdot}$ to denote averages over the DF (e.g.\ Equation \ref{eq:vmean}). The values of $N_j$, $\langle {v}_{\rm los,obs} \rangle_j $ and $\mu_{{\rm obs},j}$ calculated in this way are shown in Figures \ref{fig:apogee_02} and \ref{fig:SiO_02} for the APOGEE stars and SiO maser stars respectively.
\item For each bin $j$, we calculate the second moment of the line-of-sight velocity $\mu_{ {\rm los},i}^2$ at the on-sky position of each individual star $i$ within the bin by performing the integrals in Equations \eqref{eq:vlos3} and \eqref{eq:sigma}. Then we average these over the bin:
\begin{align} \label{eq:mudoublemean}
\mu_{ {\rm model},j}^2 \equiv \langle \mu_{\rm los}^2 \rangle_j   =  \frac{1 }{N_j}\sum_{i=0}^{N_j} \mu_{ {\rm los},i}^2 \,.
\end{align}
Given the small number of stars in each bin, the quantity \eqref{eq:mudoublemean} is a reasonable proxy of the observed quantity \eqref{eq:vlos2mean}.
\item We assume that the estimates \eqref{eq:vlos2mean} are normally distributed about their true values. Then the likelihood of a model is:
\begin{equation} \label{eq:P}
P = \exp\left( -\chi^2/2\right) \,,
\end{equation}
where
\begin{equation} \label{eq:chi2}
\chi^2 = \sum_j  \left[ \frac{  \mu_{{\rm obs}, j} - \mu_{{\rm model}, j}   }{ \Delta \mu_j}\right]^2 \,,
\end{equation}
where the sum is extended over all bins and $\Delta \mu_j$ is the error on $\mu_{{\rm obs}, j}$, which we estimate as
\begin{equation}
\Delta \mu_j =\frac{\mu_{{\rm obs},j} }{\sqrt{N_j}} \,. \end{equation}
\end{enumerate}

\subsection{Gravitational potential and density distribution} \label{sec:mass}

The Jeans equations \eqref{eq:Jz3} and \eqref{eq:JR3} require assuming a gravitational potential $\Phi(R,z)$ and a density distribution $\rho(R,z)$ in order to generate predictions for $\overline{v_z^2}(R,z)$ and $\overline{v_\phi^2}(R,z)$. Note that, as mentioned in Section \ref{sec:basic}, for equations \eqref{eq:Jz3} and \eqref{eq:JR3} to be valid it is not necessary that $\Phi(R,z)$ is the potential self-consistently generated by $\rho(R,z)$. We will consider both models in which $\Phi(R,z)$ is generated by $\rho(R,z)$ and models in which it is not.

\subsubsection{Gravitational potential} \label{sec:Phi}

At the range of Galactocentric radii considered in this paper only two components contribute significantly to the potential: the NSD, which dominates the potential at $30\pc \lesssim R\lesssim 300\pc$,  and the nuclear stellar cluster (NSC), which dominates the potential at $1\pc \lesssim R\lesssim 30\pc$ \citep[e.g.][]{Launhardt+2002,Schoedel+2014,GallegoCano+2020}. Therefore we take a gravitational potential of the following form:
\begin{equation} \label{eq:Phi}
\Phi(R,z) =  \alpha \Phi_{\rm NSD}(R,z) + \beta \Phi_{\rm NSC}(R,z) \,,
\end{equation}
where parameter $\alpha$ is the mass scaling of the NSD and will be left as a free parameter in our fitting procedure below. The value $\alpha=1$ will correspond to the normalisations of $\rho_{\rm NSD}$ as given below in this section. The parameter $\beta$ is the mass scaling of the NSC, which we keep fixed in all our fitting procedures. We will consider models with ($\beta=1$) and without ($\beta=0$) the NSC.

To calculate the potential $\Phi_{\rm NSD}(R,z)$ we consider three different NSD models, the diversity of which reflects the current large uncertainties in the mass distribution of the NSD. The first is the best-fitting model from \citet{Launhardt+2002} (see their Section 5.2; see also Equation 1 in \citealt{Li+2020}):
\begin{align}  \label{eq:NSDL02}
\rho_{\rm NSD}(R,z) 	& = \rho_1 \exp\left\{-\log(2) \left[ \left(\frac{R}{R_1}\right)^{n_R} + \left( \frac{|z|}{z_0} \right)^{n_z}\right] \right\} \nonumber \\
		& + \rho_2 \exp\left\{-\log(2) \left[ \left(\frac{R}{R_2}\right)^{n_R} + \left( \frac{|z|}{z_0} \right)^{n_z} \right] \right\} \,,
\end{align}
where $n_R=5$, $n_z=1.4$, $R_1=120\pc$, $R_2=220\pc$, $z_0=45\pc$, $\rho_1/\rho_2=3.9$, $\rho_1=15.2 \times 10^{10} \Msun \kpc^{-3}$ and $\log(2)\simeq 0.693$. \citet{Launhardt+2002} showed that this model fits well the COBE 4.9\,\micron\, emission. The vertical scale-height $z_0$ has been independently confirmed from star counts by \citet{Nishiyama+2013}.

The second NSD model we consider is the best-fitting axisymmetric model of \citet{Chatzopoulos+2015}, which has a density distribution given by (see their Equation 17):
\begin{align} \label{eq:NSDC15}  
\rho_{\rm NSD}(R,z) 	& = \frac{(3-\gamma)M}{4\pi q}\frac{a_0}{a^{\gamma} (a+a_0)^{4-\gamma}} \,,
\end{align}
where
\begin{equation} \label{eq:a}
a(R,z) = \sqrt{R^2 + \frac{z^2}{q^2}} \,,
\end{equation}
and $\gamma=0.07$, $q=0.28$, $a_0=182\pc$, and $M=6.2\times10^9 \Msun$. Note that this is only the second component from Equation (17) of \citet{Chatzopoulos+2015}, while the first component represents the NSC (see below).

\begin{figure}
	\includegraphics[width=\columnwidth]{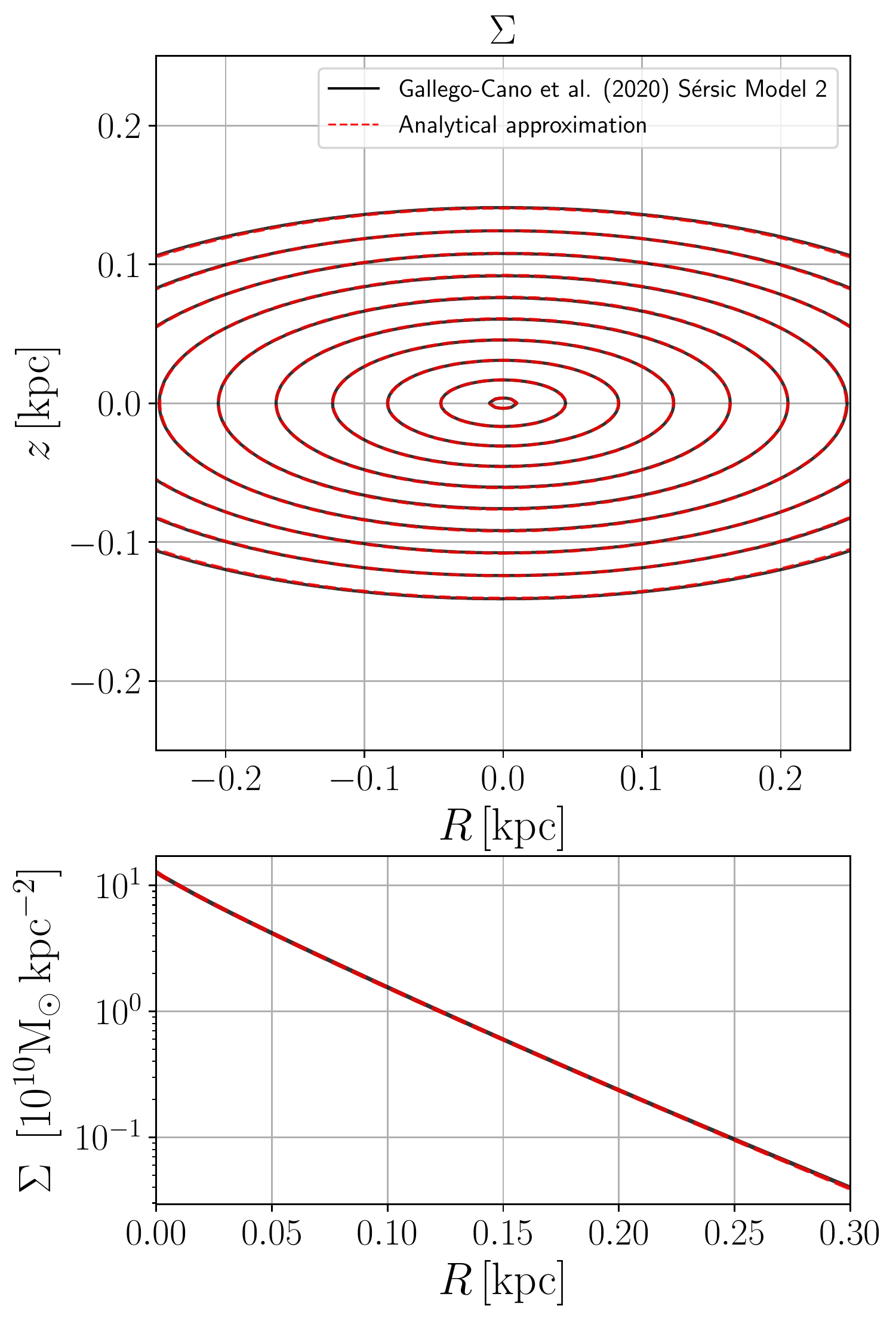}
    \caption{Comparison between S\'ersic Model 2 of \citet{GallegoCano+2020} (full black lines) and the analytical approximation given by Equation \eqref{eq:NSDGC2} (dashed red lines). \emph{Top panel}: contours of surface density $\Sigma(R,z)$. \emph{Bottom panel}: surface density radial profile in the plane $z=0$. The normalisation of both models is arbitrarily chosen so that $\Sigma=2\times 10^{10} \Msun \kpc^{-3}$ at the centre.}
    \label{fig:gc2analytic}
\end{figure}

The third NSD model we consider is obtained by deprojecting Model 2 of \citet{GallegoCano+2020} (see their Equation 3 and their Table 4). These authors have fitted a S\'ersic profile to the Spitzer/IRAC 4.5 $\micron$ stellar surface density maps of the central $300\pc \times 250\pc$. Their models gives a projected density $\Sigma(R,z)$ that can be deprojected to obtain the 3D density distribution $\rho(R,z)$. For an edge-on-system, this deprojection is unique and can be done using the Abel transform (see for example Appendix A in \citealt{Mamon+2010}). The following analytical density distribution gives an excellent approximation to the unique deprojected density distribution:
\begin{equation} \label{eq:NSDGC2}
\rho_{\rm NSD}(R,z) = \rho_1 \exp\left[ -\left(\frac{a}{R_1}\right)^{n_1} \right] + \rho_2 \exp\left[ -\left(\frac{a}{R_2}\right)^{n_2} \right] \,,
\end{equation}
where $a(R,z)$ is defined as in Equation \eqref{eq:a}, $q=0.37$, $n_1=0.72$, $n_2=0.79$, $R_1=5.06\pc$, $R_2=24.6\pc$, $\rho_1/\rho_2=1.311$ and $\rho_2=170 \times 10^{10} \Msun \kpc^{-3}$. Since \citet{GallegoCano+2020} normalise their model using observed intensity and not surface density, we choose the arbitrary normalisation $\rho_2$ by requiring that the surface density is $\Sigma=2\times 10^{10} \Msun \kpc^{-3}$ at the centre. The normalisation with respect to this value, quantified by the parameter $\alpha$, will be determined by the fitting procedure in Section \ref{sec:results}. Figure \ref{fig:gc2analytic} shows that there is excellent agreement between the surface density of Model 2 of \citet{GallegoCano+2020} and that obtained with Equation \eqref{eq:NSDGC2}.

\begin{figure}
	\includegraphics[width=\columnwidth]{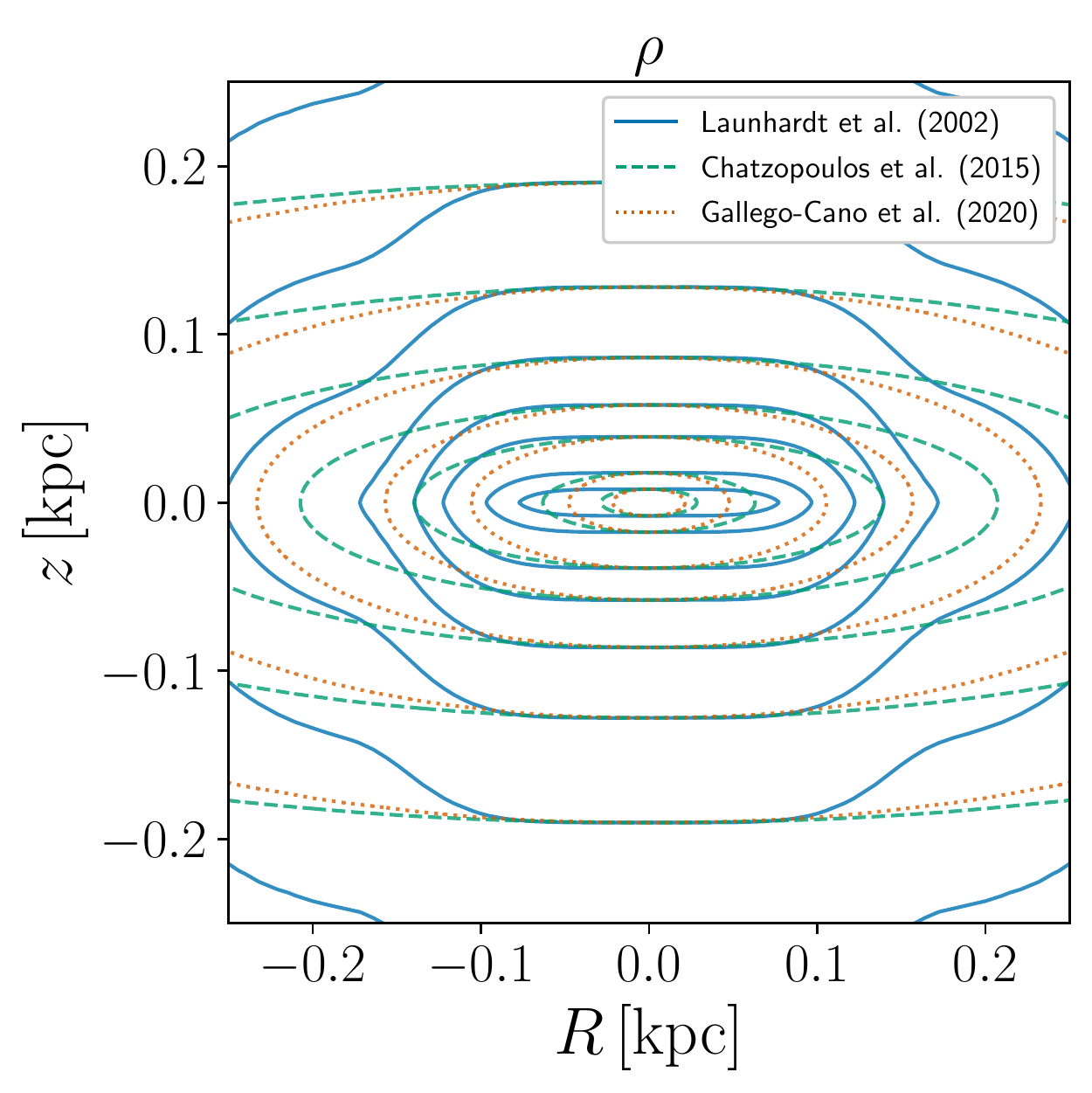}
	\includegraphics[width=\columnwidth]{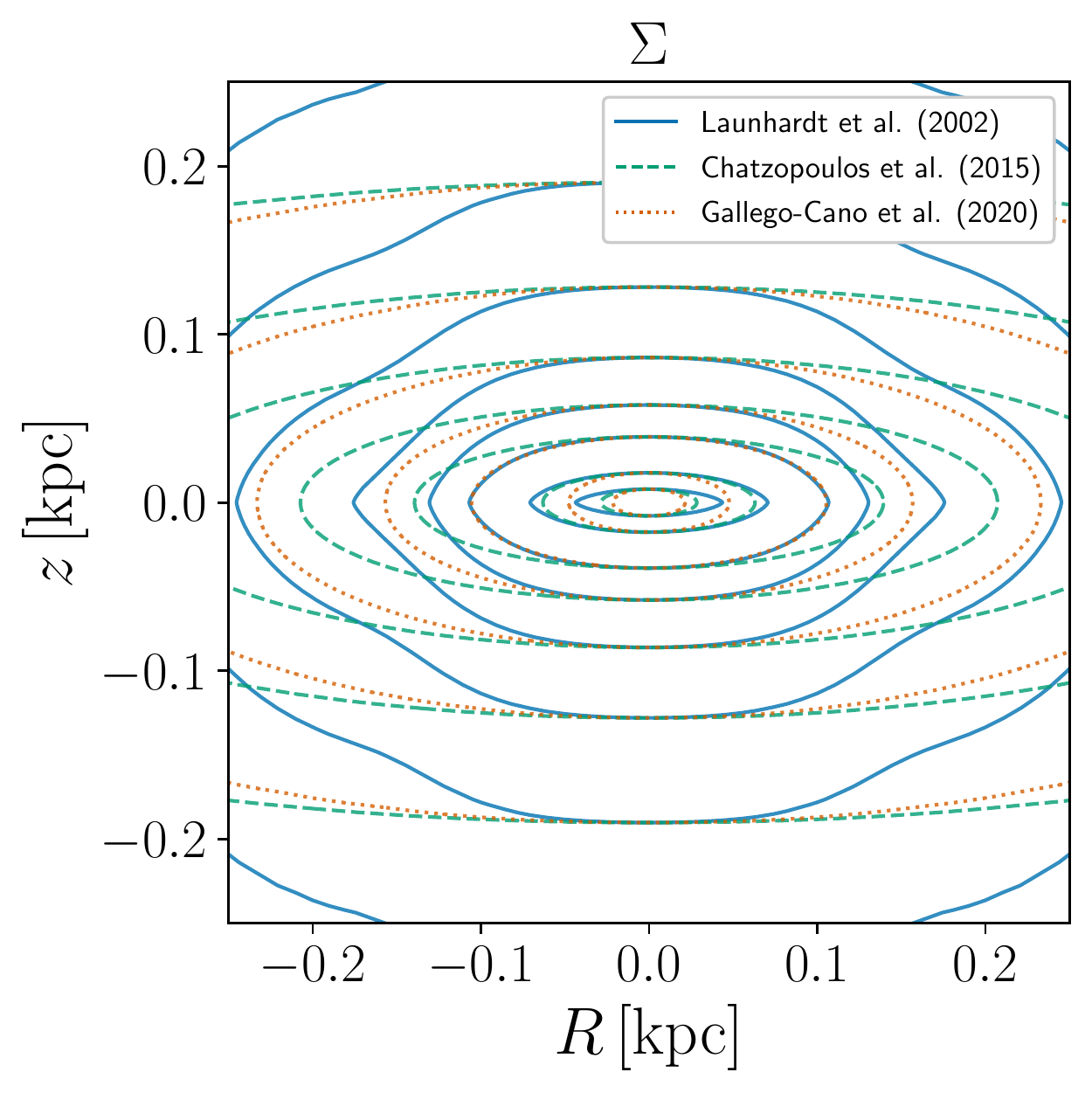}
    \caption{Comparison of the three NSD models considered in this paper (see Section \ref{sec:Phi}). \emph{Top}: density $\rho(R,z)$. \emph{Bottom}: surface density $\Sigma(R,z)$.}
    \label{fig:compareNSD}
\end{figure}

Figure \ref{fig:compareNSD} compares the three NSD models described above. It can be seen that they have rather different density contours. Since the \citet{GallegoCano+2020} NSD is obtained using data at much higher resolution than those of \citet{Launhardt+2002} and \citet{Chatzopoulos+2015}, we consider it is the most accurate of the three. We will see in Section \ref{sec:results} that the main results of this paper are not affected by the choice of the NSD model.

To calculate the potential $\Phi_{\rm NSC}(R,z)$ generated by the NSC, we adopt the mass density of the best-fitting axisymmetric model from \citet{Chatzopoulos+2015} (see their Equation 17):
\begin{align}
\rho_{\rm NSC}(R,z) 	& = \frac{(3-\gamma)M_{\rm NSC}}{4\pi q}\frac{a_0}{a^{\gamma} (a+a_0)^{4-\gamma}} \,,\label{eq:NSC}  
\end{align}
where
\begin{equation}
a(R,z) = \sqrt{R^2 + \frac{z^2}{q^2}} \,,
\end{equation}
and $\gamma=0.71$, $q=0.73$, $a_0=5.9\pc$, and $M_{\rm NSC}=6.1\times10^7 \Msun$.This corresponds to the first component from Equation (17) of \citet{Chatzopoulos+2015}, while the second component corresponds to the NSD as mentioned above.

\subsubsection{Tracer density distribution} \label{sec:rho}

For the density distribution $\rho(R,z)$ we consider two cases:
\begin{enumerate}
\item The stellar populations traced by our data (Section \ref{sec:data}) are distributed in the same way as the stars that make up most of the mass of the NSD/NSC. In this case, we take for $\rho(R,z)$ in Equations \eqref{eq:Jz3} and \eqref{eq:JR3} the density distribution that generates the potential $\Phi(R,z)$ given in Equation \eqref{eq:Phi}, i.e.
\begin{equation} \label{eq:rhofull}
\rho(R,z) = \alpha \rho_{\rm NSD}(R,z) + \beta \rho_{\rm NSC}(R,z)\,.
\end{equation}
\item The stellar populations traced by our data (Section \ref{sec:data}) are distributed differently than stars that make up most of the mass of the NSD/NSC. Indeed, the selection function of APOGEE favours younger populations of order $\sim 1 \Gyr$ of age (e.g.\ Figure 1 of \citealt{AumerSchoenrich2015}), while the SiO maser stars have estimated ages of $0.2-2\Gyr$ \citep{Habing+2006}, which may be distributed differently than the older stars that are believed to make up most of the NSD mass \citep{NoguerasLara+2020}. Since the gas in the CMZ is believed to have a ring-like morphology \citep[e.g.][]{Molinari+2011,Kruijssen+2015,Sormani+2018a,Tress+2020}, and that the distribution of relatively young stars traced by our data might still reflect the gas distribution, we consider a ring-like stellar density distribution given by:
\begin{equation} \label{eq:rho_ring}
\rho(R,z)= \rho_0 \exp \left[ - \eta \left(\frac{R_0}{R} + \frac{R}{R_0} \right) - \frac{|z|}{z_0} \right]\,,
\end{equation}
where $R_0=100\pc$ is the radius at which the density is maximum, $\eta=1$ is a parameter that controls the width of the ring, and for the vertical scale-height we take $z_0=45\pc$ as in Equation \eqref{eq:NSDL02}.
\end{enumerate}

\begin{table*}
\centering
\rowcolors{2}{gray!20}{white}
\begin{tabular}{lllllllllc}  
\toprule
model			& fitted to 			& filter			& $\Phi(R,z)$	& $\rho(R,z)$	& NSD 					& $\alpha$ 	& $b$	& $\chi^2$		& $M(r<100\pc)$ 		\\
\midrule
1				& APOGEE		& standard		& NSD+NSC 	& NSD+NSC 	& \citealt{Launhardt+2002}	& 0.8			& 0.475	& 11.74		& 4.3 				\\
2				& APOGEE		& standard		& NSD+NSC 	& NSD+NSC 	& \citealt{Chatzopoulos+2015}	& 0.875		& 0.45	& 10.79		& 5.0  				\\
3 (fiducial)			& APOGEE		& standard		& NSD+NSC 	& NSD+NSC 	& \citealt{GallegoCano+2020}	& 0.9		      	& 0.4  	& 10.73 		& 3.9  				\\
4				& SiO masers		& standard		& NSD+NSC  	& NSD+NSC 	& \citealt{Launhardt+2002}	& 0.675		& 0.8		& 0.82		& 3.7  				\\
5				& SiO masers		& standard		& NSD+NSC  	& NSD+NSC 	& \citealt{Chatzopoulos+2015}	& 0.675		& 0.925	& 0.74		& 4.0  				\\
6				& SiO masers		& standard		& NSD+NSC  	& NSD+NSC 	& \citealt{GallegoCano+2020}	& 0.85		& 0.725	& 0.80		& 3.7  				\\
\midrule
7				& APOGEE		& restrictive		& NSD+NSC 	& NSD+NSC 	& \citealt{Launhardt+2002}	& 0.85		& 0.425	& 11.91		& 4.5 	 			\\
8				& APOGEE		& restrictive		& NSD+NSC  	& NSD+NSC 	& \citealt{Chatzopoulos+2015}	& 0.925		& 0.375	& 11.84		& 5.2  				\\
9				& APOGEE		& restrictive		& NSD+NSC  	& NSD+NSC 	& \citealt{GallegoCano+2020}	& 0.875		& 0.4		& 12.79		& 3.8 				\\
10				& APOGEE		& $(l,v)$ cut		& NSD+NSC  	& NSD+NSC 	&\citealt{Launhardt+2002}		& 0.75		& 0.225	& 4.19		& 4.0 	 			\\
11				& APOGEE		& $(l,v)$ cut		& NSD+NSC  	& NSD+NSC 	& \citealt{Chatzopoulos+2015}	& 0.85		& 0.175	& 2.42		& 4.9	 				\\
12				& APOGEE		& $(l,v)$ cut		& NSD+NSC 	& NSD+NSC 	& \citealt{GallegoCano+2020}	& 0.675 		& 0.125	& 1.83		& 3.1		 			\\
\midrule
13				& APOGEE		& standard		& NSD only 	& NSD only 	& \citealt{Launhardt+2002}	& 0.925		& 0.6		& 13.22		& 4.3   				\\
14				& APOGEE		& standard		& NSD only  	& NSD only 	& \citealt{Chatzopoulos+2015}	& 0.975		& 0.625 	& 12.71		& 4.9			  		\\
15				& APOGEE		& standard		& NSD only 	& NSD only 	& \citealt{GallegoCano+2020}	& 1.175		& 0.65	& 11.80		& 4.4			  		\\
16				& APOGEE		& standard		& NSD+NSC 	& ring		& \citealt{Launhardt+2002}	& 0.95		& 0.825	& 13.47		& 5.0   				\\
17				& APOGEE		& standard		& NSD+NSC  	& ring		& \citealt{Chatzopoulos+2015}	& 0.875		& 0.625	& 13.37		& 5.0				  	\\
18				& APOGEE		& standard		& NSD+NSC  	& ring		& \citealt{GallegoCano+2020}	& 1.45		& 0.625	& 13.04		& 6.0		  			\\
\bottomrule
\end{tabular}
\caption{Best-fitting parameters for all the models considered in this paper. Each row corresponds to a panel in Figures \ref{fig:chi2_1}, \ref{fig:chi2_2} and \ref{fig:chi2_3}. Columns are defined as follows. \emph{Filter:} selection criteria used to filter the data. ``standard'' denotes the criteria as described in Sections \ref{sec:apogee} and \ref{sec:SiO} (see red dashed lines in Figures \ref{fig:apogee_01} and \ref{fig:SiO_01}). ``restrictive'' is the same as standard, but using a more restrictive color-magnitude cut which is shifted by 0.2 magnitudes (see blue lines in the third panels of Figures \ref{fig:apogee_01} and \ref{fig:SiO_01}). ``$(l,v)$ cut'' is the same as standard, but with an additional cut that excludes all stars outside the parallelogram shown in blue dashed lines in the second panels of Figures \ref{fig:apogee_01} and \ref{fig:SiO_01}. \emph{$\Phi(R,z)$}: employed gravitational potential. ``NSD+NSC'' and ``NSD only'' mean that the potential is calculated using Equation \eqref{eq:Phi} with fixed $\beta=1$ and $\beta=0$ respectively. \emph{$\rho(R,z)$}: employed tracer density distribution, which can be either the same as the one that generates the potential $\Phi(R,z)$ (Equation \ref{eq:rhofull}) or the ring distribution (Equation \ref{eq:rho_ring}). NSD: mass model used to calculate $\Phi_{\rm NSD}$ (see Section \ref{sec:Phi}).   \emph{$\alpha$}: best-fitting mass scaling of the NSD relative to the normalisation as given in Section \ref{sec:Phi}. \emph{b}: best fitting anisotropy parameter. \emph{$\chi^2$}: value for the best fitting model. \emph{$M(r<100\pc)$:} mass contained within spherical radius $r=100\pc$ in units of $10^8 \Msun$.}
\label{tab:1}
\end{table*}

\section{Results} \label{sec:results}

\begin{figure*}
	\centering 
	\includegraphics[width=0.33\textwidth]{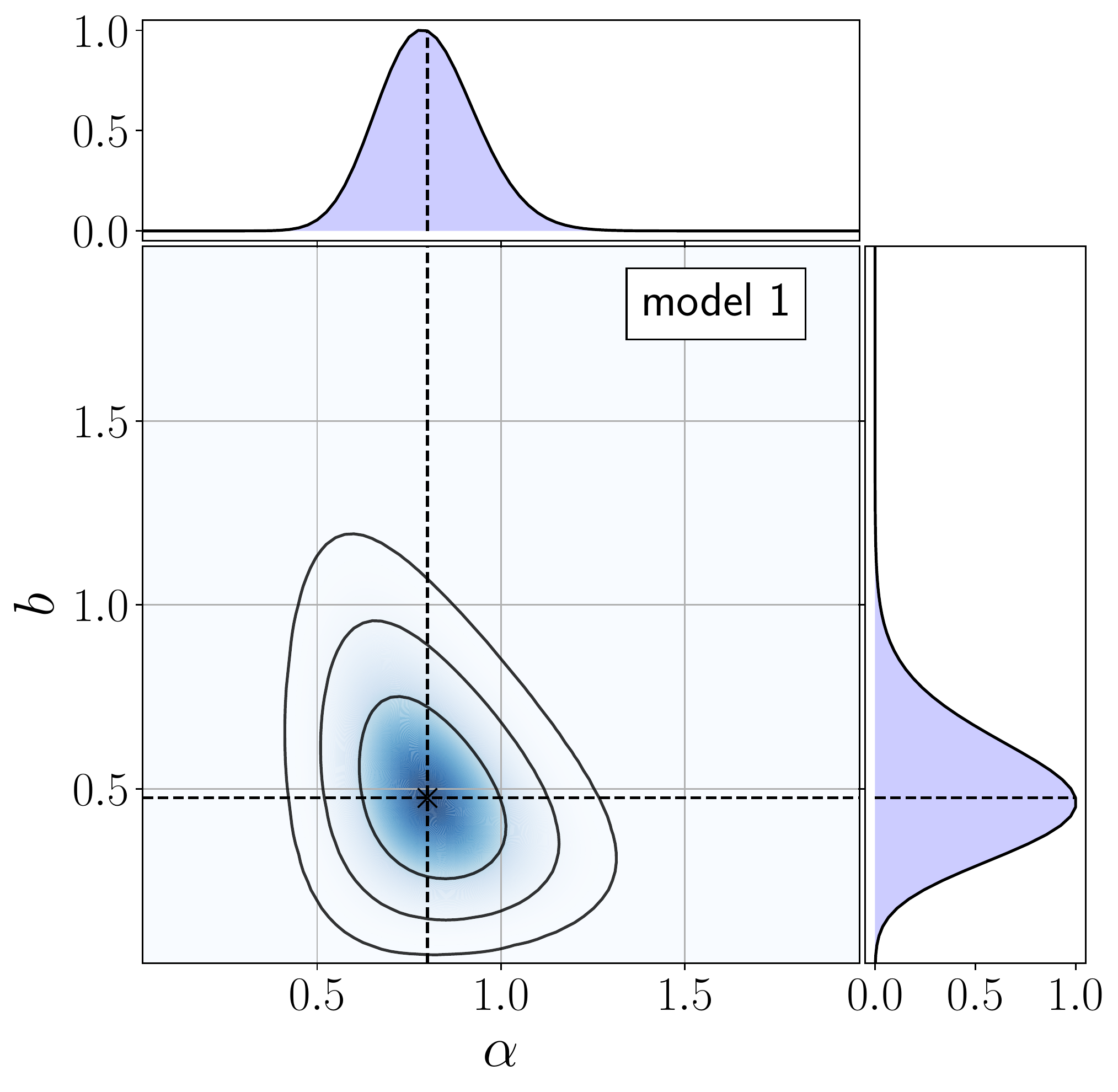}\includegraphics[width=0.33\textwidth]{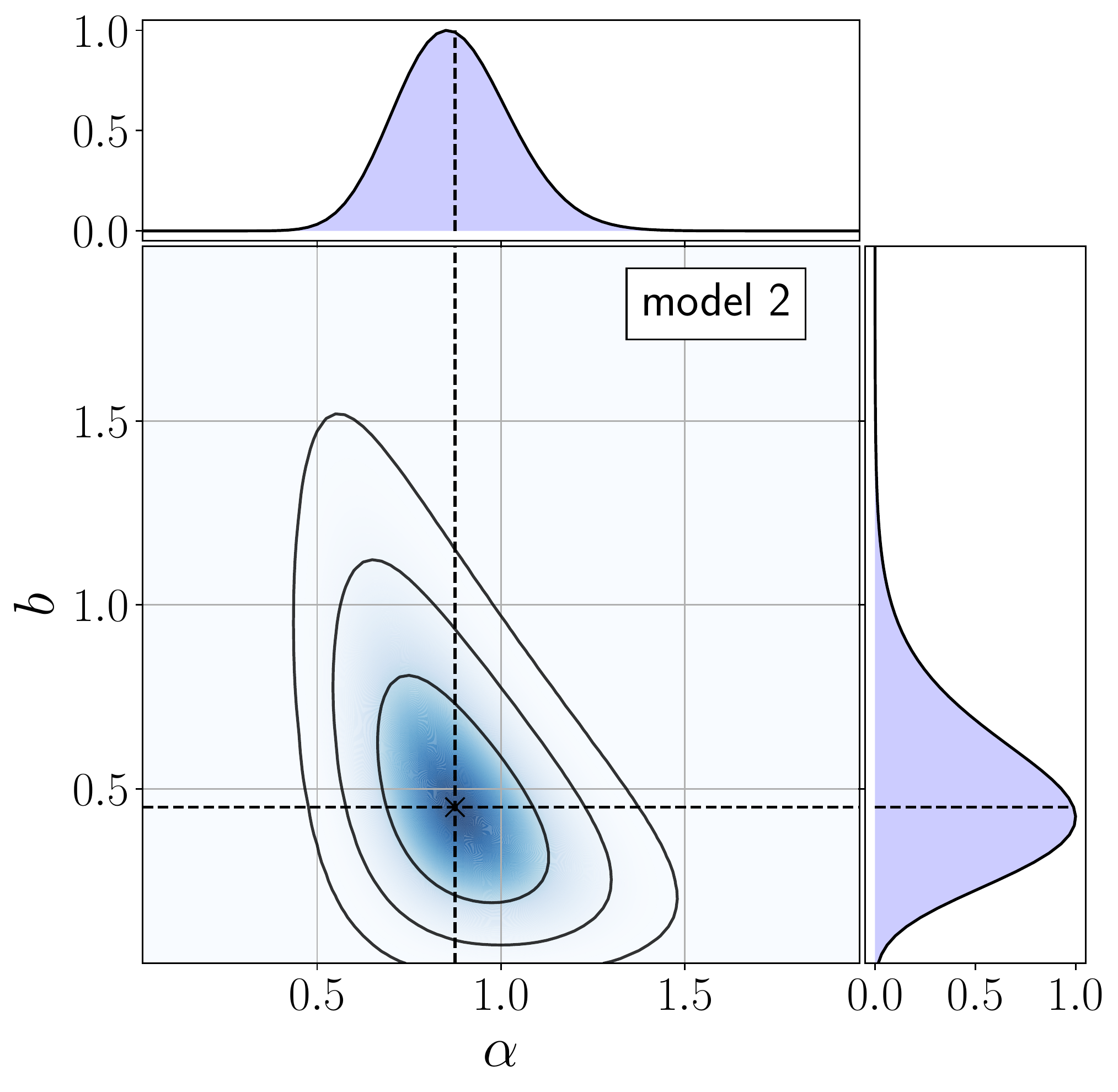}\includegraphics[width=0.33\textwidth]{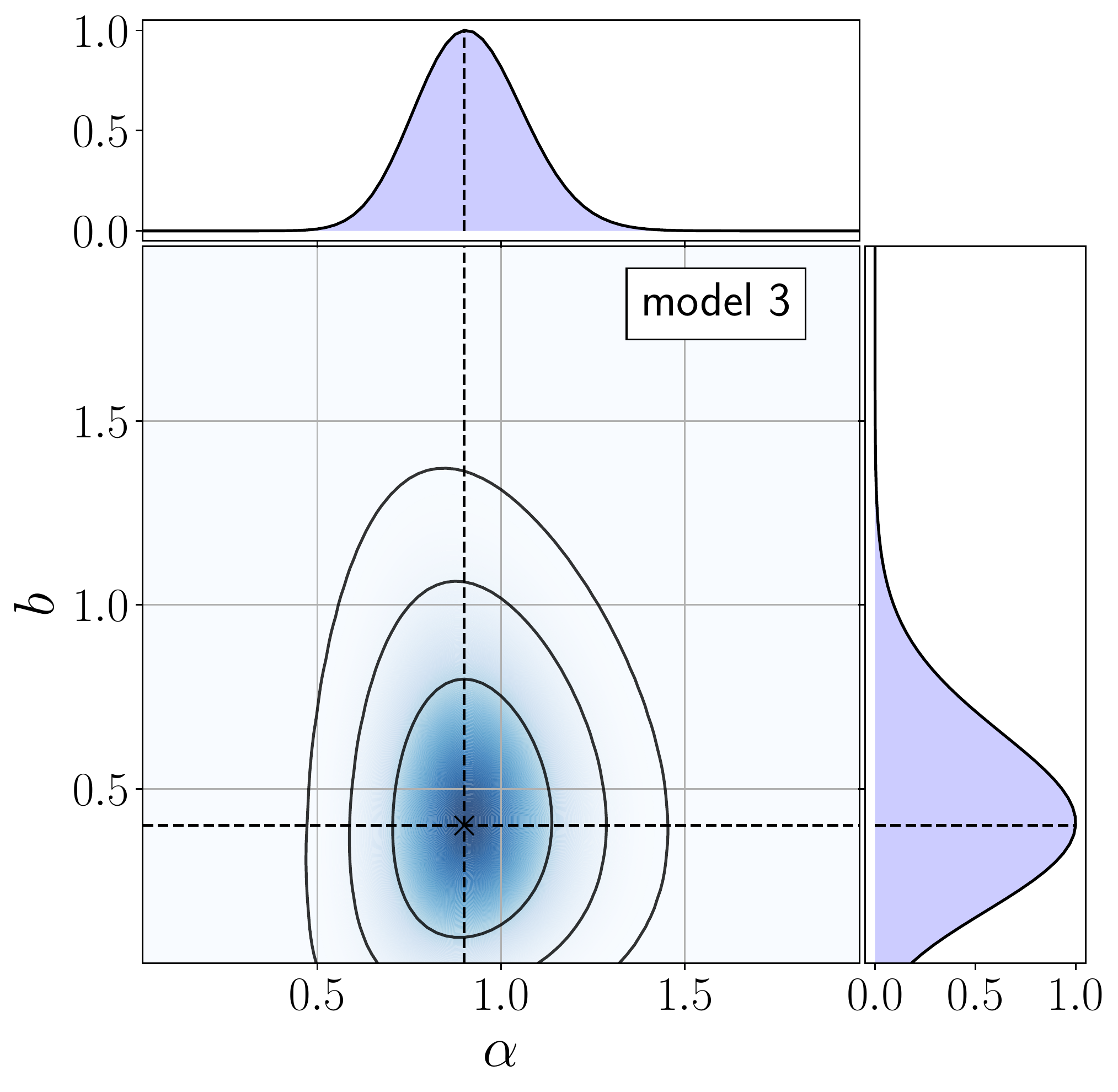}
	\includegraphics[width=0.33\textwidth]{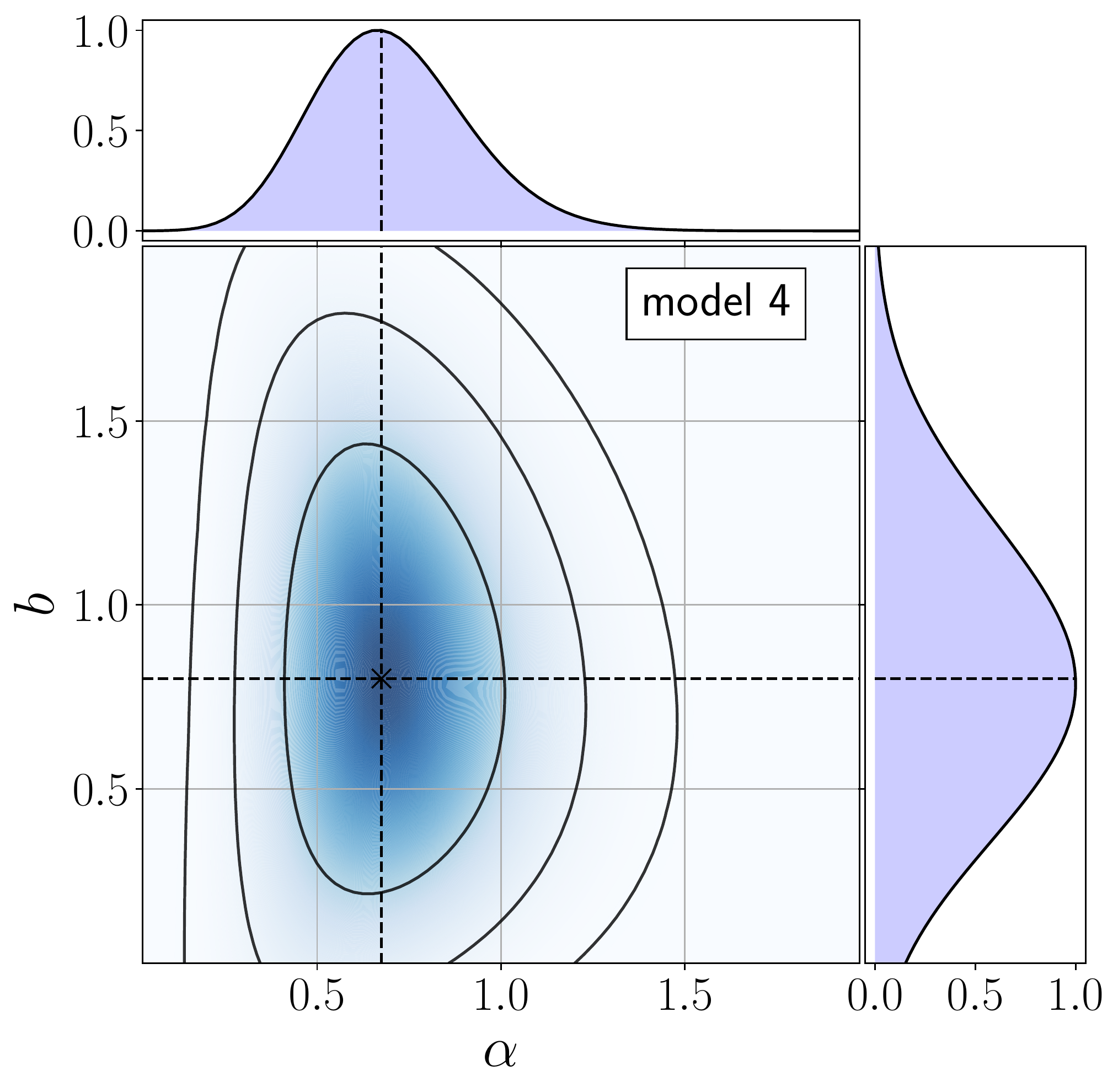}\includegraphics[width=0.33\textwidth]{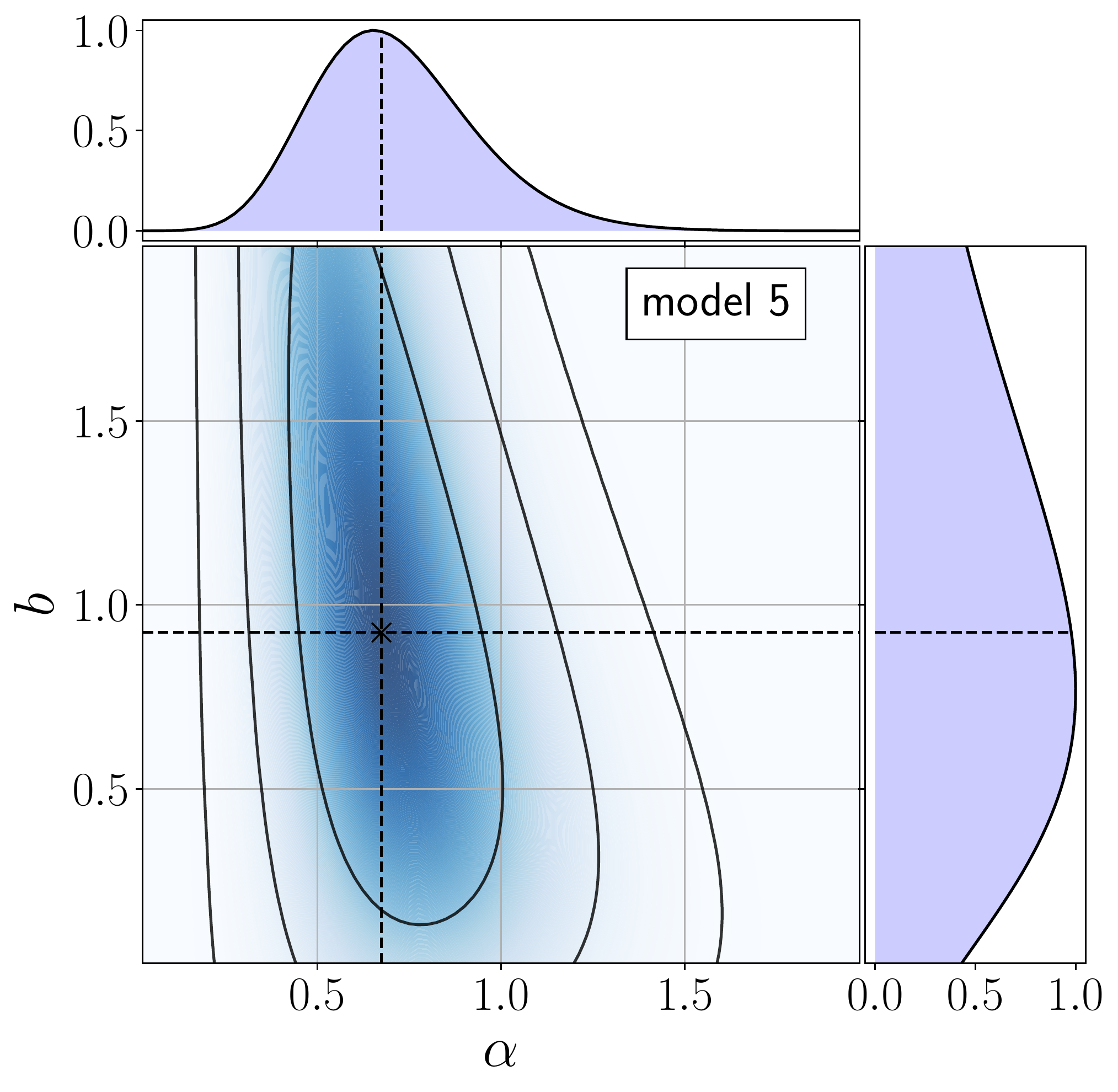}\includegraphics[width=0.33\textwidth]{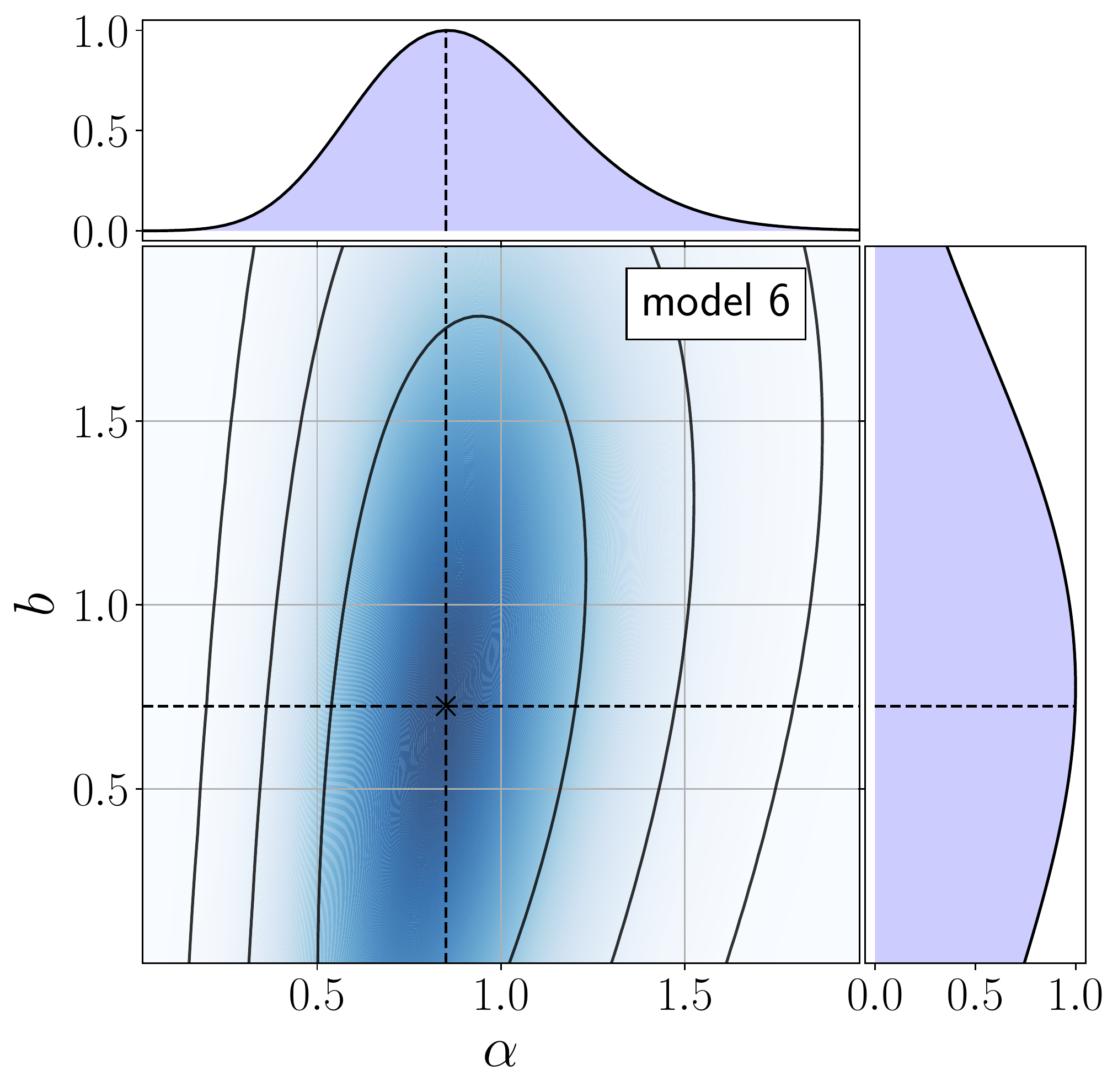}
    
    \caption{Probability distributions $P(\alpha,b)$ calculated using Equation \eqref{eq:P} for models 1-6 (see Table \ref{tab:1}). Contours show the 1-,2- and 3-$\sigma$ contours that contain $68\%,95\%,99.7\%$ of the total probability respectively. The parameter $\alpha$ is the mass normalisation of the NSD relative to the normalisations as given in Section~\ref{sec:Phi}. The parameter $b=\sigma_R^2/\sigma_z^2$ is the anisotropy parameter introduced by \citet{Cappellari2008}. The lateral panels represent the marginalised distributions $P(\alpha)=\int P(\alpha,b)\di b$ and $P(b)=\int P(\alpha,b)\di \alpha$. The crosses mark the maximum of $P$ in the 2D distributions, which correspond to the values reported in Table~\ref{tab:1}.}
    \label{fig:chi2_1}
\end{figure*}

All the models considered in this paper are listed in Table \ref{tab:1}. Figure \ref{fig:chi2_1} shows the probability distributions for models 1-6. The left, middle and right columns differ for the NSD models employed, and correspond to those of \citet{Launhardt+2002}, \citet{Chatzopoulos+2015} and \citet{GallegoCano+2020}, respectively. The top row corresponds to models which are fitted to APOGEE data, while the bottom row corresponds to models which are fitted to SiO maser data.

The top row in Figure \ref{fig:chi2_1} shows that although the three NSD models have rather different density distributions (see Figure \ref{fig:compareNSD}), they all give similar values for $M(r<100\pc)$ and for the anisotropy parameter $b$ when fitted to APOGEE data (Table \ref{tab:1}). The best fitting value for our (fiducial) model 3 is $\alpha=0.9 \pm 0.2$ which corresponds to a mass $M(r<100\pc)= (3.9\pm1) \times 10^8 \Msun$ and a total NSD mass of $M_{\rm NSD}= (6.9 \pm 2) \times 10^8 \Msun$. The anisotropy parameter is consistently $b\sim0.5$ for all models. Fitting the same model to SiO masers (bottom-left panel) gives results that are consistent with the fit to APOGEE data, but with significantly larger uncertainty (as is expected given the smaller number of stars in the SiO masers data).

To assess the impact of our data selection criteria, we follow a strategy similar to that of \citet{NoguerasLara+2020} and repeat the fits using different cuts. Models 7-9 are identical to models 1-3 except that we use a more restrictive color-magnitude cut which is shifted by 0.2 magnitudes with respect to the standard cut (see blue dashed line in Figure \ref{fig:apogee_01}). This excludes 30 additional APOGEE stars from the sample, leaving 243. Table \ref{tab:1} and Figure \ref{fig:chi2_2} show that this does not affect the results significantly. 

The second panel in Figure \ref{fig:apogee_01} displays several stars at negative (positive) longitude that have large positive (negative) line-of-sight velocities and therefore naively appear to be counter-rotating. Such stars are most likely stars on elongated $x_1$-like orbits that belong to the Galactic bar \citep{Molloy+2015,AumerSchoenrich2015}, and indeed occupy the same area in the $(l,v)$ plane as the so called ``forbidden velocity'' gas, which has a similar interpretation \citep{Binney+1991,Fux1999,SBM2015c}. Also visible in the second panel of Figure \ref{fig:apogee_01} are stars with very high line-of-sight velocities ($v_{\rm los}\geq 200\kms$), which are also most likely stars on $x_1$-type bar orbits \citep{Molloy+2015,Habing2016} and also have a gas counterpart as ``high-velocity peaks'' in the $(l,v)$ plane \citep{Binney+1991,SBM2015c}. In order to assess the potential impact of such contamination from the Galactic bar, models 10-12 repeat the fits excluding all the stars outside the blue parallelogram in the second panel of Figure \ref{fig:apogee_01}. This removes 54 APOGEE stars from the sample, leaving 219. Table \ref{tab:1} and Figure \ref{fig:chi2_2} show that this does not affect the mass normalisation significantly, but it tends to give even lower values for the anisotropy parameter $b$.

To assess the impact of including the NSC component, which is important only for $R\lesssim 30\pc$ ($|l|\lesssim0.2\degree$), we now consider models which only include the NSD potential and density. Models 13-15 are identical to models 1-3 except that we exclude the NSC by setting its normalisation to $\beta=0$ (see Equations \ref{eq:Phi} and \ref{eq:rhofull}). Table \ref{tab:1} and Figure \ref{fig:chi2_3} shows that this favours a slightly larger mass and anisotropy parameter $b$ than the NSD+NSC models, but are consistent within the uncertainties. Thus, the inclusion of the NSC does not affect the results significantly, which is reasonable given the small number of datapoints at $|l|\lesssim0.2\degree$ (Figures \ref{fig:apogee_02} and \ref{fig:SiO_02}). The best fitting NSD only model fits the data comparably well as the best NSD+NSC model. This confirms that our approach to keep the NSC mass fixed to the value determined by \citet{Chatzopoulos+2015} in our fitting procedure is reasonable. The inclusion of the NSC will be more important when better data will be available in the future.

As mentioned in Section \ref{sec:rho}, the stellar populations traced by our data might be distributed differently than stars that make up most of the mass of the NSD/NSC. In particular, given that both the selection functions of APOGEE and SiO maser stars tend to favour relatively young stellar populations with ages $\sim 1\mhyphen 2\Gyr$ (see references in Section \ref{sec:rho}), we consider a ring-like density distribution that might reflect more closely the gas distribution in the CMZ. Table \ref{tab:1} and the bottom row of Figure \ref{fig:chi2_3} show the result of fitting the ring models, which have the same potential as the NSD+NSC models but a ring-like density distribution given by Equation \eqref{eq:rho_ring}, to APOGEE data. As for the NSD only models, the ring models favour a slightly larger mass scaling parameter $\alpha$ and an anisotropy a bit higher and closer to $b\sim1$. We continue the discussion on this point in Section \ref{sec:discussion_b}.

\begin{figure*}
	\includegraphics[width=0.5\textwidth]{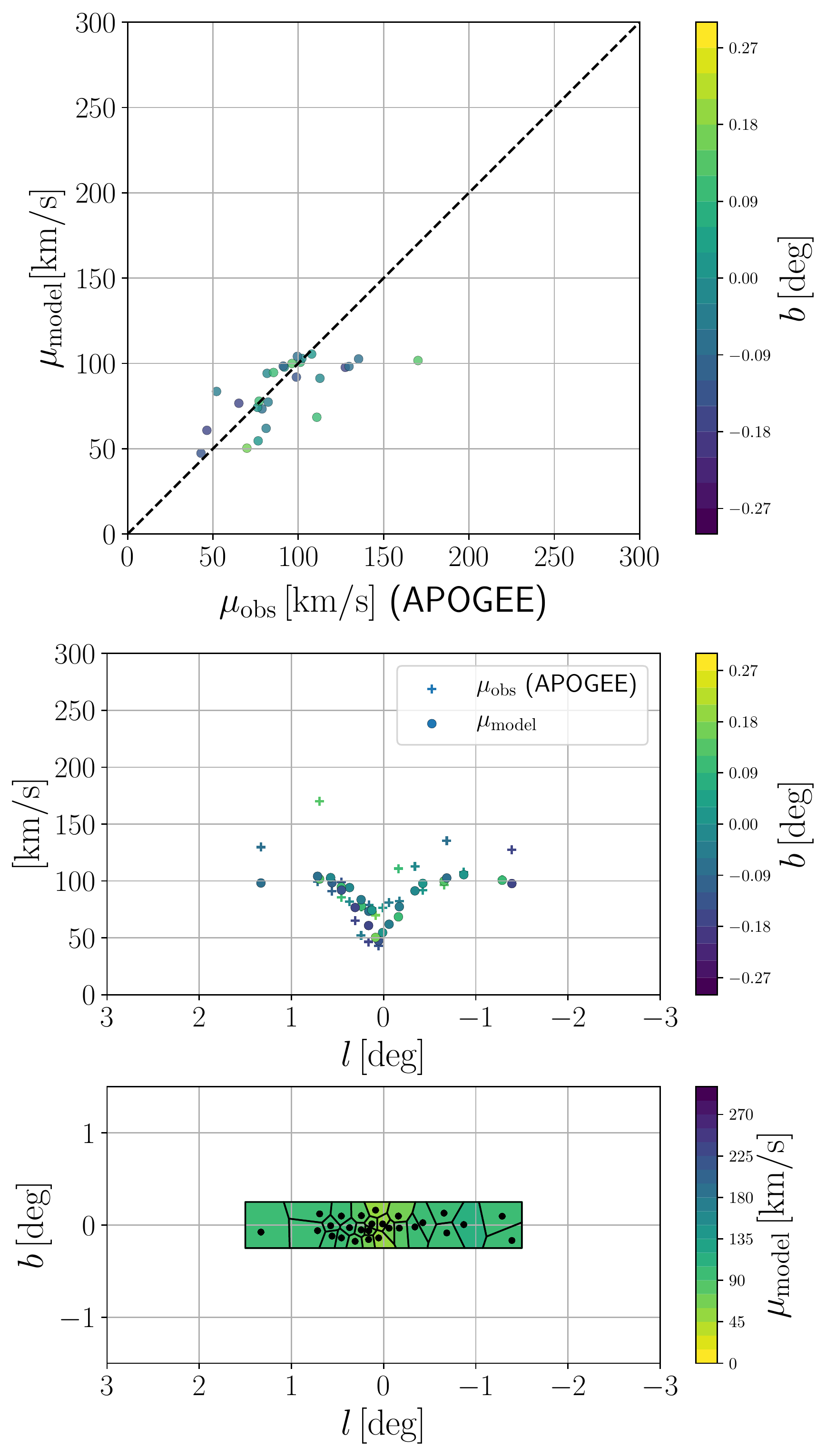}\includegraphics[width=0.5\textwidth]{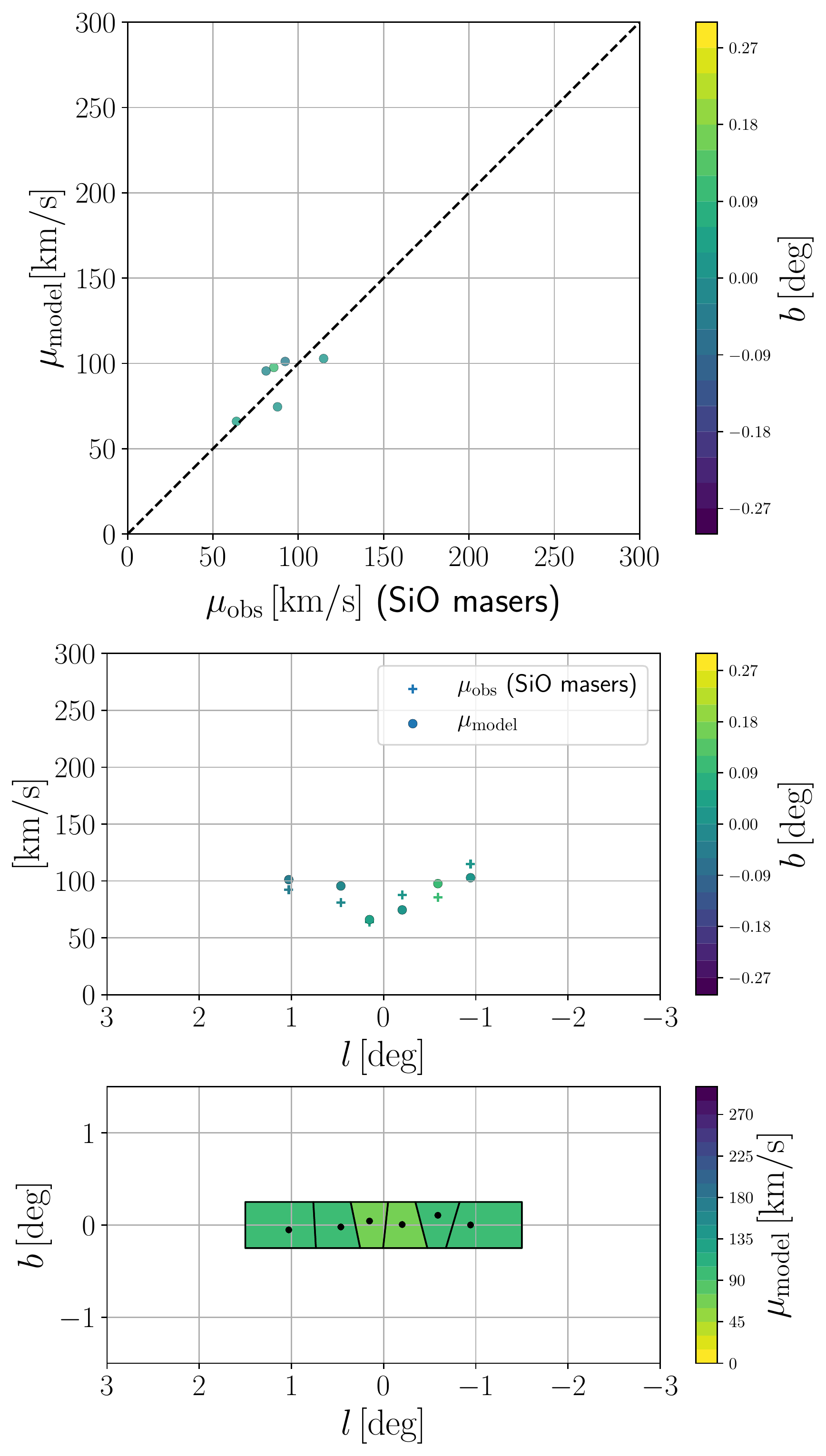}
    \caption{Comparison between observational data and model 3 (see Table \ref{tab:1}). \emph{Left}: APOGEE data. \emph{Right}: SiO maser data. \emph{Top}: observed second moment of the line-of-sight velocity (Equation \ref{eq:vlos2mean}) vs predictions (Equation \ref{eq:mudoublemean}). Each point on the left (right) panel represents one of the bins shown in Figure \ref{fig:apogee_02} (Figure \ref{fig:SiO_02}). \emph{Middle}: second moment of the line-of-sight velocity as a function of longitude for observational data (crosses) and best-fitting model (circles). \emph{Bottom}: predictions for the second moment of the line-of-sight velocity (compare with bottom panels in Figures \ref{fig:apogee_02} and \ref{fig:SiO_02}).}
    \label{fig:bestmodel_01}
\end{figure*}

Figure \ref{fig:bestmodel_01} compares our fiducial model 3 to both APOGEE and SiO maser data. The model and data show a reasonably good agreement given the quality of the data. Models 1 and 2, which employ a different NSD  mass distribution, offer comparably good representations of the data. A similar consideration applies to the NSD only models. This makes clear that the limiting factor in our analysis is the quality of the data, and not the assumed potential/density distribution. Trying to refine the potential/density distribution would not make sense until better data become available.

\begin{figure}
	\includegraphics[width=0.5\textwidth]{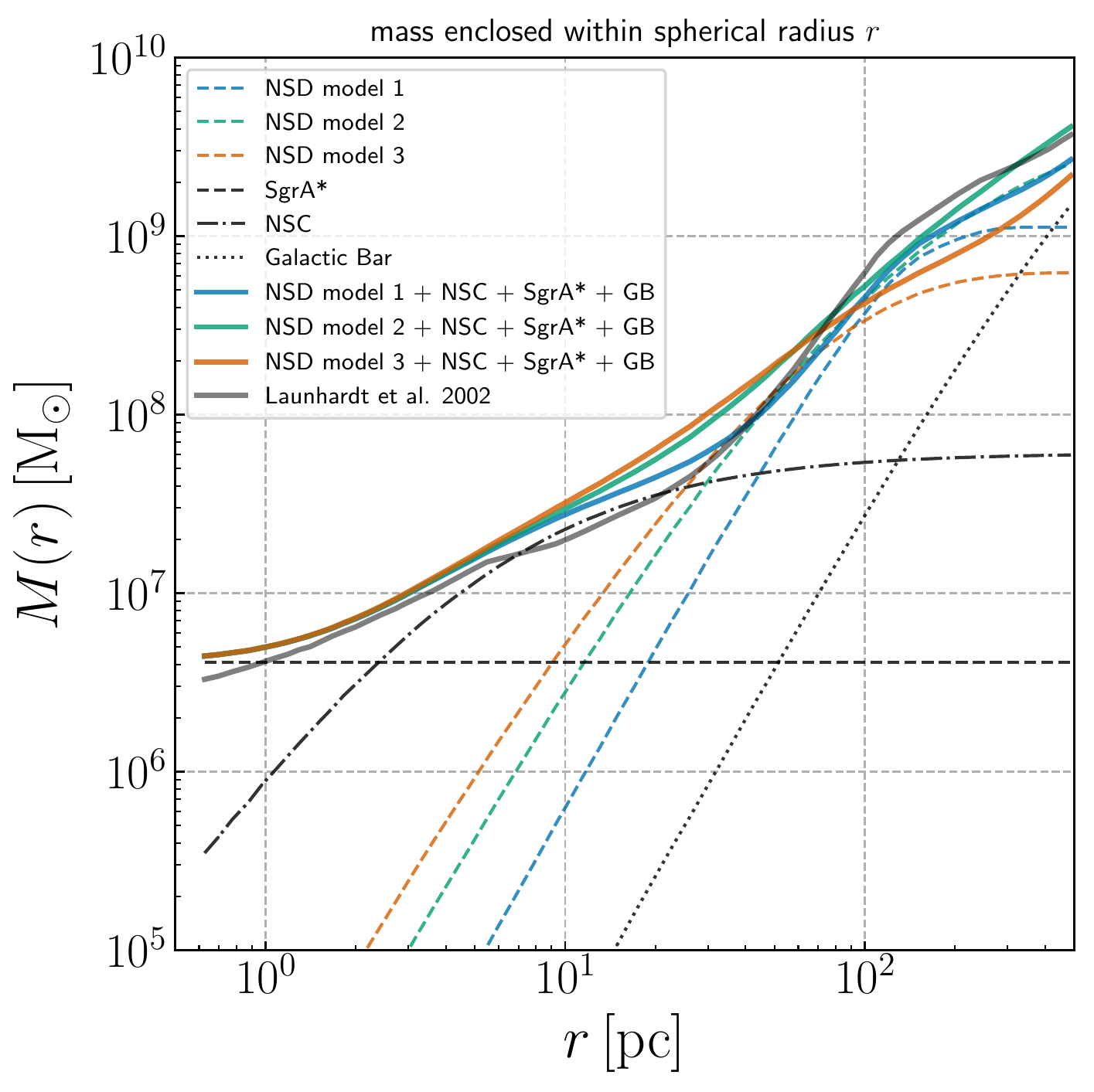}
    \caption{Mass enclosed within spherical radius $r$. \emph{Orange dashed}: best-fitting NSD of our (fiducial) model 3. \emph{Blue dashed}: best-fitting NSD of model 1. \emph{Blue dashed}: best-fitting NSD of model 1. \emph{Green dashed}: best-fitting NSD of model 2. \emph{Black dashed}: mass of the central black hole SgrA* \citep{Gravity2019}. \emph{Dash-dotted line:} NSC (Equation \ref{eq:NSC}). \emph{Dotted line:} Galactic bulge/bar (GB) model from Section 4.2 of \citet{Launhardt+2002}. \emph{Orange solid:} sum NSD from model 3+NSC+SgrA*+GB. \emph{Blue thick solid:} sum NSD from model 1+NSC+SgrA*+GB. \emph{Green thick solid:} sum NSD from model 2+NSC+SgrA*+GB. \emph{Grey thick solid:} best-fitting photometric mass from \citet{Launhardt+2002}.}
    \label{fig:Menc}
\end{figure}

\begin{figure*}
	\includegraphics[width=0.25\textwidth]{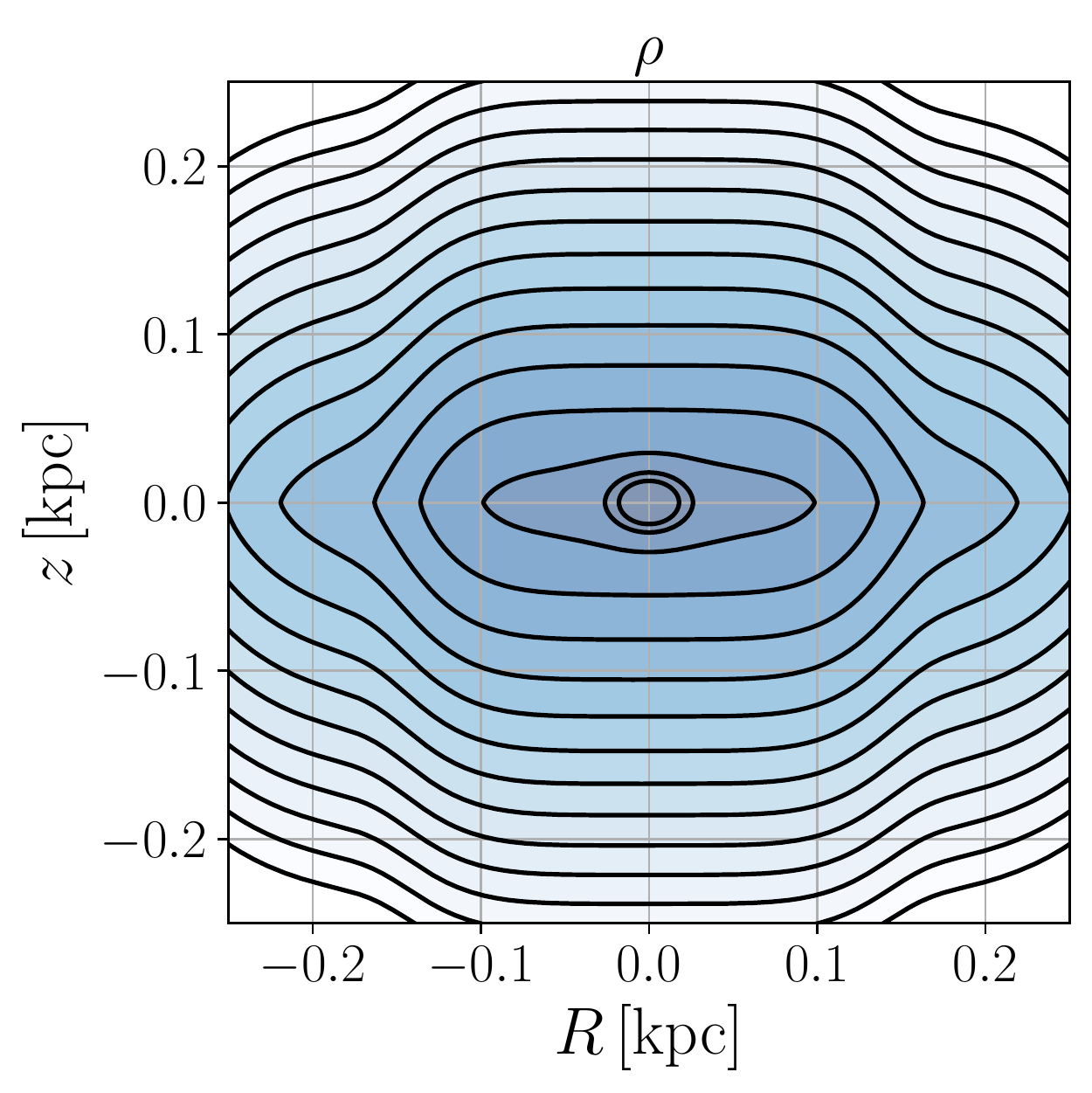}\includegraphics[width=0.25\textwidth]{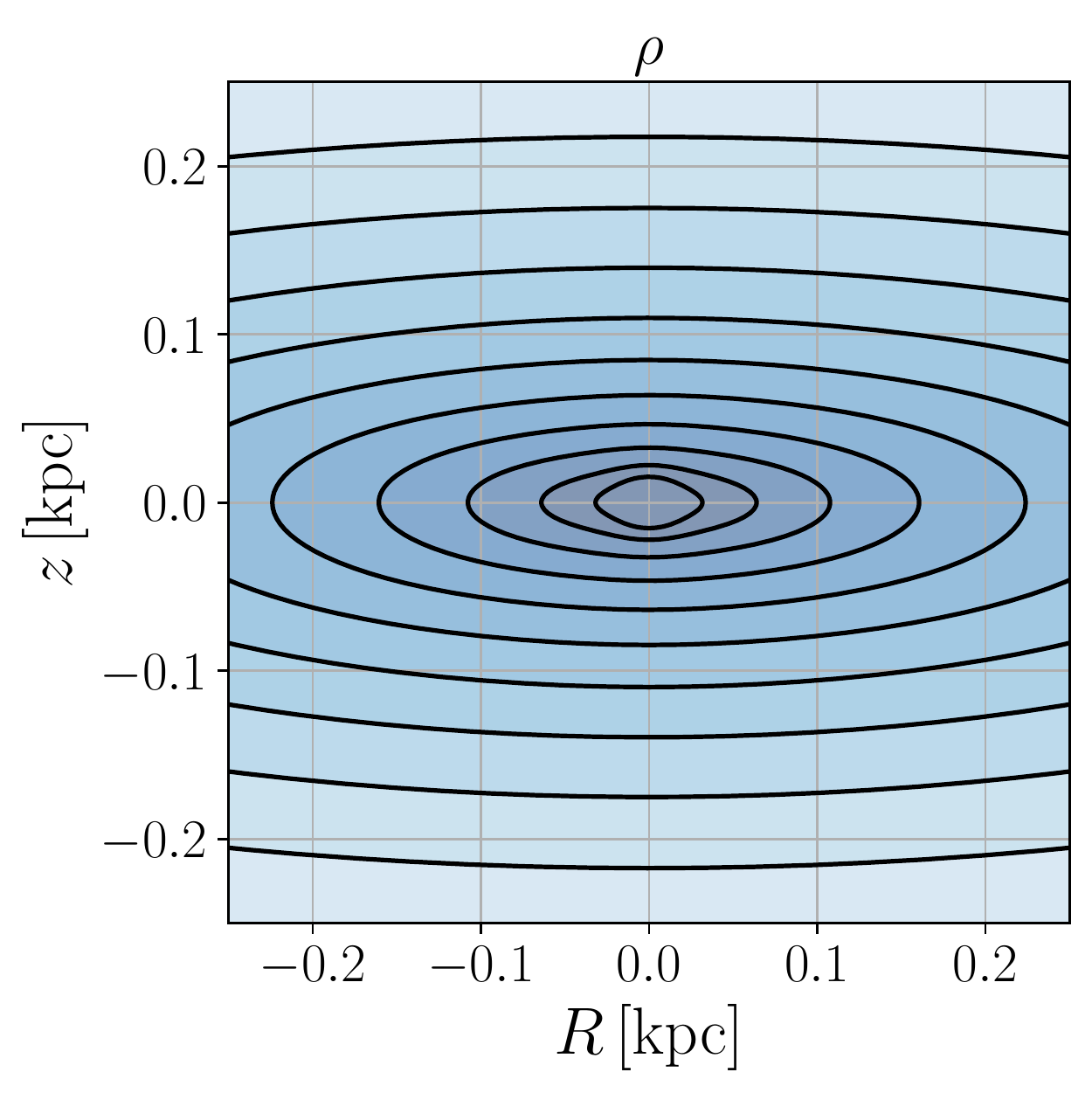}\includegraphics[width=0.25\textwidth]{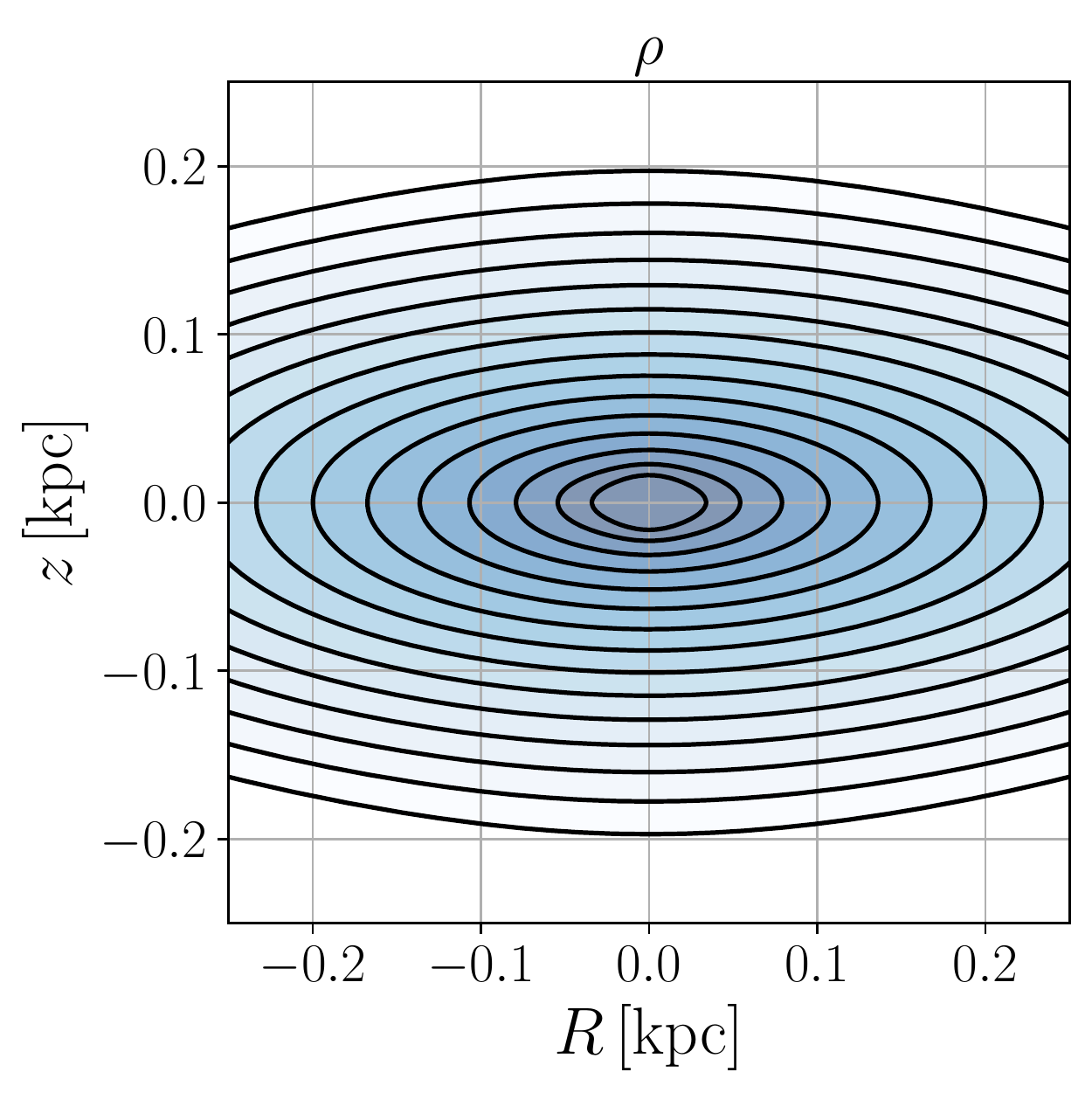}\includegraphics[width=0.25\textwidth]{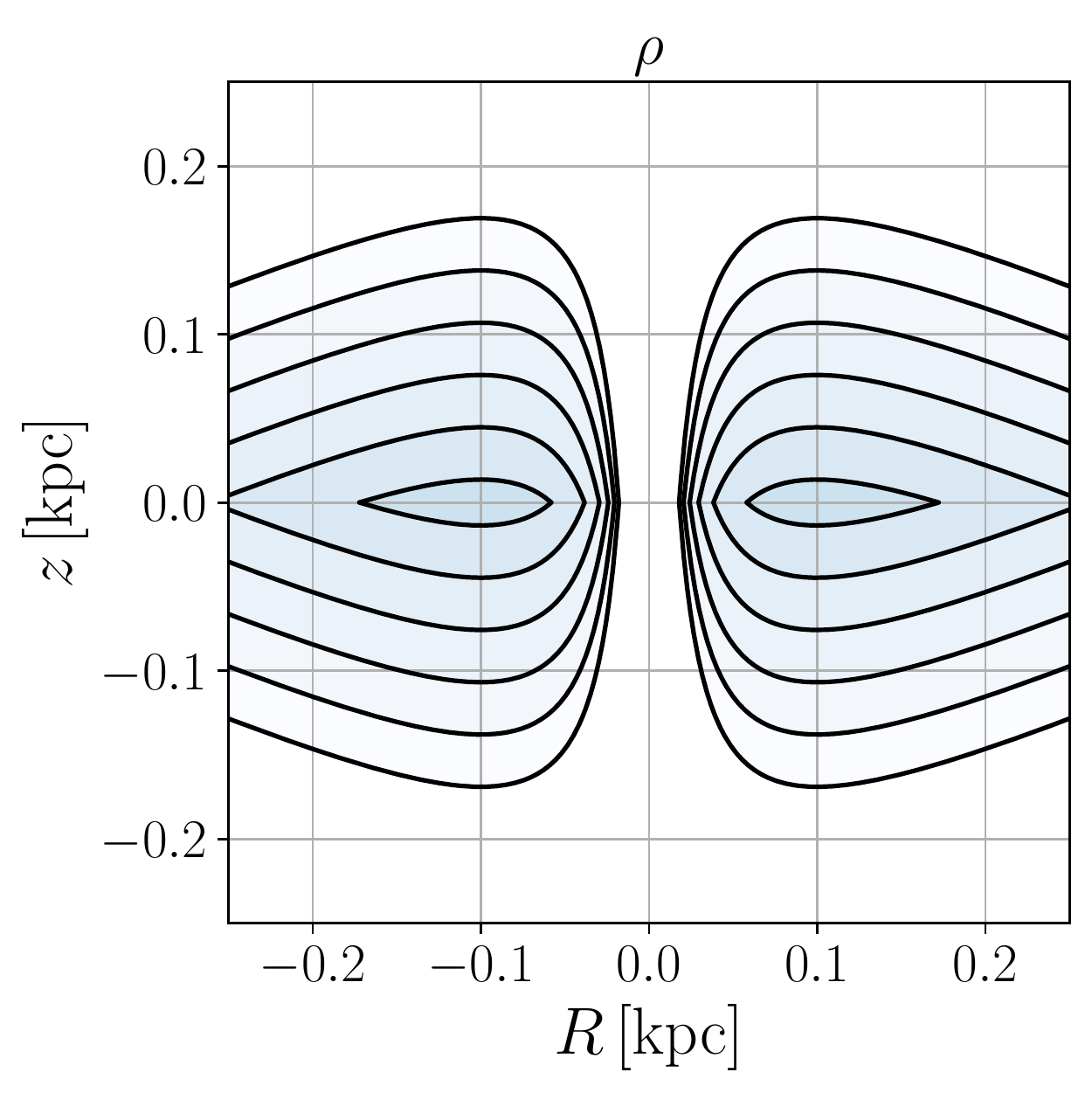}
		\includegraphics[width=0.25\textwidth]{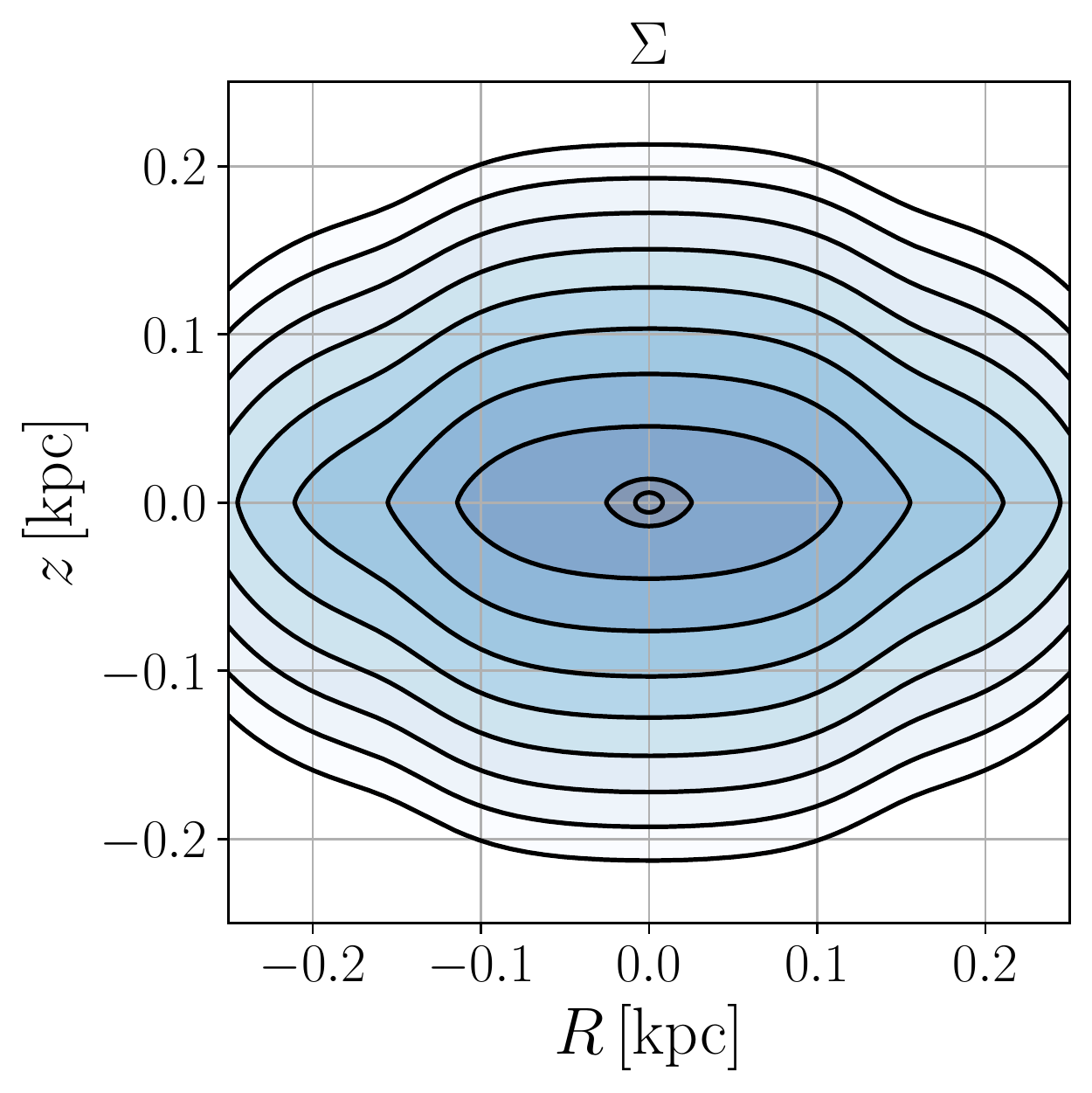}\includegraphics[width=0.25\textwidth]{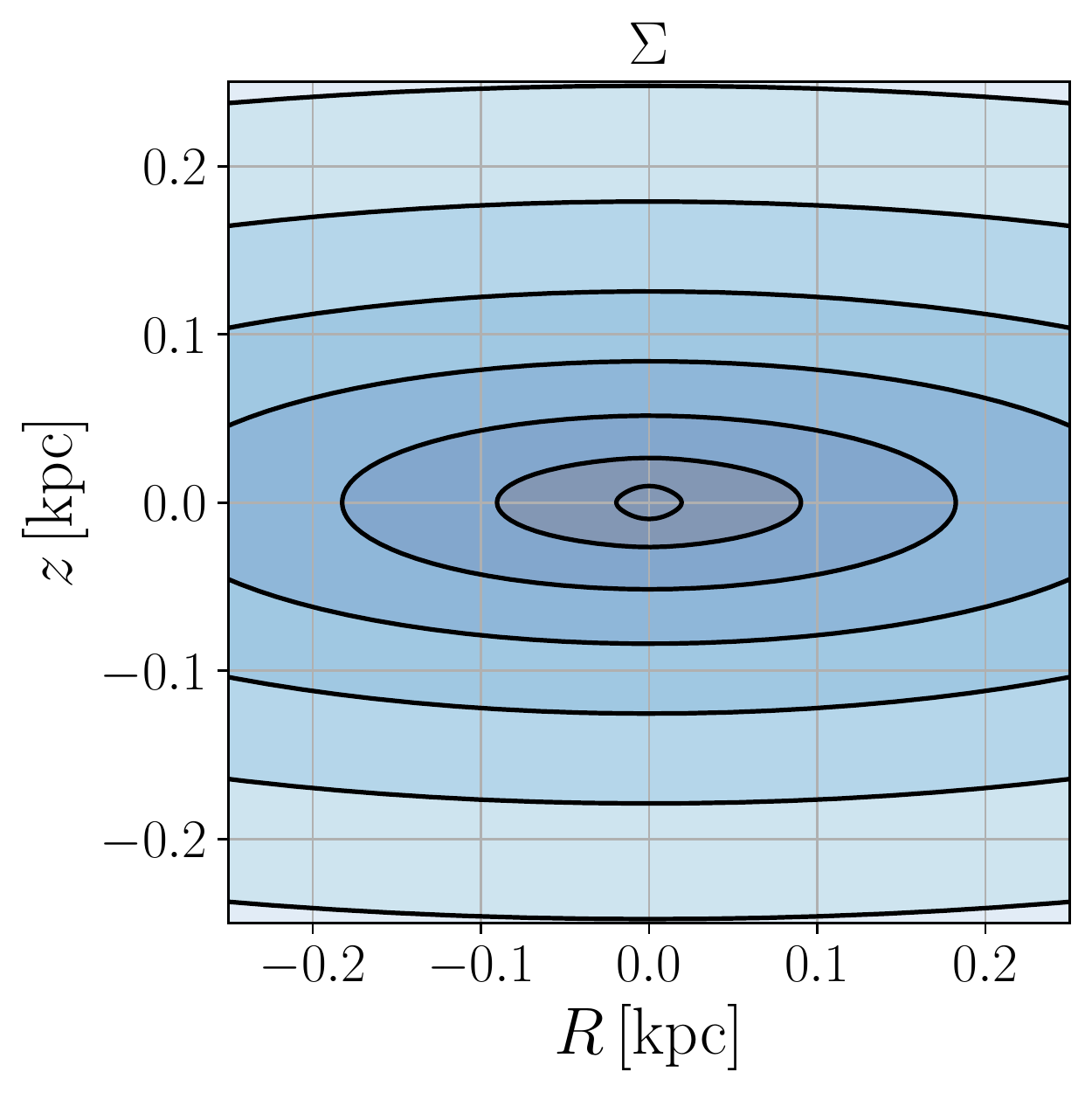}\includegraphics[width=0.25\textwidth]{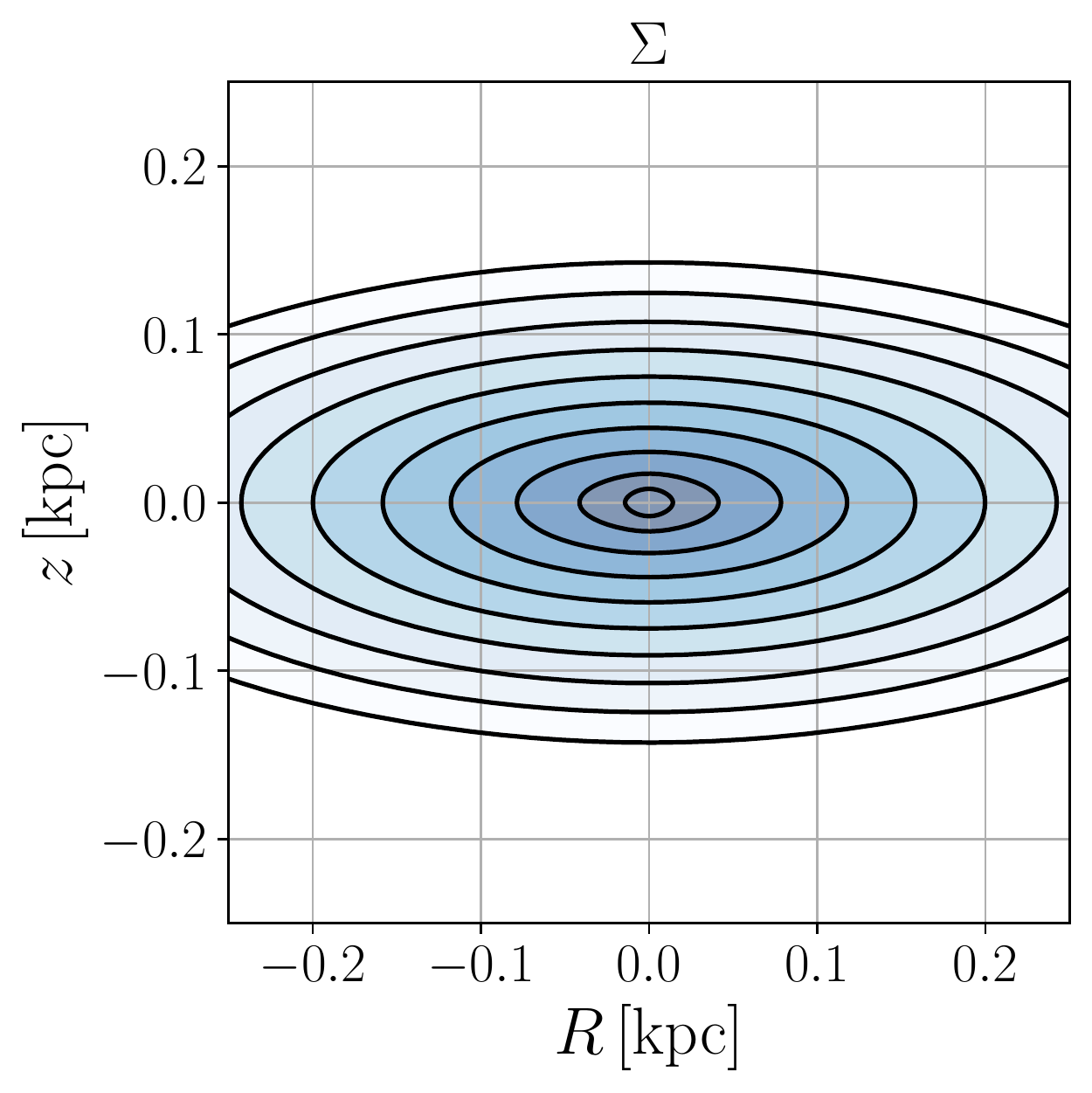}\includegraphics[width=0.25\textwidth]{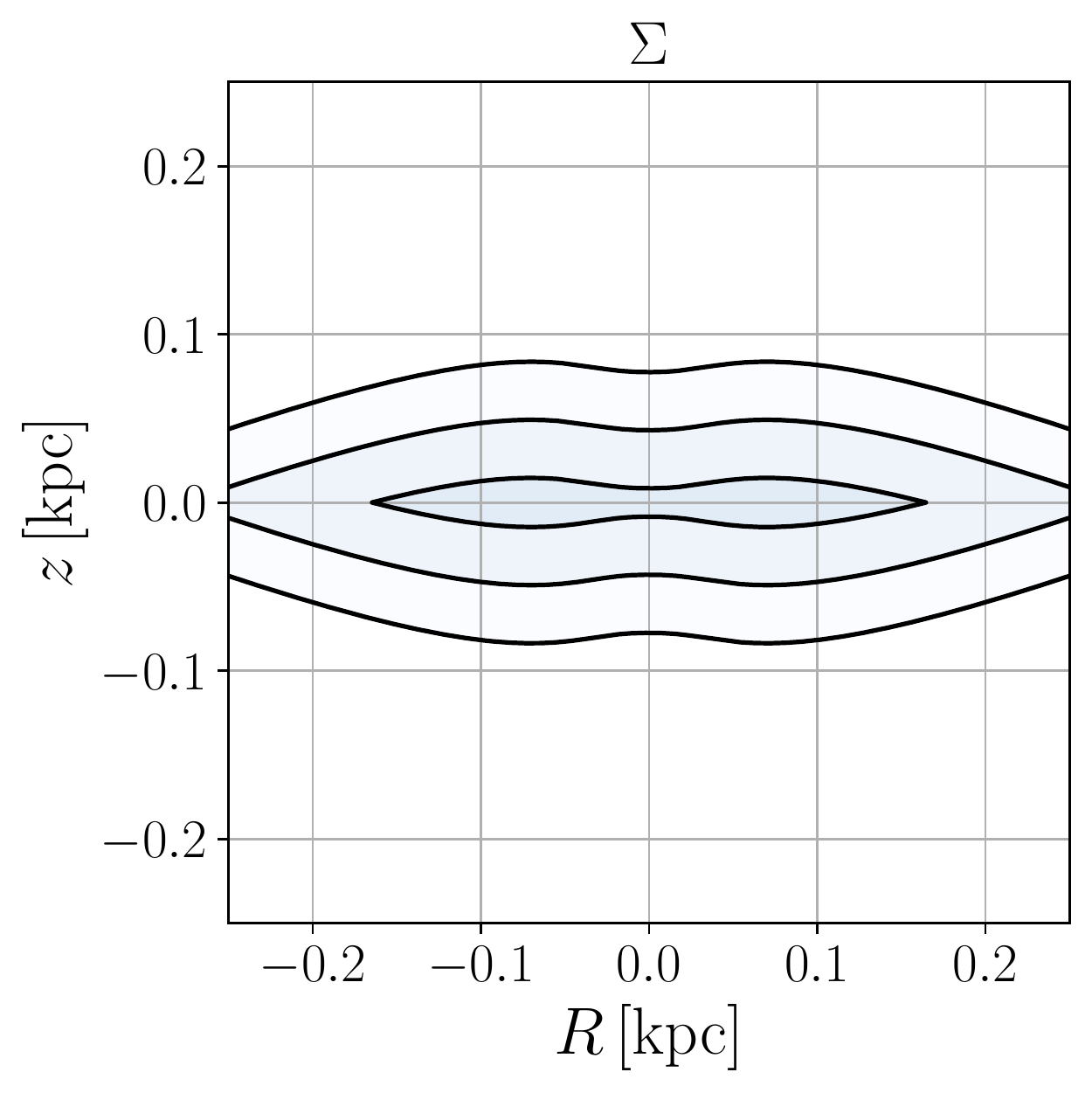}
    \caption{Contours of constant density $\rho(R,z)$ (top) and surface density $\Sigma(R,z)$ (bottom). From left to right: model 1, 2, 3 (see Table \ref{tab:1} and Equation \ref{eq:rhofull}) and ring model (see Equation \ref{eq:rho_ring}). The ring model is normalised with an arbitrary density scaling $\rho_0= 10^{10} \Msun \kpc^{-3}$ (this scaling does not enter the fitting procedure since it simplifies in the calculations of the second moments \ref{eq:mudoublemean}). The lowest contour in the top panels corresponds to a density of $\rho=3.2 \times 10^ 7 \Msun \kpc^{-3}$ and contours are geometrically spaced every $0.3\rm\, dex$. The lowest contour in the top panels corresponds to a surface density of $\Sigma= 10^8 \Msun \kpc^{-2}$ and contours are geometrically spaced every $0.33\rm\, dex$.}
    \label{fig:bestmodel_02}
\end{figure*}

\begin{figure}
	\includegraphics[width=0.5\textwidth]{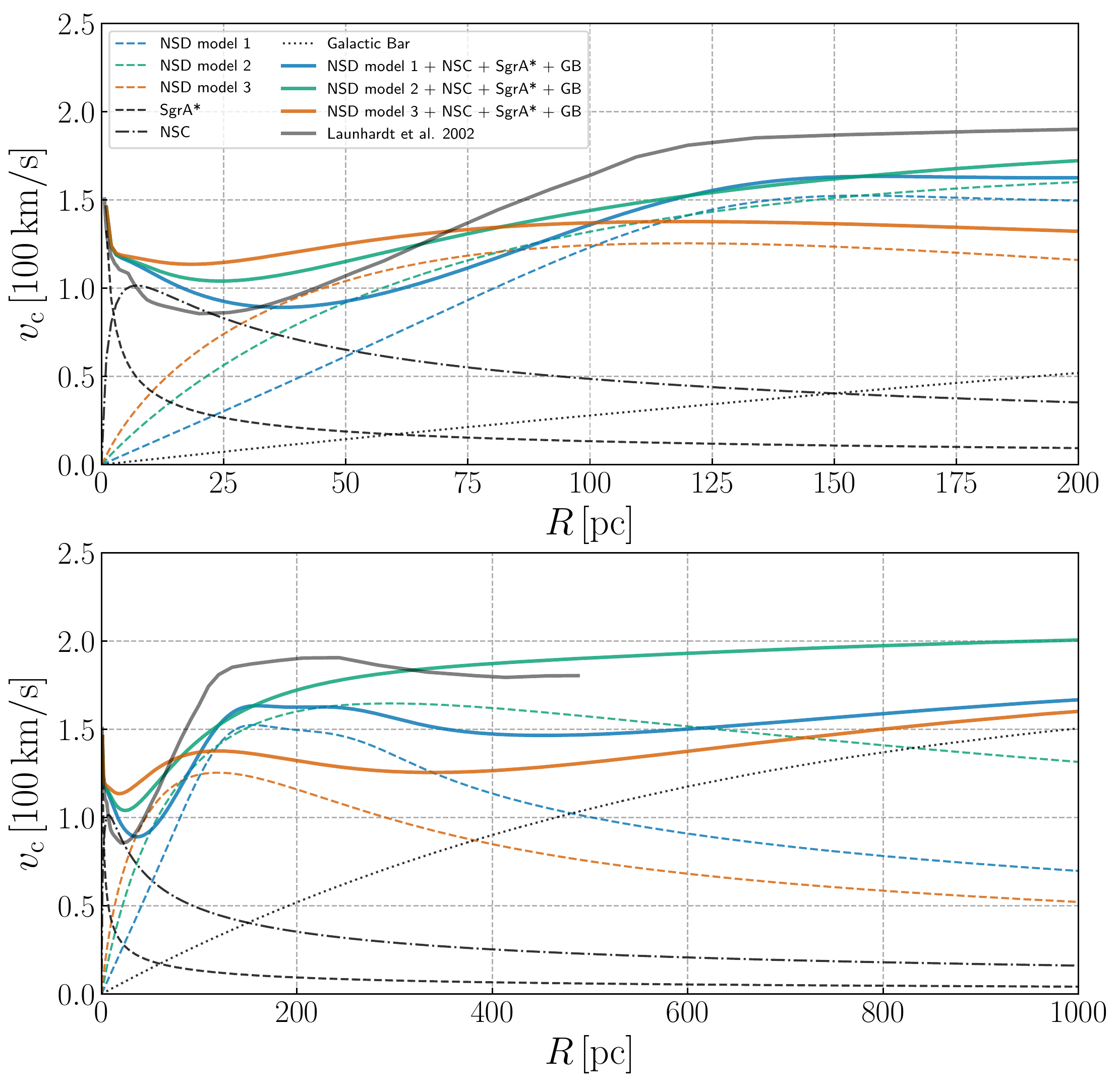}
    \caption{Circular velocity curves in the plane $z=0$. The top panel is a zoom in the innermost $200\pc$ of the bottom panel. The style/colour scheme is the same as in Figure \ref{fig:Menc}. The circular velocity curve of \citet{Launhardt+2002} is calculated from the corresponding line in Figure \ref{fig:Menc} assuming spherical symmetry. For all the other components the circular velocity curves is calculated as $v_{\rm circ} = \sqrt{R \di \Phi/\di R}$. For the Galactic bar component, the velocity curve is calculated after azimuthally-averaging the triaxial density profile given in \citet{Launhardt+2002}.} 
    \label{fig:vcirc}
\end{figure}

\section{Discussion} \label{sec:discussion}

\subsection{The mass of the nuclear stellar disc} \label{sec:discussion_mass}

We have seen in Section \ref{sec:results} that our models favour a mass $M(r<100\pc) = (4\pm 1) \times 10^8 \Msun$ which is consistent with but lower than the best fitting value  $M(r<100\pc) \simeq (6\pm 2) \times 10^8 \Msun$ of \citet{Launhardt+2002}. The two determinations are to a large extent independent since ours is based on the line-of-sight kinematics while the one of \citet{Launhardt+2002} is purely based on the photometry.

As mentioned in Section \ref{sec:introduction}, the size of the CMZ in simulations of gas flow in Milky Way-like barred potentials depends on the mass of the NSD. \citet{Li+2020} use this fact to constrain the mass of the NSD. They run several simulations with different NSD mass until the size of the simulated CMZ matches the size of the observed CMZ. While there are several uncertainties in this approach related to the fact that the size of the simulated CMZ also depends on the assumed equation of state of the gas \citep[e.g.][]{SBM2015a} and on the details of the assumed Galactic bar potential \cite[e.g.][]{SBM2015c}, they also found a NSD mass which is on the lower side of the range indicated by \citet{Launhardt+2002} (see Figure 7 in \citealt{Li+2020}), consistent with our result.

\citet{NoguerasLara+2020} used the GALACTICNUCLEUS survey \citep{NoguerasLara+2019b} to create de-reddened $K_s$ luminosity functions and fit them using theoretical stellar evolution models, and estimated the mass contained in a cylinder of $R\lesssim 45 \pc$ and $|z|\lesssim 20 \pc$ to be $M = 6.5\pm0.4 \times 10^7 \Msun$. As shown in Figure 7 of \citet{Li+2020}, this mass would also be consistent with a mass slightly lower than that of \citet{Launhardt+2002}. 

The mass estimation of \cite{Launhardt+2002} involves assuming a mass-to-infrared-light ratio, which carries rather large uncertainties (see their Section 5.4). For the NSD, they assumed a rather large value of $\Upsilon=2$. Assuming a value closer to the more common $\Upsilon=0.6$ \citep[e.g.][]{Meidt+2004,Schoedel+2014} would lower their mass estimate considerably. 

Given that all our models consistently suggest a mass that is on the lower side of the mass estimated by \citet{Launhardt+2002} for all the combination of potential/density/dataset/filters we considered, and given the large uncertainties in the mass estimation of the latter stemming from the mass-to-infrared-light ratio, we conclude that it is likely the mass of the NSD is lower than the mass estimated by \citet{Launhardt+2002}.

Figure \ref{fig:Menc} compares the mass enclosed within spherical radius $r$ of models 1-3, which differ in the assumed NSD mass distributions. While at $r \lesssim 100\pc$ the three models agree well with each other, they diverge at larger radii. This is because the three models have very different extensions as can be seen in Figure \ref{fig:bestmodel_02}. Model 3 is the least extended of the three, while model 2 is by far the most extended. As a result, the total mass of the NSD in models 1,2 and 3 are $M_{\rm NSD1}= 1.2 \times 10^9 \Msun$,  $M_{\rm NSD2}= 5.3 \times 10^9 \Msun$ and  $M_{\rm NSD3}= 0.69 \times 10^9 \Msun$ respectively. The first is the easiest to compare with \citet{Launhardt+2002} ($M_{\rm NSD} = 1.4 \pm 0.6 \times 10^9 \Msun$) since it assumes the same underlying NSD mass distribution. The model of \citet{Chatzopoulos+2015} is most likely too extended and gives an unrealistically high total mass. This is not too surprising since these authors were mostly concerned with fitting the innermost few tens of parsecs and not the larger scales considered here. Our fiducial model 3 gives the lowest mass of the three, and is probably the most accurate at least out to $r\simeq 150\pc$ given that it is based on the highest resolution data and that the subtraction of the Galactic bulge/bar is made with exactly the same model as \citet{Launhardt+2002}.

Figure \ref{fig:vcirc} shows the rotation curves for models 1-3. The rotation curves show significant differences. Since all the three models are all plausible models of the NSD, the scatter can be taken as a measure of the uncertainty in the rotation curve of the Galaxy in the innermost few hundred parsec. Note however that all the rotation curves are significantly lower than the rotation curve implied by \citet{Launhardt+2002}. 

\subsection{A vertically biased disc?} \label{sec:discussion_b}

All our models favour a value of the anisotropy parameter $1/\sqrt{b} = \sigma_z/\sigma_R>1$ (see Table \ref{tab:1} and Figures \ref{fig:chi2_1}, \ref{fig:chi2_2} and \ref{fig:chi2_3}). This means that vertical oscillations are stronger than radial oscillations, which is unusual for a disc system. For example, the Galactic disc in the solar neighbourhood has values ranging from $\sigma_z/\sigma_R\simeq0.4$ for the youngest populations to $\simeq0.8$ for the oldest \citep[e.g.][]{Holmberg+2009,Martig+2016,Mackereth+2019,MariaSelina+2020}. Modelling of the kinematics of external galaxies hints at a loose correlation between $\sigma_z/\sigma_R$ and Hubble type \citep[e.g][]{vanderKruitdeGrijs1999,GerssenShapiroGriffin2012,Pinna+2018}, with $\sigma_z/\sigma_R$ decreasing from about 1.0 in early types (lenticulars) to about 0.4 in late types (Sd). \citet{Gentile+2015} find $\sigma_z/\sigma_R=1.2\pm0.2$ for the Sb galaxy NGC~3223, which is one of the highest values found in any other
galaxy.  Our value of $\sigma_z/\sigma_R \sim 1.5$ for the NSD is much larger than any of these.  We note, however, that these other measurements are all for large, kpc-scale discs, not for a compact NSD.  On much smaller scales, \citet{BrownMagorrian2013} fit unusually large vertical oscillations in their model of the eccentric disc at the centre of M31.

In order to test whether the finding that $\sigma_z/\sigma_R>1$ depends on our assumption that the velocity ellipsoid is aligned on cylindrical coordinates, we have repeated our analysis assuming that the velocity ellipsoid is aligned on spherical coordinates (see Section 2.4 of \citealt{Cappellari2020}). We found that models with $\sigma_\theta/\sigma_r>1$ are clearly favoured, which on the plane $z=0$ corresponds to $\sigma_z/\sigma_R>1$. Thus, the alignment of the velocity ellipsoid does not affect the results discussed here.

\begin{figure*}
	\includegraphics[width=0.5\textwidth]{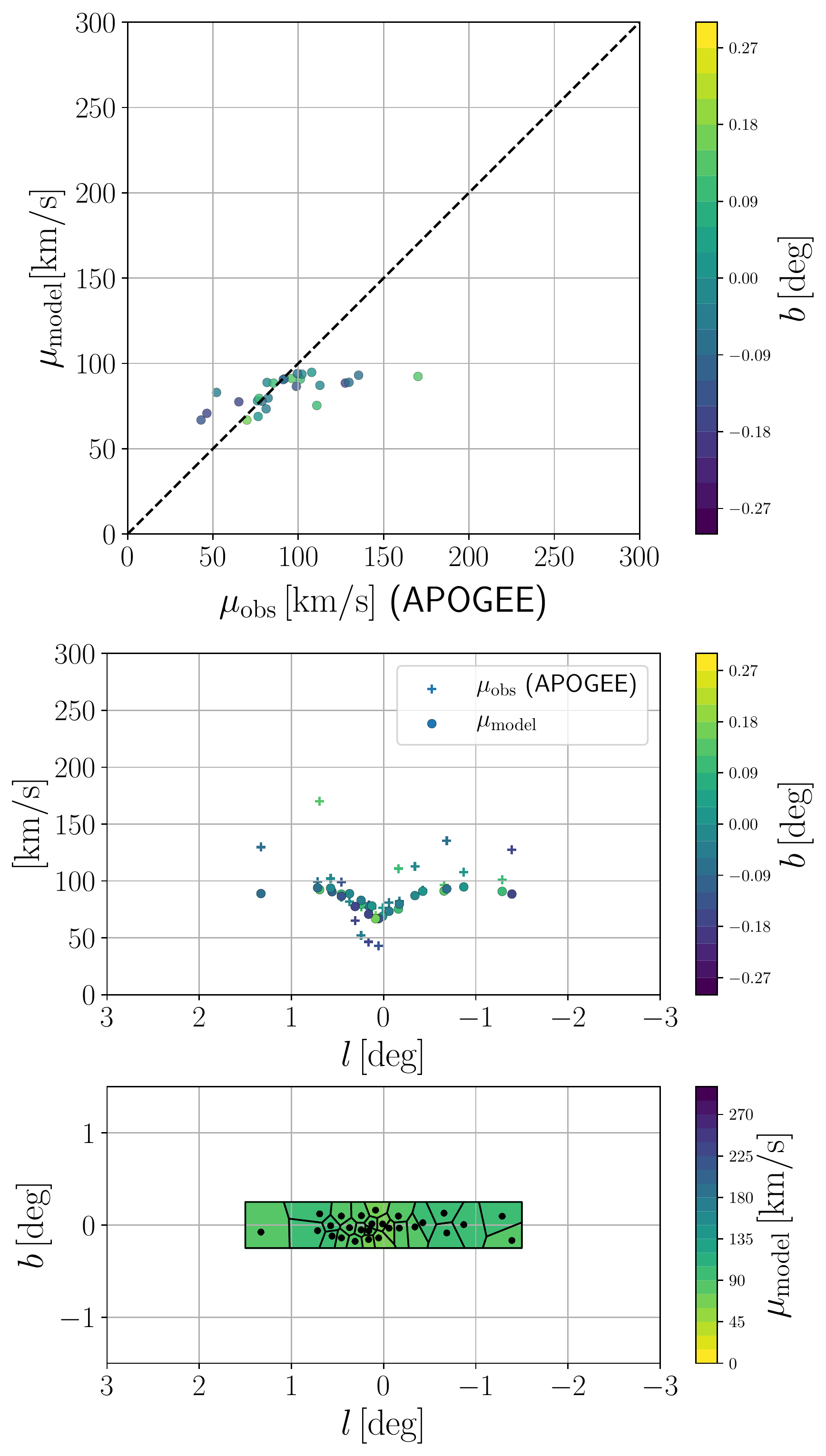}\includegraphics[width=0.5\textwidth]{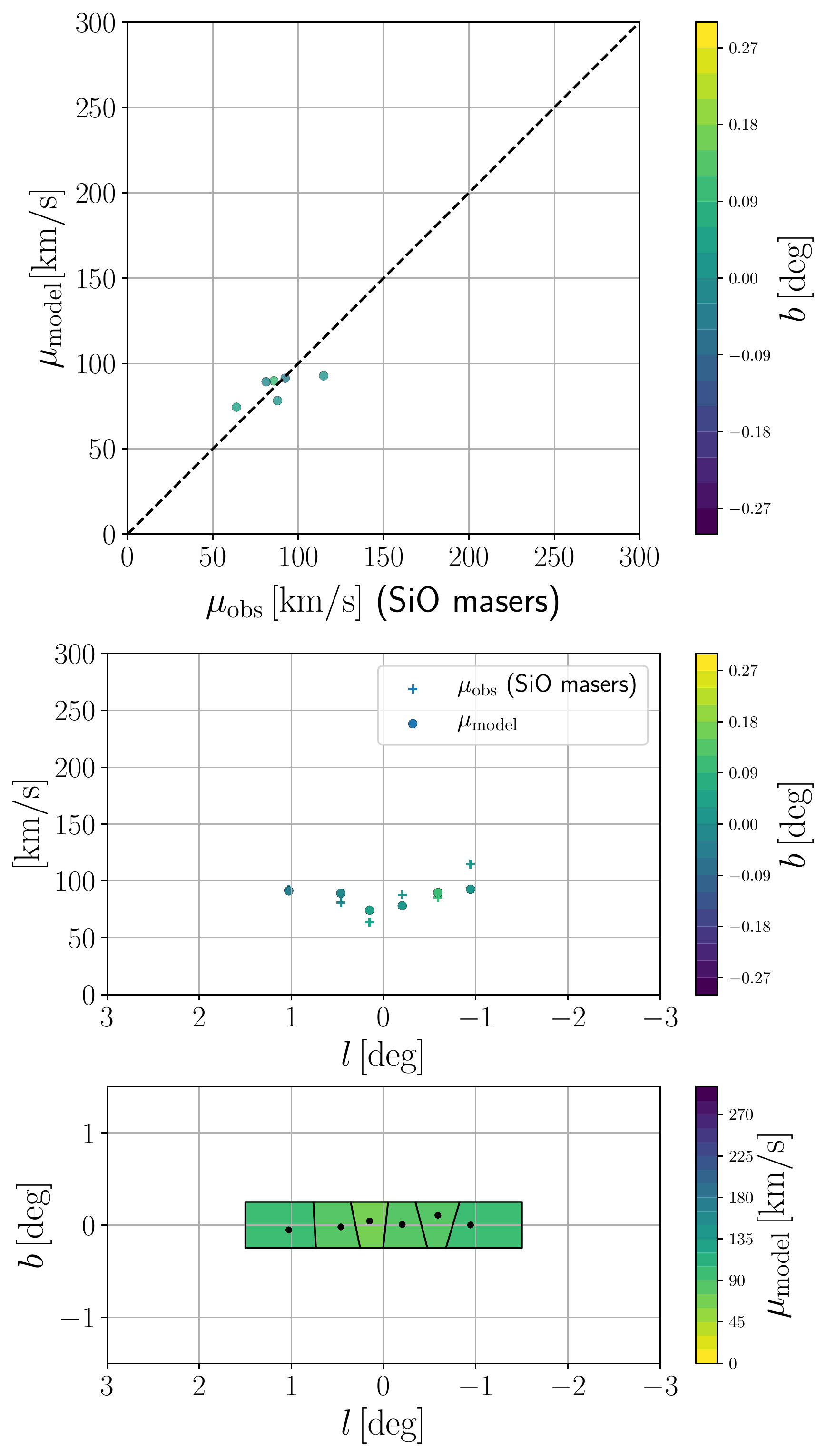}
    \caption{Same as Figure \ref{fig:bestmodel_01}, but for a model that is identical to model 3 except that $b=1$. Note that the drop in second moment velocities at $|l|\lesssim0.5\degree$ is reproduced less well than in the model shown in Figure \ref{fig:bestmodel_01}.}
    \label{fig:anothermodel}
\end{figure*}

There are two questions in relation to our finding that $\sigma_z/\sigma_R>1$. The first is why are such values favoured by our models? Comparison of Figure \ref{fig:bestmodel_01} with Figure \ref{fig:anothermodel} shows that the reason is that a small value of $b$ (i.e. a large $\sigma_z/\sigma_R$) is needed to reproduce the drop in the observed $\mu_{\rm obs}$ near the centre ($|l|\lesssim 0.5$), which is present both in the APOGEE data and the SiO maser data (see middle row in Figure \ref{fig:bestmodel_01}). In Figure \ref{fig:bestmodel_01}, which shows model 3 with $b=0.45$, the drop is well reproduced, while in Figure \ref{fig:anothermodel}, which shows the same model but for $b=1$, the drop is not well reproduced. The ``NSD only'' models favour a slightly larger value of $b$ compared to the ``NSD+NSC'' models because the absence of the NSC component in the middle lowers the velocity dispersion in the central regions compared to the outer parts. The ring models also favour a larger value of $b$ (consistent with $b\sim1$, see Figure \ref{fig:chi2_3}) because the density is essentially zero for $R\lesssim 50\pc$, and therefore those regions do not contribute to the integrals in Equation \eqref{eq:Jz3} and \eqref{eq:JR3}. Assuming that the drop in the data is real (which ought to be confirmed with better data), this suggests that indeed $b<1$, or that our assumed tracer density population is not representative of the population from which the kinematics are drawn.
We note that the observed kinematics directly constrain only the $v_R$ and $v_\phi$ components of velocity: our constraints on the $v_z$ component come from the integral~\eqref{eq:Jz3} of $\rho$ and $\Phi$ along~$z$.  If the potential or density were much flatter than we have assumed then the $\sigma_z$ given by~\eqref{eq:Jz3} would increase and we could fit the observed kinematics with larger values of $b$. It is currently unclear whether such a strong flattening would be detectable given the extreme and strong differential extinction.

Assuming that $\sigma_z/\sigma_R>1$ then the second question is how would stars get such large vertical oscillations?  In the solar neighbourhood stars are formed from gas clouds that move on almost closed orbits, beginning their lives with random velocities of the order of a few $\kms$.  There are a number of dynamical processes that inevitably cause these random velocities to increase over time (see \citealt{Sellwood2014} for a recent review). Each of these heating mechanisms has a different effect on the ratio $\sigma_z/\sigma_R$. For example, spiral density waves tend to increase $\sigma_R$, but have little effect on $\sigma_z$.  We note, however, that the NSD is probably hot enough that any spiral waves are weak. Two-body scattering of stars by other stars or by giant molecular clouds produces more isotropic heating \citep{JenkinsBinney1990,SBA2016}, but still limited to $\sigma_z/\sigma_R \lesssim0.6$ \citep{Ida+1993}, much smaller than we find in the NSD.  The most promising mechanism for producing $\sigma_z/\sigma_R>1$ from an initially cold stellar population is probably from bending instabilities caused by the presence of a counterrotating population \citep{KhoperskovBertin2017}.

An alternative explanation is to relax the assumption that NSD stars were born from a kinematically cold gas disc.   Interestingly, observations show that the dense and star-forming molecular gas in the CMZ is currently concentrated into streams that possess strong vertical oscillations of the order of $\Delta z \simeq30\pc$ (see for example Figure 4 in \citealt{Molinari+2011} and Figure 5 of \citealt{Purcell+2012}). This value is similar to the NSD scale-height determined by \citet{Nishiyama+2013}. Moreover, \citet{Tress+2020} argue that these large vertical oscillations in the CMZ gas are induced by the large-scale bar-driven accretion and are quite typical in that region based on a combination of observations and hydrodynamical simulations (see their Section 6.4). This suggests that NSD stars might already possess large vertically oscillations at birth.

That leaves open the question of whether this mechanism would produce vertical oscillations that are so much stronger than the radial ones.   The currently observed scale-height of the NSD is $\simeq45\pc$. Assuming that this is similar to the typical vertical excursions of stars in the NSD, it implies that stars oscillate between $z_{\rm min}  \simeq -45\pc$ and $z_{\rm max} \simeq 45\pc$. Assuming $\sigma_z/\sigma_R =1/\sqrt{b} \simeq 1.5$ (Table \ref{tab:1}) and a mean radius of $R\simeq120\pc$, it implies typical radial oscillations between $R_{\rm min} \simeq 90\pc$ and $R_{\rm max} \simeq 150\pc$, which is roughly consistent with the expected eccentricities of the $x_2$ orbits on which the CMZ gas is believed to be flowing on \citep[e.g.][]{Binney+1991,EnglmaierGerhard1999,SBM2015a,Tress+2020,Sormani+2020}, and is also consistent with the eccentricity of the ballistic model of \citet{Kruijssen+2015} (see their Table 1). However, large uncertainties remain, and the question should be re-addressed in the future when better data become available.

We conclude that the NSD might be the first example of a vertically biased stellar disc ($\sigma_z/\sigma_R>1$). We propose that the large vertical dispersion might be already imprinted at stellar birth by the star-forming molecular gas in the CMZ.

\section{Conclusions} \label{sec:conclusion}

We have constructed axisymmetric Jeans models of the nuclear stellar disc and have fitted them to the line-of-sight kinematic data of APOGEE and SiO maser stars. We adopted three rather different mass distributions which have been previously shown to be consistent with near/mid-infrared photometry and star counts. Our main results are as follows:
\begin{enumerate}
\item All our models indicate that the mass of the NSD is lower than, but consistent with, the value determined independently from near-infrared photometry by \citet{Launhardt+2002} (see Figure \ref{fig:Menc}). Our fiducial model, based on the recent analysis of high-resolution mid-infrared Spitzer data by \citet{GallegoCano+2020}, has a mass contained within spherical radius $r=100\pc$ of $M(r<100\pc) = 3.9 \pm 1 \times 10^8 \Msun$ and a total mass of $M_{\rm NSD} = 6.9 \pm 2 \times 10^8 \Msun$. If instead we assume the same underlying mass distribution of the NSD as \citet{Launhardt+2002}, which is more spatially extended than our fiducial model, we obtain $M_{\rm NSD}=1.2 \pm 2 \times 10^ 9 \Msun$, still lower side than the original determination of \citet{Launhardt+2002}. The absence/presence of the nuclear stellar cluster in our models and switching between a disc- or ring-like density distribution for the tracer population do not affect these results significantly.
\item We find evidence that the NSD is vertically biased, i.e. $\sigma_z/\sigma_R>1$. If true, the NSD would be the first example of a vertically biased disc system. Observations and theoretical models of the dense star-forming molecular gas in the CMZ suggest that large vertical velocity dispersions may be already imprinted at stellar birth. However, we caution that the finding $\sigma_z/\sigma_R>1$ depends on many assumptions, and in particular on the observed drop in the second moment of the line-of-sight velocity in the innermost parts, on our assumptions of axisymmetry/that the anisotropy is spatially constant, on whether the stellar populations traced by APOGEE and SiO maser data follow a disc- or ring-like density distribution, and, more generally, on our assumption that the available corrected starcount data provide good estimates of the underlying light and mass distribution. All of these need to be established by future observations and/or modelling.
\item The rotation curves implied by our models are shown in Figure \ref{fig:vcirc}. The rotation curve of our fiducial model 3 is the most accurate to date for the innermost $500\pc$ of our Galaxy. The scatter between model 1,2 and 3 can be taken as a measure of the current uncertainty of the rotation curve in this region.
\end{enumerate}
While Jeans models provide useful constraints and insight into the dynamics of the NSD, they are limited as there is no guarantee that they correspond to a physical DF which is everywhere positive ($f>0$). Therefore, a worthwhile direction of future investigation is to produce DF-based models that can overcome the shortcomings of Jeans modelling.

\section*{Acknowledgements}

We thank the referee for a constructive report that improved the quality of the paper. MCS thanks James Binney, Francesca Fragkoudi, Dimitri Gadotti, Ryan Leaman, Zhi Li, Lorenzo Posti and Robin Tress for useful comments and discussions. MCS, FNL, NN and RSK acknowledge financial support from the German Research Foundation (DFG) via the collaborative research center (SFB 881, Project-ID 138713538) ``The Milky Way System'' (subprojects A1, B1, B2, and B8). RSK furthermore thanks for financial support from the European Research Council via the ERC Synergy Grant ``ECOGAL -- Understanding our Galactic ecosystem: from the disk of the Milky Way to the formation sites of stars and planets'' (grant 855130). JM acknowledges funding from the UK Science and Technology Facilities Council under grant number ST/S000488/1. RS acknowledges funding by the Royal Society via a University Research Fellowship.

\section*{Data availability statement}

The datasets used in this article are publicly available. For APOGEE stars: \url{https://www.sdss.org/dr16/irspec/}. For SiO masers: \url{http://vizier.u-strasbg.fr/viz-bin/VizieR?-source=J/A+A/393/115}, \url{http://vizier.u-strasbg.fr/viz-bin/VizieR?-source=J/A+A/418/103} and \url{http://vizier.u-strasbg.fr/viz-bin/VizieR?-source=J/A+A/435/575}.




\bibliographystyle{mnras}
\bibliography{bibliography} 

\begin{thebibliography}{}
\makeatletter
\relax
\def\mn@urlcharsother{\let\do\@makeother \do\$\do\&\do\#\do\^\do\_\do\%\do\~}
\def\mn@doi{\begingroup\mn@urlcharsother \@ifnextchar [ {\mn@doi@}
  {\mn@doi@[]}}
\def\mn@doi@[#1]#2{\def\@tempa{#1}\ifx\@tempa\@empty \href
  {http://dx.doi.org/#2} {doi:#2}\else \href {http://dx.doi.org/#2} {#1}\fi
  \endgroup}
\def\mn@eprint#1#2{\mn@eprint@#1:#2::\@nil}
\def\mn@eprint@arXiv#1{\href {http://arxiv.org/abs/#1} {{\tt arXiv:#1}}}
\def\mn@eprint@dblp#1{\href {http://dblp.uni-trier.de/rec/bibtex/#1.xml}
  {dblp:#1}}
\def\mn@eprint@#1:#2:#3:#4\@nil{\def\@tempa {#1}\def\@tempb {#2}\def\@tempc
  {#3}\ifx \@tempc \@empty \let \@tempc \@tempb \let \@tempb \@tempa \fi \ifx
  \@tempb \@empty \def\@tempb {arXiv}\fi \@ifundefined
  {mn@eprint@\@tempb}{\@tempb:\@tempc}{\expandafter \expandafter \csname
  mn@eprint@\@tempb\endcsname \expandafter{\@tempc}}}

\bibitem[\protect\citeauthoryear{{Ahumada} et~al.,}{{Ahumada}
  et~al.}{2019}]{Ahumada+2019}
{Ahumada} R.,  et~al., 2019, arXiv e-prints, \href
  {https://ui.adsabs.harvard.edu/abs/2019arXiv191202905A} {p. arXiv:1912.02905}

\bibitem[\protect\citeauthoryear{{Alard}}{{Alard}}{2001}]{Alard2001}
{Alard} C.,  2001, \mn@doi [\aap] {10.1051/0004-6361:20011487}, \href
  {http://adsabs.harvard.edu/abs/2001A%26A...379L..44A} {379, L44}

\bibitem[\protect\citeauthoryear{{Aumer} \& {Sch{\"o}nrich}}{{Aumer} \&
  {Sch{\"o}nrich}}{2015}]{AumerSchoenrich2015}
{Aumer} M.,  {Sch{\"o}nrich} R.,  2015, \mn@doi [\mnras]
  {10.1093/mnras/stv2252}, \href
  {https://ui.adsabs.harvard.edu/abs/2015MNRAS.454.3166A} {454, 3166}

\bibitem[\protect\citeauthoryear{{Aumer}, {Binney}  \& {Sch{\"o}nrich}}{{Aumer}
  et~al.}{2016}]{SBA2016}
{Aumer} M.,  {Binney} J.,   {Sch{\"o}nrich} R.,  2016, \mn@doi [\mnras]
  {10.1093/mnras/stw1639}, \href
  {https://ui.adsabs.harvard.edu/abs/2016MNRAS.462.1697A} {462, 1697}

\bibitem[\protect\citeauthoryear{{Baba} \& {Kawata}}{{Baba} \&
  {Kawata}}{2020}]{BabaKawata2020}
{Baba} J.,  {Kawata} D.,  2020, \mn@doi [\mnras] {10.1093/mnras/staa140}, \href
  {https://ui.adsabs.harvard.edu/abs/2020MNRAS.492.4500B} {492, 4500}

\bibitem[\protect\citeauthoryear{{Binney} \& {Tremaine}}{{Binney} \&
  {Tremaine}}{1987}]{BT1987}
{Binney} J.,  {Tremaine} S.,  1987, {Galactic dynamics}.
Princeton series in astrophysics, Princeton University Press

\bibitem[\protect\citeauthoryear{{Binney} \& {Tremaine}}{{Binney} \&
  {Tremaine}}{2008}]{BT2008}
{Binney} J.,  {Tremaine} S.,  2008, {Galactic Dynamics: Second Edition}.
Princeton University Press

\bibitem[\protect\citeauthoryear{{Binney}, {Gerhard}, {Stark}, {Bally}  \&
  {Uchida}}{{Binney} et~al.}{1991}]{Binney+1991}
{Binney} J.,  {Gerhard} O.~E.,  {Stark} A.~A.,  {Bally} J.,   {Uchida} K.~I.,
  1991, \mnras, \href {http://adsabs.harvard.edu/abs/1991MNRAS.252..210B} {252,
  210}

\bibitem[\protect\citeauthoryear{{Binney} et~al.,}{{Binney}
  et~al.}{2014}]{Binney+2014}
{Binney} J.,  et~al., 2014, \mn@doi [\mnras] {10.1093/mnras/stt2367}, \href
  {https://ui.adsabs.harvard.edu/abs/2014MNRAS.439.1231B} {439, 1231}

\bibitem[\protect\citeauthoryear{{Bland-Hawthorn} \&
  {Gerhard}}{{Bland-Hawthorn} \& {Gerhard}}{2016}]{BlandHawthornGerhard2016}
{Bland-Hawthorn} J.,  {Gerhard} O.,  2016, \mn@doi [\araa]
  {10.1146/annurev-astro-081915-023441}, \href
  {http://adsabs.harvard.edu/abs/2016ARA%26A..54..529B} {54, 529}

\bibitem[\protect\citeauthoryear{{Bovy} et~al.,}{{Bovy}
  et~al.}{2014}]{Bovy+2014}
{Bovy} J.,  et~al., 2014, \mn@doi [\apj] {10.1088/0004-637X/790/2/127}, \href
  {https://ui.adsabs.harvard.edu/abs/2014ApJ...790..127B} {790, 127}

\bibitem[\protect\citeauthoryear{{Bovy}, {Rix}, {Green}, {Schlafly}  \&
  {Finkbeiner}}{{Bovy} et~al.}{2016}]{Bovy+2016a}
{Bovy} J.,  {Rix} H.-W.,  {Green} G.~M.,  {Schlafly} E.~F.,   {Finkbeiner}
  D.~P.,  2016, \mn@doi [\apj] {10.3847/0004-637X/818/2/130}, \href
  {https://ui.adsabs.harvard.edu/abs/2016ApJ...818..130B} {818, 130}

\bibitem[\protect\citeauthoryear{{Brown} \& {Magorrian}}{{Brown} \&
  {Magorrian}}{2013}]{BrownMagorrian2013}
{Brown} C.~K.,  {Magorrian} J.,  2013, \mn@doi [\mnras] {10.1093/mnras/stt104},
  \href {https://ui.adsabs.harvard.edu/abs/2013MNRAS.431...80B} {431, 80}

\bibitem[\protect\citeauthoryear{{Cappellari}}{{Cappellari}}{2008}]{Cappellari2008}
{Cappellari} M.,  2008, \mn@doi [\mnras] {10.1111/j.1365-2966.2008.13754.x},
  \href {https://ui.adsabs.harvard.edu/abs/2008MNRAS.390...71C} {390, 71}

\bibitem[\protect\citeauthoryear{{Cappellari}}{{Cappellari}}{2020}]{Cappellari2020}
{Cappellari} M.,  2020, \mn@doi [\mnras] {10.1093/mnras/staa959}, \href
  {https://ui.adsabs.harvard.edu/abs/2020MNRAS.494.4819C} {494, 4819}

\bibitem[\protect\citeauthoryear{{Cappellari} \& {Copin}}{{Cappellari} \&
  {Copin}}{2003}]{CappellariCopin2003}
{Cappellari} M.,  {Copin} Y.,  2003, \mn@doi [\mnras]
  {10.1046/j.1365-8711.2003.06541.x}, \href
  {https://ui.adsabs.harvard.edu/abs/2003MNRAS.342..345C} {342, 345}

\bibitem[\protect\citeauthoryear{{Catchpole}, {Whitelock}  \&
  {Glass}}{{Catchpole} et~al.}{1990}]{Catchpole+1990}
{Catchpole} R.~M.,  {Whitelock} P.~A.,   {Glass} I.~S.,  1990, \mnras, \href
  {https://ui.adsabs.harvard.edu/abs/1990MNRAS.247..479C} {247, 479}

\bibitem[\protect\citeauthoryear{{Chatzopoulos}, {Fritz}, {Gerhard},
  {Gillessen}, {Wegg}, {Genzel}  \& {Pfuhl}}{{Chatzopoulos}
  et~al.}{2015}]{Chatzopoulos+2015}
{Chatzopoulos} S.,  {Fritz} T.~K.,  {Gerhard} O.,  {Gillessen} S.,  {Wegg} C.,
  {Genzel} R.,   {Pfuhl} O.,  2015, \mn@doi [\mnras] {10.1093/mnras/stu2452},
  \href {https://ui.adsabs.harvard.edu/abs/2015MNRAS.447..948C} {447, 948}

\bibitem[\protect\citeauthoryear{{Cole}, {Debattista}, {Erwin}, {Earp}  \&
  {Ro{\v{s}}kar}}{{Cole} et~al.}{2014}]{Cole+2014}
{Cole} D.~R.,  {Debattista} V.~P.,  {Erwin} P.,  {Earp} S. W.~F.,
  {Ro{\v{s}}kar} R.,  2014, \mn@doi [\mnras] {10.1093/mnras/stu1985}, \href
  {https://ui.adsabs.harvard.edu/abs/2014MNRAS.445.3352C} {445, 3352}

\bibitem[\protect\citeauthoryear{{Englmaier} \& {Gerhard}}{{Englmaier} \&
  {Gerhard}}{1999}]{EnglmaierGerhard1999}
{Englmaier} P.,  {Gerhard} O.,  1999, \mn@doi [\mnras]
  {10.1046/j.1365-8711.1999.02280.x}, \href
  {http://adsabs.harvard.edu/abs/1999MNRAS.304..512E} {304, 512}

\bibitem[\protect\citeauthoryear{{Evans} \& {de Zeeuw}}{{Evans} \& {de
  Zeeuw}}{1994}]{EvansDeZeeuw1994}
{Evans} N.~W.,  {de Zeeuw} P.~T.,  1994, \mn@doi [\mnras]
  {10.1093/mnras/271.1.202}, \href
  {https://ui.adsabs.harvard.edu/abs/1994MNRAS.271..202E} {271, 202}

\bibitem[\protect\citeauthoryear{{Everall}, {Evans}, {Belokurov}  \&
  {Sch{\"o}nrich}}{{Everall} et~al.}{2019}]{Everall+2019}
{Everall} A.,  {Evans} N.~W.,  {Belokurov} V.,   {Sch{\"o}nrich} R.,  2019,
  \mn@doi [\mnras] {10.1093/mnras/stz2217}, \href
  {https://ui.adsabs.harvard.edu/abs/2019MNRAS.489..910E} {489, 910}

\bibitem[\protect\citeauthoryear{{Feldmeier} et~al.,}{{Feldmeier}
  et~al.}{2014}]{Feldmeier+2014}
{Feldmeier} A.,  et~al., 2014, \mn@doi [\aap] {10.1051/0004-6361/201423777},
  \href {https://ui.adsabs.harvard.edu/abs/2014A&A...570A...2F} {570, A2}

\bibitem[\protect\citeauthoryear{{Fux}}{{Fux}}{1999}]{Fux1999}
{Fux} R.,  1999, \aap, \href
  {http://adsabs.harvard.edu/abs/1999A%26A...345..787F} {345, 787}

\bibitem[\protect\citeauthoryear{{Gadotti} et~al.,}{{Gadotti}
  et~al.}{2019}]{Gadotti+2019}
{Gadotti} D.~A.,  et~al., 2019, \mn@doi [\mnras] {10.1093/mnras/sty2666}, \href
  {https://ui.adsabs.harvard.edu/abs/2019MNRAS.482..506G} {482, 506}

\bibitem[\protect\citeauthoryear{{Gadotti} et~al.,}{{Gadotti}
  et~al.}{2020}]{Gadotti+2020}
{Gadotti} D.~A.,  et~al., 2020, arXiv e-prints, \href
  {https://ui.adsabs.harvard.edu/abs/2020arXiv200901852G} {p. arXiv:2009.01852}

\bibitem[\protect\citeauthoryear{{Gallego-Cano}, {Sch{\"o}del},
  {Nogueras-Lara}, {Dong}, {Shahzamanian}, {Fritz}, {Gallego-Calvente}  \&
  {Neumayer}}{{Gallego-Cano} et~al.}{2020}]{GallegoCano+2020}
{Gallego-Cano} E.,  {Sch{\"o}del} R.,  {Nogueras-Lara} F.,  {Dong} H.,
  {Shahzamanian} B.,  {Fritz} T.~K.,  {Gallego-Calvente} A.~T.,   {Neumayer}
  N.,  2020, \mn@doi [\aap] {10.1051/0004-6361/201935303}, \href
  {https://ui.adsabs.harvard.edu/abs/2020A&A...634A..71G} {634, A71}

\bibitem[\protect\citeauthoryear{{Gentile} et~al.,}{{Gentile}
  et~al.}{2015}]{Gentile+2015}
{Gentile} G.,  et~al., 2015, \mn@doi [\aap] {10.1051/0004-6361/201425279},
  \href {https://ui.adsabs.harvard.edu/abs/2015A&A...576A..57G} {576, A57}

\bibitem[\protect\citeauthoryear{{Gerhard} \& {Martinez-Valpuesta}}{{Gerhard}
  \& {Martinez-Valpuesta}}{2012}]{GerhardMartinezValpuesta2012}
{Gerhard} O.,  {Martinez-Valpuesta} I.,  2012, \mn@doi [\apjl]
  {10.1088/2041-8205/744/1/L8}, \href
  {https://ui.adsabs.harvard.edu/abs/2012ApJ...744L...8G} {744, L8}

\bibitem[\protect\citeauthoryear{{Gerssen} \& {Shapiro Griffin}}{{Gerssen} \&
  {Shapiro Griffin}}{2012}]{GerssenShapiroGriffin2012}
{Gerssen} J.,  {Shapiro Griffin} K.,  2012, \mn@doi [\mnras]
  {10.1111/j.1365-2966.2012.21078.x}, \href
  {https://ui.adsabs.harvard.edu/abs/2012MNRAS.423.2726G} {423, 2726}

\bibitem[\protect\citeauthoryear{{Gravity Collaboration} et~al.,}{{Gravity
  Collaboration} et~al.}{2019}]{Gravity2019}
{Gravity Collaboration} et~al., 2019, \mn@doi [\aap]
  {10.1051/0004-6361/201935656}, \href
  {https://ui.adsabs.harvard.edu/abs/2019A&A...625L..10G} {625, L10}

\bibitem[\protect\citeauthoryear{{Habing}}{{Habing}}{2016}]{Habing2016}
{Habing} H.~J.,  2016, \mn@doi [\aap] {10.1051/0004-6361/201526083}, \href
  {https://ui.adsabs.harvard.edu/abs/2016A&A...587A.140H} {587, A140}

\bibitem[\protect\citeauthoryear{{Habing}, {Sevenster}, {Messineo}, {van de
  Ven}  \& {Kuijken}}{{Habing} et~al.}{2006}]{Habing+2006}
{Habing} H.~J.,  {Sevenster} M.~N.,  {Messineo} M.,  {van de Ven} G.,
  {Kuijken} K.,  2006, \mn@doi [\aap] {10.1051/0004-6361:20054480}, \href
  {https://ui.adsabs.harvard.edu/abs/2006A&A...458..151H} {458, 151}

\bibitem[\protect\citeauthoryear{{Henshaw} et~al.,}{{Henshaw}
  et~al.}{2016}]{Henshaw+2016}
{Henshaw} J.~D.,  et~al., 2016, \mn@doi [\mnras] {10.1093/mnras/stw121}, \href
  {https://ui.adsabs.harvard.edu/abs/2016MNRAS.457.2675H} {457, 2675}

\bibitem[\protect\citeauthoryear{{Holmberg}, {Nordstr{\"o}m}  \&
  {Andersen}}{{Holmberg} et~al.}{2009}]{Holmberg+2009}
{Holmberg} J.,  {Nordstr{\"o}m} B.,   {Andersen} J.,  2009, \mn@doi [\aap]
  {10.1051/0004-6361/200811191}, \href
  {https://ui.adsabs.harvard.edu/abs/2009A&A...501..941H} {501, 941}

\bibitem[\protect\citeauthoryear{{Ida}, {Kokubo}  \& {Makino}}{{Ida}
  et~al.}{1993}]{Ida+1993}
{Ida} S.,  {Kokubo} E.,   {Makino} J.,  1993, \mn@doi [\mnras]
  {10.1093/mnras/263.4.875}, \href
  {https://ui.adsabs.harvard.edu/abs/1993MNRAS.263..875I} {263, 875}

\bibitem[\protect\citeauthoryear{{Jenkins} \& {Binney}}{{Jenkins} \&
  {Binney}}{1990}]{JenkinsBinney1990}
{Jenkins} A.,  {Binney} J.,  1990, \mnras, \href
  {https://ui.adsabs.harvard.edu/abs/1990MNRAS.245..305J} {245, 305}

\bibitem[\protect\citeauthoryear{{Khoperskov} \& {Bertin}}{{Khoperskov} \&
  {Bertin}}{2017}]{KhoperskovBertin2017}
{Khoperskov} S.,  {Bertin} G.,  2017, \mn@doi [\aap]
  {10.1051/0004-6361/201629032}, \href
  {https://ui.adsabs.harvard.edu/abs/2017A&A...597A.103K} {597, A103}

\bibitem[\protect\citeauthoryear{{Kruijssen}, {Dale}  \&
  {Longmore}}{{Kruijssen} et~al.}{2015}]{Kruijssen+2015}
{Kruijssen} J.~M.~D.,  {Dale} J.~E.,   {Longmore} S.~N.,  2015, \mn@doi
  [\mnras] {10.1093/mnras/stu2526}, \href
  {http://adsabs.harvard.edu/abs/2015MNRAS.447.1059K} {447, 1059}

\bibitem[\protect\citeauthoryear{{Launhardt}, {Zylka}  \& {Mezger}}{{Launhardt}
  et~al.}{2002}]{Launhardt+2002}
{Launhardt} R.,  {Zylka} R.,   {Mezger} P.~G.,  2002, \mn@doi [\aap]
  {10.1051/0004-6361:20020017}, \href
  {http://adsabs.harvard.edu/abs/2002A%26A...384..112L} {384, 112}

\bibitem[\protect\citeauthoryear{{Li}, {Shen}  \& {Schive}}{{Li}
  et~al.}{2020}]{Li+2020}
{Li} Z.,  {Shen} J.,   {Schive} H.-Y.,  2020, \mn@doi [\apj]
  {10.3847/1538-4357/ab6598}, \href
  {https://ui.adsabs.harvard.edu/abs/2020ApJ...889...88L} {889, 88}

\bibitem[\protect\citeauthoryear{{Lindqvist}, {Habing}  \&
  {Winnberg}}{{Lindqvist} et~al.}{1992}]{Lindqvist+1992}
{Lindqvist} M.,  {Habing} H.~J.,   {Winnberg} A.,  1992, \aap, \href
  {https://ui.adsabs.harvard.edu/abs/1992A&A...259..118L} {259, 118}

\bibitem[\protect\citeauthoryear{{Mackereth} et~al.,}{{Mackereth}
  et~al.}{2019}]{Mackereth+2019}
{Mackereth} J.~T.,  et~al., 2019, \mn@doi [\mnras] {10.1093/mnras/stz1521},
  \href {https://ui.adsabs.harvard.edu/abs/2019MNRAS.489..176M} {489, 176}

\bibitem[\protect\citeauthoryear{{Majewski} et~al.,}{{Majewski}
  et~al.}{2017}]{Majewski+2017}
{Majewski} S.~R.,  et~al., 2017, \mn@doi [\aj] {10.3847/1538-3881/aa784d},
  \href {https://ui.adsabs.harvard.edu/abs/2017AJ....154...94M} {154, 94}

\bibitem[\protect\citeauthoryear{{Mamon} \& {Bou{\'e}}}{{Mamon} \&
  {Bou{\'e}}}{2010}]{Mamon+2010}
{Mamon} G.~A.,  {Bou{\'e}} G.,  2010, \mn@doi [\mnras]
  {10.1111/j.1365-2966.2009.15817.x}, \href
  {https://ui.adsabs.harvard.edu/abs/2010MNRAS.401.2433M} {401, 2433}

\bibitem[\protect\citeauthoryear{{Martig}, {Minchev}, {Ness}, {Fouesneau}  \&
  {Rix}}{{Martig} et~al.}{2016}]{Martig+2016}
{Martig} M.,  {Minchev} I.,  {Ness} M.,  {Fouesneau} M.,   {Rix} H.-W.,  2016,
  \mn@doi [\apj] {10.3847/0004-637X/831/2/139}, \href
  {https://ui.adsabs.harvard.edu/abs/2016ApJ...831..139M} {831, 139}

\bibitem[\protect\citeauthoryear{{Matsunaga} et~al.,}{{Matsunaga}
  et~al.}{2015}]{Matsunaga+2015}
{Matsunaga} N.,  et~al., 2015, \mn@doi [\apj] {10.1088/0004-637X/799/1/46},
  \href {https://ui.adsabs.harvard.edu/abs/2015ApJ...799...46M} {799, 46}

\bibitem[\protect\citeauthoryear{{Meidt} et~al.,}{{Meidt}
  et~al.}{2014}]{Meidt+2004}
{Meidt} S.~E.,  et~al., 2014, \mn@doi [\apj] {10.1088/0004-637X/788/2/144},
  \href {https://ui.adsabs.harvard.edu/abs/2014ApJ...788..144M} {788, 144}

\bibitem[\protect\citeauthoryear{{Messineo}, {Habing}, {Sjouwerman}, {Omont}
  \& {Menten}}{{Messineo} et~al.}{2002}]{Messineo+2002}
{Messineo} M.,  {Habing} H.~J.,  {Sjouwerman} L.~O.,  {Omont} A.,   {Menten}
  K.~M.,  2002, \mn@doi [\aap] {10.1051/0004-6361:20021017}, \href
  {https://ui.adsabs.harvard.edu/abs/2002A&A...393..115M} {393, 115}

\bibitem[\protect\citeauthoryear{{Messineo}, {Habing}, {Menten}, {Omont}  \&
  {Sjouwerman}}{{Messineo} et~al.}{2004}]{Messineo+2004}
{Messineo} M.,  {Habing} H.~J.,  {Menten} K.~M.,  {Omont} A.,   {Sjouwerman}
  L.~O.,  2004, \mn@doi [\aap] {10.1051/0004-6361:20034488}, \href
  {https://ui.adsabs.harvard.edu/abs/2004A&A...418..103M} {418, 103}

\bibitem[\protect\citeauthoryear{{Messineo}, {Habing}, {Menten}, {Omont},
  {Sjouwerman}  \& {Bertoldi}}{{Messineo} et~al.}{2005}]{Messineo+2005}
{Messineo} M.,  {Habing} H.~J.,  {Menten} K.~M.,  {Omont} A.,  {Sjouwerman}
  L.~O.,   {Bertoldi} F.,  2005, \mn@doi [\aap] {10.1051/0004-6361:20040533},
  \href {https://ui.adsabs.harvard.edu/abs/2005A&A...435..575M} {435, 575}

\bibitem[\protect\citeauthoryear{{Molinari} et~al.,}{{Molinari}
  et~al.}{2011}]{Molinari+2011}
{Molinari} S.,  et~al., 2011, \mn@doi [\apjl] {10.1088/2041-8205/735/2/L33},
  \href {http://adsabs.harvard.edu/abs/2011ApJ...735L..33M} {735, L33}

\bibitem[\protect\citeauthoryear{{Molloy}, {Smith}, {Evans}  \&
  {Shen}}{{Molloy} et~al.}{2015}]{Molloy+2015}
{Molloy} M.,  {Smith} M.~C.,  {Evans} N.~W.,   {Shen} J.,  2015, \mn@doi [\apj]
  {10.1088/0004-637X/812/2/146}, \href
  {https://ui.adsabs.harvard.edu/abs/2015ApJ...812..146M} {812, 146}

\bibitem[\protect\citeauthoryear{{Nishiyama}, {Nagata}, {Tamura}, {Kand ori},
  {Hatano}, {Sato}  \& {Sugitani}}{{Nishiyama} et~al.}{2008}]{Nishiyama+2008}
{Nishiyama} S.,  {Nagata} T.,  {Tamura} M.,  {Kand ori} R.,  {Hatano} H.,
  {Sato} S.,   {Sugitani} K.,  2008, \mn@doi [\apj] {10.1086/587791}, \href
  {https://ui.adsabs.harvard.edu/abs/2008ApJ...680.1174N} {680, 1174}

\bibitem[\protect\citeauthoryear{{Nishiyama} et~al.,}{{Nishiyama}
  et~al.}{2013}]{Nishiyama+2013}
{Nishiyama} S.,  et~al., 2013, \mn@doi [\apjl] {10.1088/2041-8205/769/2/L28},
  \href {https://ui.adsabs.harvard.edu/abs/2013ApJ...769L..28N} {769, L28}

\bibitem[\protect\citeauthoryear{{Nitschai}, {Cappellari}  \&
  {Neumayer}}{{Nitschai} et~al.}{2020}]{MariaSelina+2020}
{Nitschai} M.~S.,  {Cappellari} M.,   {Neumayer} N.,  2020, \mn@doi [\mnras]
  {10.1093/mnras/staa1128}, \href
  {https://ui.adsabs.harvard.edu/abs/2020MNRAS.494.6001N} {494, 6001}

\bibitem[\protect\citeauthoryear{{Nogueras-Lara} et~al.,}{{Nogueras-Lara}
  et~al.}{2018a}]{NoguerasLara+2018a}
{Nogueras-Lara} F.,  et~al., 2018a, \mn@doi [\aap]
  {10.1051/0004-6361/201732002}, \href
  {https://ui.adsabs.harvard.edu/abs/2018A&A...610A..83N} {610, A83}

\bibitem[\protect\citeauthoryear{{Nogueras-Lara} et~al.,}{{Nogueras-Lara}
  et~al.}{2018b}]{NoguerasLara+2018b}
{Nogueras-Lara} F.,  et~al., 2018b, \mn@doi [\aap]
  {10.1051/0004-6361/201833518}, \href
  {https://ui.adsabs.harvard.edu/abs/2018A&A...620A..83N} {620, A83}

\bibitem[\protect\citeauthoryear{{Nogueras-Lara}, {Sch{\"o}del}, {Najarro},
  {Gallego-Calvente}, {Gallego-Cano}, {Shahzamanian}  \&
  {Neumayer}}{{Nogueras-Lara} et~al.}{2019a}]{NoguerasLara+2019a}
{Nogueras-Lara} F.,  {Sch{\"o}del} R.,  {Najarro} F.,  {Gallego-Calvente}
  A.~T.,  {Gallego-Cano} E.,  {Shahzamanian} B.,   {Neumayer} N.,  2019a,
  \mn@doi [\aap] {10.1051/0004-6361/201936322}, \href
  {https://ui.adsabs.harvard.edu/abs/2019A&A...630L...3N} {630, L3}

\bibitem[\protect\citeauthoryear{{Nogueras-Lara} et~al.,}{{Nogueras-Lara}
  et~al.}{2019b}]{NoguerasLara+2019b}
{Nogueras-Lara} F.,  et~al., 2019b, \mn@doi [\aap]
  {10.1051/0004-6361/201936263}, \href
  {https://ui.adsabs.harvard.edu/abs/2019A&A...631A..20N} {631, A20}

\bibitem[\protect\citeauthoryear{{Nogueras-Lara}, {Sch{\"o}del}, {Neumayer},
  {Gallego-Cano}, {Shahzamanian}, {Gallego-Calvente}  \&
  {Najarro}}{{Nogueras-Lara} et~al.}{2020a}]{NoguerasLara+2020b}
{Nogueras-Lara} F.,  {Sch{\"o}del} R.,  {Neumayer} N.,  {Gallego-Cano} E.,
  {Shahzamanian} B.,  {Gallego-Calvente} A.~T.,   {Najarro} F.,  2020a, \aap,
  \href {https://ui.adsabs.harvard.edu/abs/2020arXiv200704401N} {p.
  arXiv:2007.04401}

\bibitem[\protect\citeauthoryear{{Nogueras-Lara} et~al.,}{{Nogueras-Lara}
  et~al.}{2020b}]{NoguerasLara+2020}
{Nogueras-Lara} F.,  et~al., 2020b, \mn@doi [Nature Astronomy]
  {10.1038/s41550-019-0967-9}, \href
  {https://ui.adsabs.harvard.edu/abs/2020NatAs...4..377N} {4, 377}

\bibitem[\protect\citeauthoryear{{Pinna}, {Falc{\'o}n-Barroso}, {Martig},
  {Mart{\'\i}nez-Valpuesta}, {M{\'e}ndez-Abreu}, {van de Ven}, {Leaman}  \&
  {Lyubenova}}{{Pinna} et~al.}{2018}]{Pinna+2018}
{Pinna} F.,  {Falc{\'o}n-Barroso} J.,  {Martig} M.,  {Mart{\'\i}nez-Valpuesta}
  I.,  {M{\'e}ndez-Abreu} J.,  {van de Ven} G.,  {Leaman} R.,   {Lyubenova} M.,
   2018, \mn@doi [\mnras] {10.1093/mnras/stx3331}, \href
  {https://ui.adsabs.harvard.edu/abs/2018MNRAS.475.2697P} {475, 2697}

\bibitem[\protect\citeauthoryear{{Pizzella}, {Corsini}, {Morelli}, {Sarzi},
  {Scarlata}, {Stiavelli}  \& {Bertola}}{{Pizzella}
  et~al.}{2002}]{Pizzella+2002}
{Pizzella} A.,  {Corsini} E.~M.,  {Morelli} L.,  {Sarzi} M.,  {Scarlata} C.,
  {Stiavelli} M.,   {Bertola} F.,  2002, \mn@doi [\apj] {10.1086/340486}, \href
  {https://ui.adsabs.harvard.edu/abs/2002ApJ...573..131P} {573, 131}

\bibitem[\protect\citeauthoryear{{Purcell} et~al.,}{{Purcell}
  et~al.}{2012}]{Purcell+2012}
{Purcell} C.~R.,  et~al., 2012, \mn@doi [\mnras]
  {10.1111/j.1365-2966.2012.21800.x}, \href
  {http://adsabs.harvard.edu/abs/2012MNRAS.426.1972P} {426, 1972}

\bibitem[\protect\citeauthoryear{{Rodriguez-Fernandez} \&
  {Combes}}{{Rodriguez-Fernandez} \& {Combes}}{2008}]{RFC2008}
{Rodriguez-Fernandez} N.~J.,  {Combes} F.,  2008, \mn@doi [\aap]
  {10.1051/0004-6361:200809644}, \href
  {http://adsabs.harvard.edu/abs/2008A%26A...489..115R} {489, 115}

\bibitem[\protect\citeauthoryear{{Sch{\"o}del}, {Najarro}, {Muzic}  \&
  {Eckart}}{{Sch{\"o}del} et~al.}{2010}]{Schoedel+2010}
{Sch{\"o}del} R.,  {Najarro} F.,  {Muzic} K.,   {Eckart} A.,  2010, \mn@doi
  [\aap] {10.1051/0004-6361/200913183}, \href
  {https://ui.adsabs.harvard.edu/abs/2010A&A...511A..18S} {511, A18}

\bibitem[\protect\citeauthoryear{{Sch{\"o}del}, {Feldmeier}, {Kunneriath},
  {Stolovy}, {Neumayer}, {Amaro-Seoane}  \& {Nishiyama}}{{Sch{\"o}del}
  et~al.}{2014}]{Schoedel+2014}
{Sch{\"o}del} R.,  {Feldmeier} A.,  {Kunneriath} D.,  {Stolovy} S.,  {Neumayer}
  N.,  {Amaro-Seoane} P.,   {Nishiyama} S.,  2014, \mn@doi [\aap]
  {10.1051/0004-6361/201423481}, \href
  {https://ui.adsabs.harvard.edu/abs/2014A&A...566A..47S} {566, A47}

\bibitem[\protect\citeauthoryear{{Sch{\"o}nrich}, {Aumer}  \&
  {Sale}}{{Sch{\"o}nrich} et~al.}{2015}]{Schoenrich+2015}
{Sch{\"o}nrich} R.,  {Aumer} M.,   {Sale} S.~E.,  2015, \mn@doi [\apjl]
  {10.1088/2041-8205/812/2/L21}, \href
  {https://ui.adsabs.harvard.edu/abs/2015ApJ...812L..21S} {812, L21}

\bibitem[\protect\citeauthoryear{{Sellwood}}{{Sellwood}}{2014}]{Sellwood2014}
{Sellwood} J.~A.,  2014, \mn@doi [Reviews of Modern Physics]
  {10.1103/RevModPhys.86.1}, \href
  {https://ui.adsabs.harvard.edu/abs/2014RvMP...86....1S} {86, 1}

\bibitem[\protect\citeauthoryear{{Siebert} et~al.,}{{Siebert}
  et~al.}{2008}]{Siebert+2008}
{Siebert} A.,  et~al., 2008, \mn@doi [\mnras]
  {10.1111/j.1365-2966.2008.13912.x}, \href
  {https://ui.adsabs.harvard.edu/abs/2008MNRAS.391..793S} {391, 793}

\bibitem[\protect\citeauthoryear{{Skrutskie} et~al.,}{{Skrutskie}
  et~al.}{2006}]{2MASS}
{Skrutskie} M.~F.,  et~al., 2006, \mn@doi [\aj] {10.1086/498708}, \href
  {http://adsabs.harvard.edu/abs/2006AJ....131.1163S} {131, 1163}

\bibitem[\protect\citeauthoryear{{Sormani}, {Binney}  \& {Magorrian}}{{Sormani}
  et~al.}{2015a}]{SBM2015a}
{Sormani} M.~C.,  {Binney} J.,   {Magorrian} J.,  2015a, \mn@doi [\mnras]
  {10.1093/mnras/stv441}, \href
  {http://adsabs.harvard.edu/abs/2015MNRAS.449.2421S} {449, 2421}

\bibitem[\protect\citeauthoryear{{Sormani}, {Binney}  \& {Magorrian}}{{Sormani}
  et~al.}{2015b}]{SBM2015c}
{Sormani} M.~C.,  {Binney} J.,   {Magorrian} J.,  2015b, \mn@doi [\mnras]
  {10.1093/mnras/stv2067}, \href
  {http://adsabs.harvard.edu/abs/2015MNRAS.454.1818S} {454, 1818}

\bibitem[\protect\citeauthoryear{{Sormani}, {Tre{\ss}}, {Ridley}, {Glover},
  {Klessen}, {Binney}, {Magorrian}  \& {Smith}}{{Sormani}
  et~al.}{2018}]{Sormani+2018a}
{Sormani} M.~C.,  {Tre{\ss}} R.~G.,  {Ridley} M.,  {Glover} S.~C.~O.,
  {Klessen} R.~S.,  {Binney} J.,  {Magorrian} J.,   {Smith} R.,  2018, \mn@doi
  [\mnras] {10.1093/mnras/stx3258}, \href
  {http://adsabs.harvard.edu/abs/2018MNRAS.475.2383S} {475, 2383}

\bibitem[\protect\citeauthoryear{{Sormani}, {Tress}, {Glover}, {Klessen},
  {Battersby}, {Clark}, {Hatchfield}  \& {Smith}}{{Sormani}
  et~al.}{2020}]{Sormani+2020}
{Sormani} M.~C.,  {Tress} R.~G.,  {Glover} S. C.~O.,  {Klessen} R.~S.,
  {Battersby} C.~D.,  {Clark} P.~C.,  {Hatchfield} H.~P.,   {Smith} R.~J.,
  2020, \mn@doi [Paper II (submitted)] {10.1093/mnras/stz2054}, \href
  {https://ui.adsabs.harvard.edu/abs/2019MNRAS.488.4663S} {}

\bibitem[\protect\citeauthoryear{{Tress}, {Sormani}, {Glover}, {Klessen},
  {Battersby}, {Clark}, {Hatchfield}  \& {Smith}}{{Tress}
  et~al.}{2020}]{Tress+2020}
{Tress} R.~G.,  {Sormani} M.~C.,  {Glover} S. C.~O.,  {Klessen} R.~S.,
  {Battersby} C.~D.,  {Clark} P.~C.,  {Hatchfield} H.~P.,   {Smith} R.~J.,
  2020, arXiv e-prints, \href
  {https://ui.adsabs.harvard.edu/abs/2020arXiv200406724T} {p. arXiv:2004.06724}

\bibitem[\protect\citeauthoryear{{Wright} et~al.,}{{Wright}
  et~al.}{2010}]{Wright+2010}
{Wright} E.~L.,  et~al., 2010, \mn@doi [\aj] {10.1088/0004-6256/140/6/1868},
  \href {https://ui.adsabs.harvard.edu/abs/2010AJ....140.1868W} {140, 1868}

\bibitem[\protect\citeauthoryear{{Zasowski} et~al.,}{{Zasowski}
  et~al.}{2013}]{Zasowski+2013}
{Zasowski} G.,  et~al., 2013, \mn@doi [\aj] {10.1088/0004-6256/146/4/81}, \href
  {https://ui.adsabs.harvard.edu/abs/2013AJ....146...81Z} {146, 81}

\bibitem[\protect\citeauthoryear{{van der Kruit} \& {de Grijs}}{{van der Kruit}
  \& {de Grijs}}{1999}]{vanderKruitdeGrijs1999}
{van der Kruit} P.~C.,  {de Grijs} R.,  1999, \aap, \href
  {https://ui.adsabs.harvard.edu/abs/1999A&A...352..129V} {352, 129}

\makeatother
\end{thebibliography}



\appendix

\section{Probability distributions for models 7-18}

\begin{figure*}
	\centering 
	\includegraphics[width=0.33\textwidth]{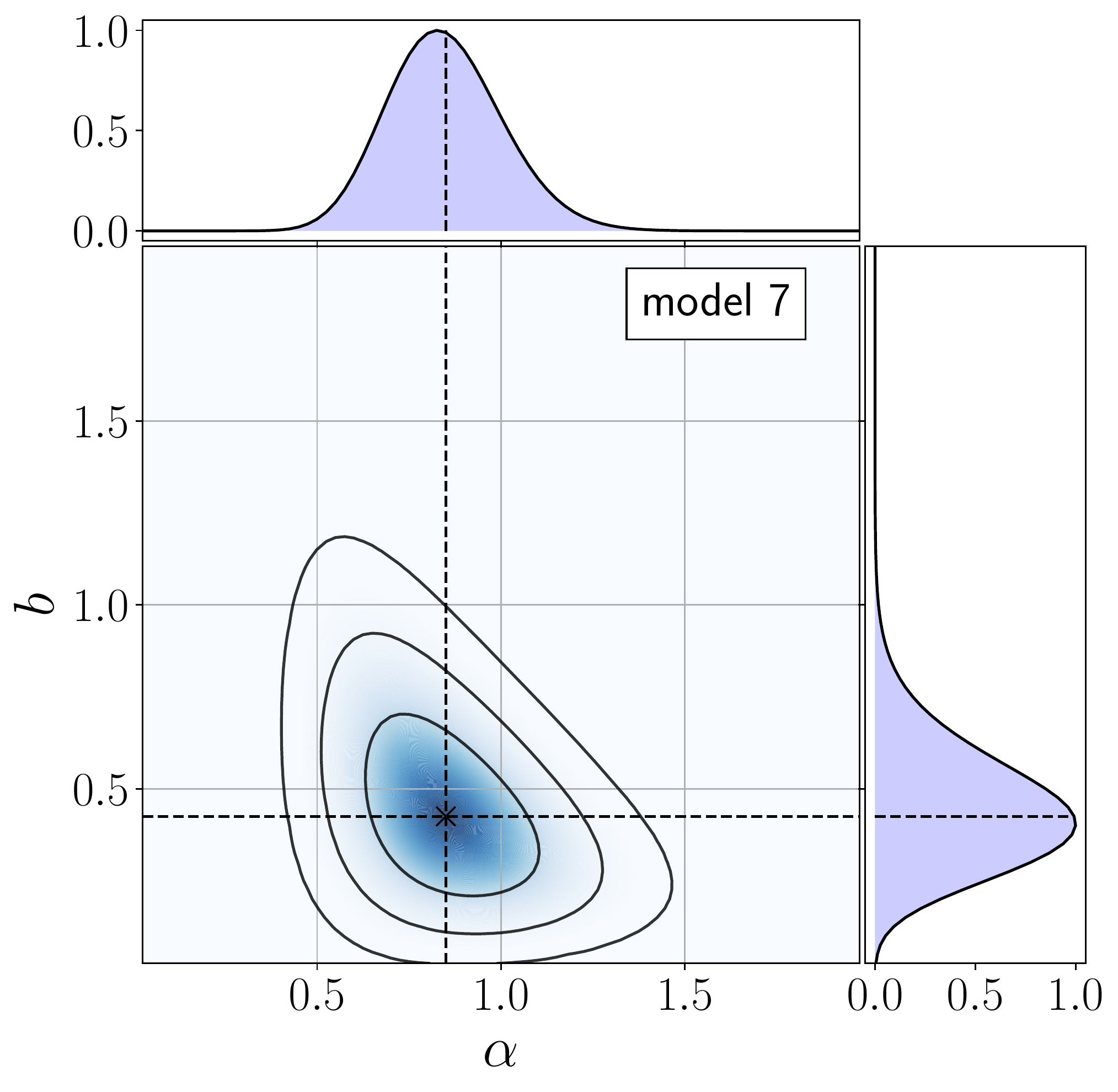}\includegraphics[width=0.33\textwidth]{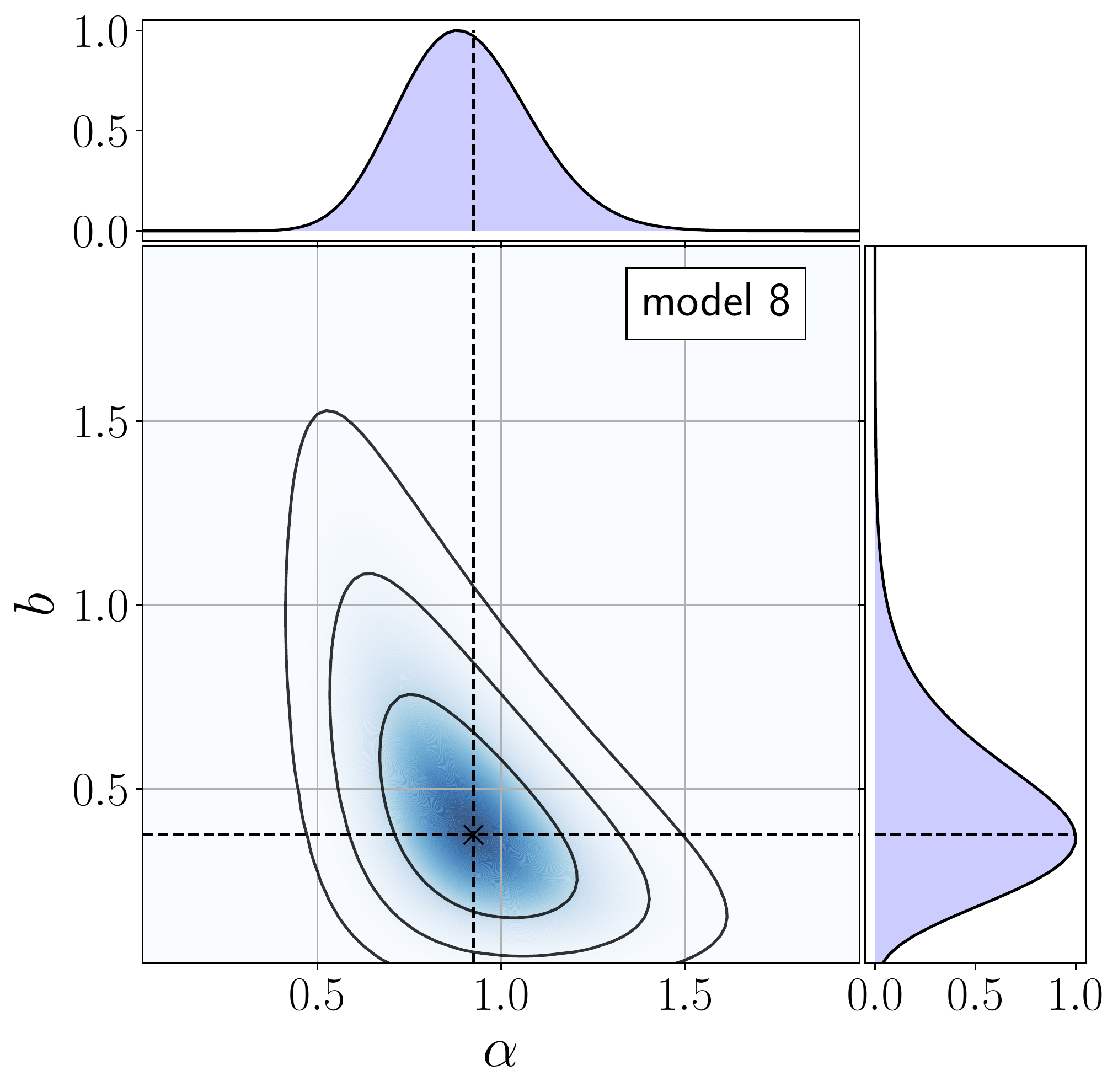}\includegraphics[width=0.33\textwidth]{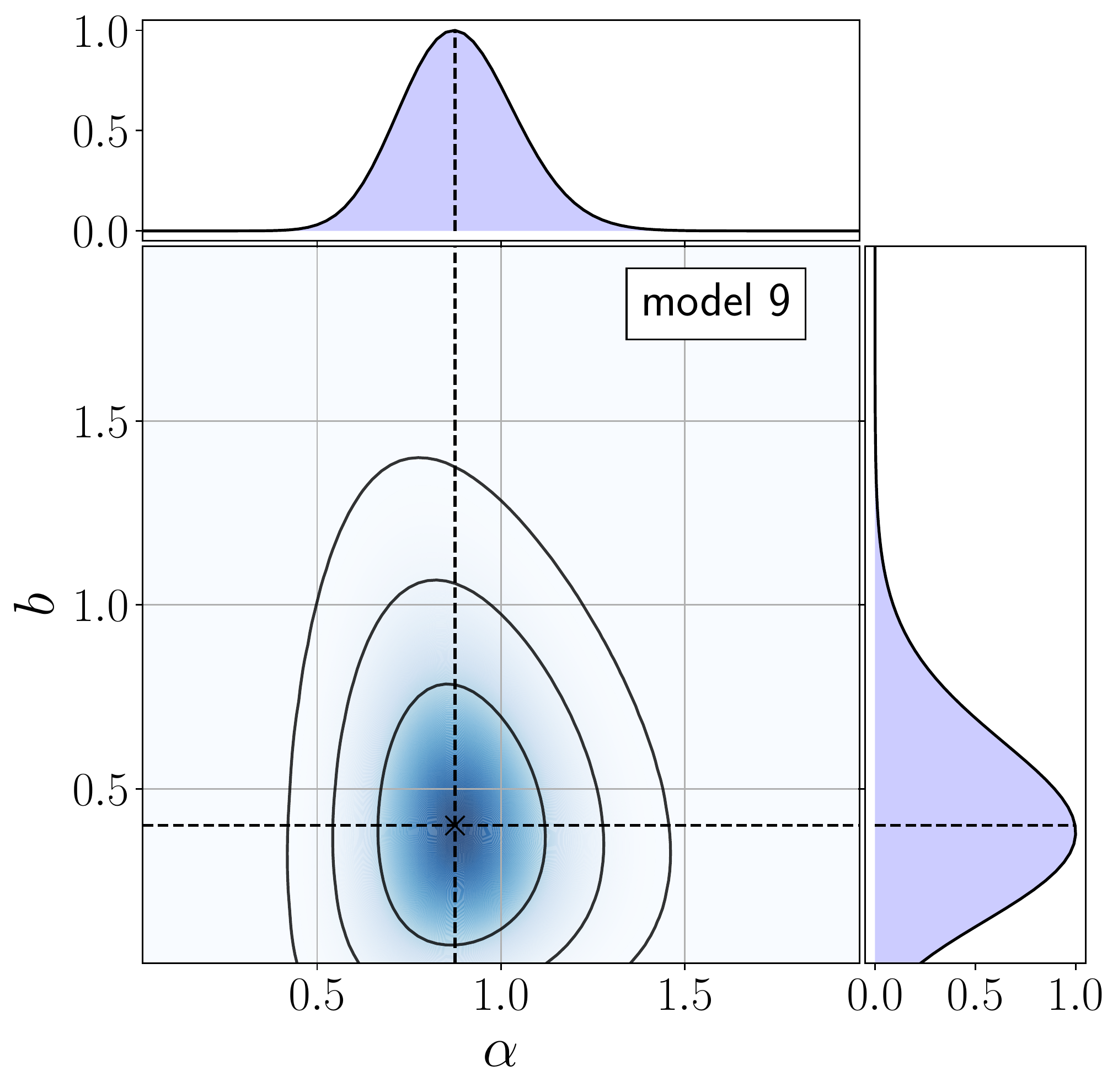}
	\includegraphics[width=0.33\textwidth]{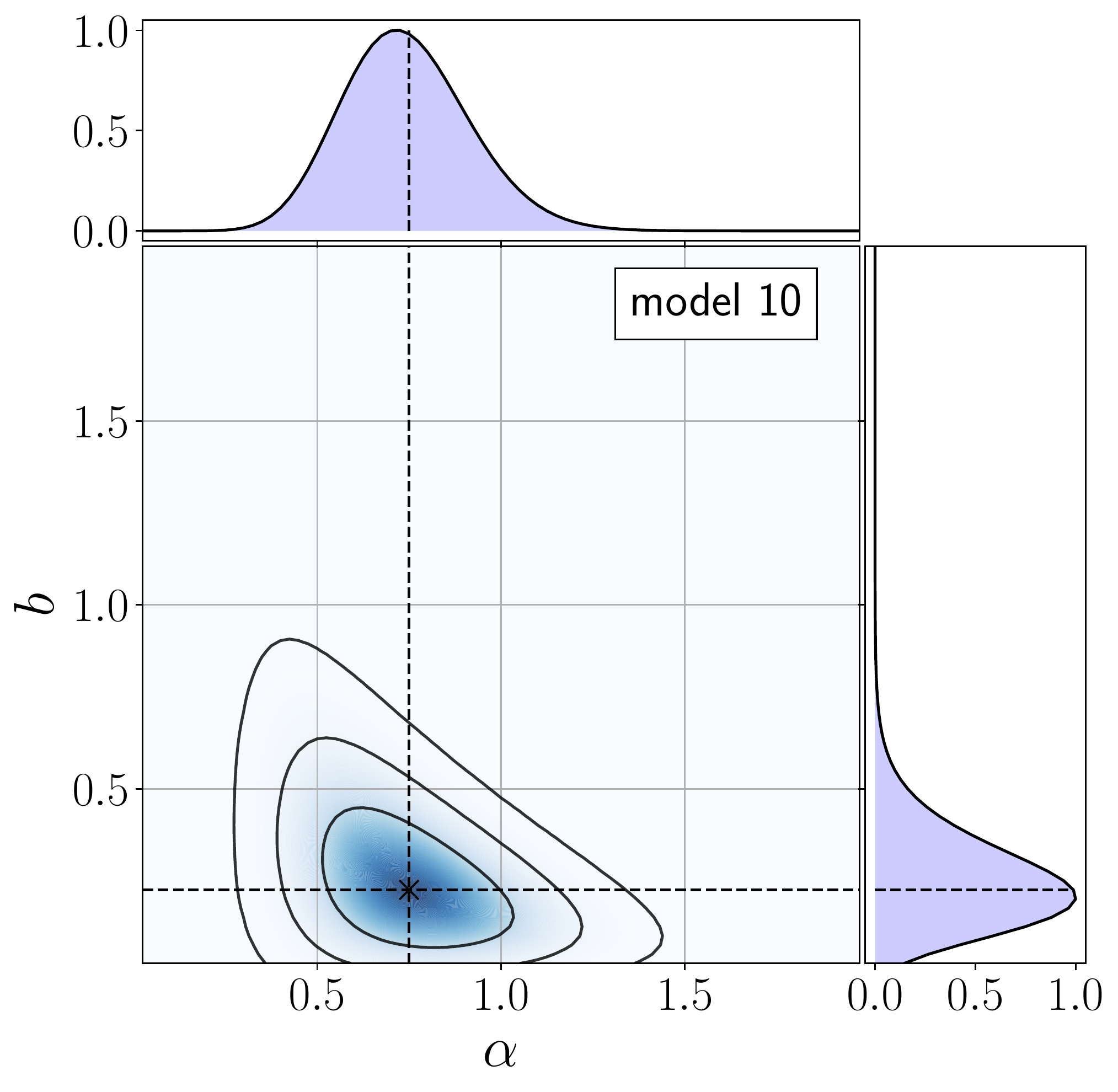}\includegraphics[width=0.33\textwidth]{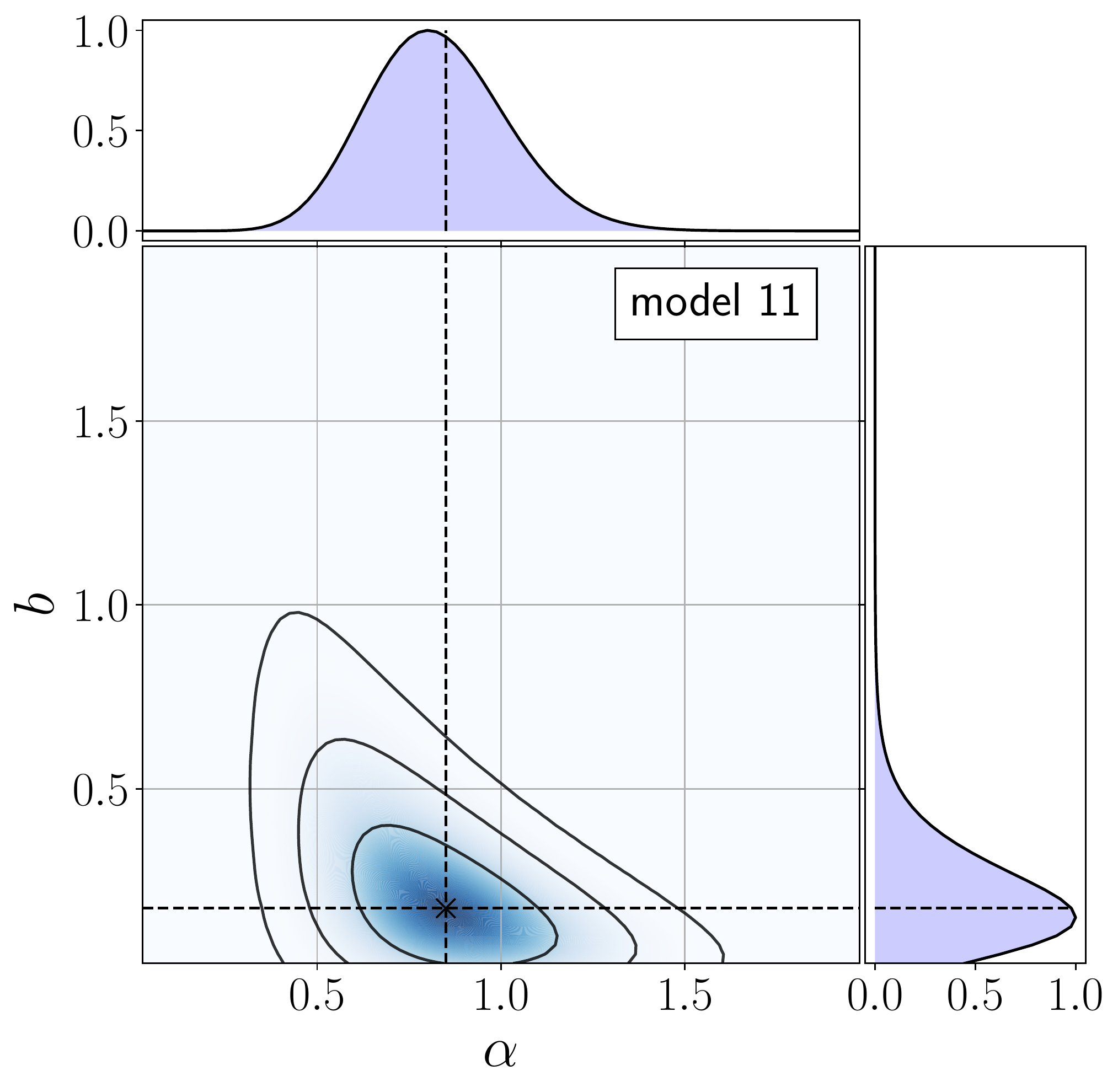}\includegraphics[width=0.33\textwidth]{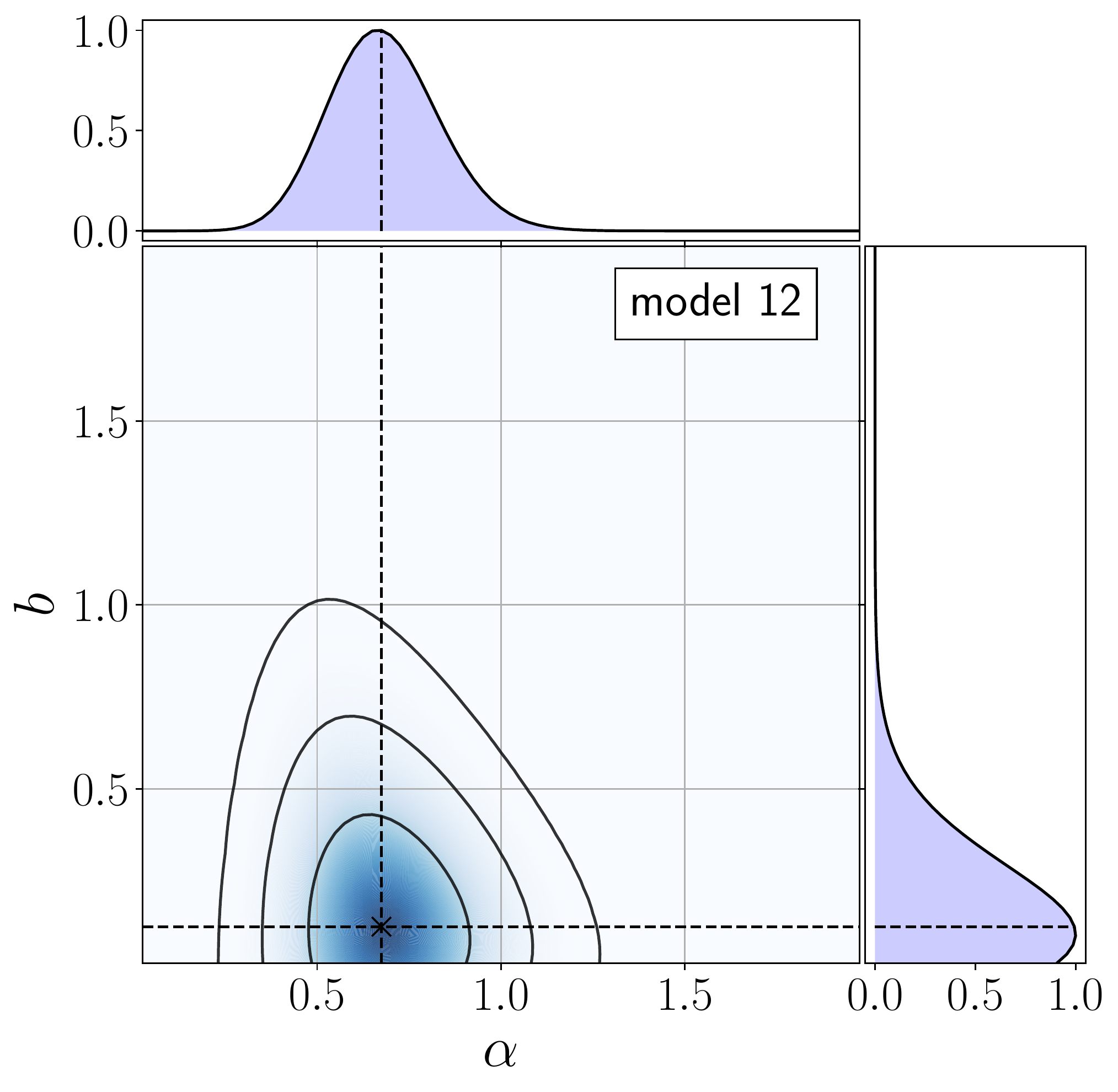}
    \caption{Same as Figure \ref{fig:chi2_1}, but for models 7-12.}
    \label{fig:chi2_2}
\end{figure*}

\begin{figure*}
	\centering 
	\includegraphics[width=0.33\textwidth]{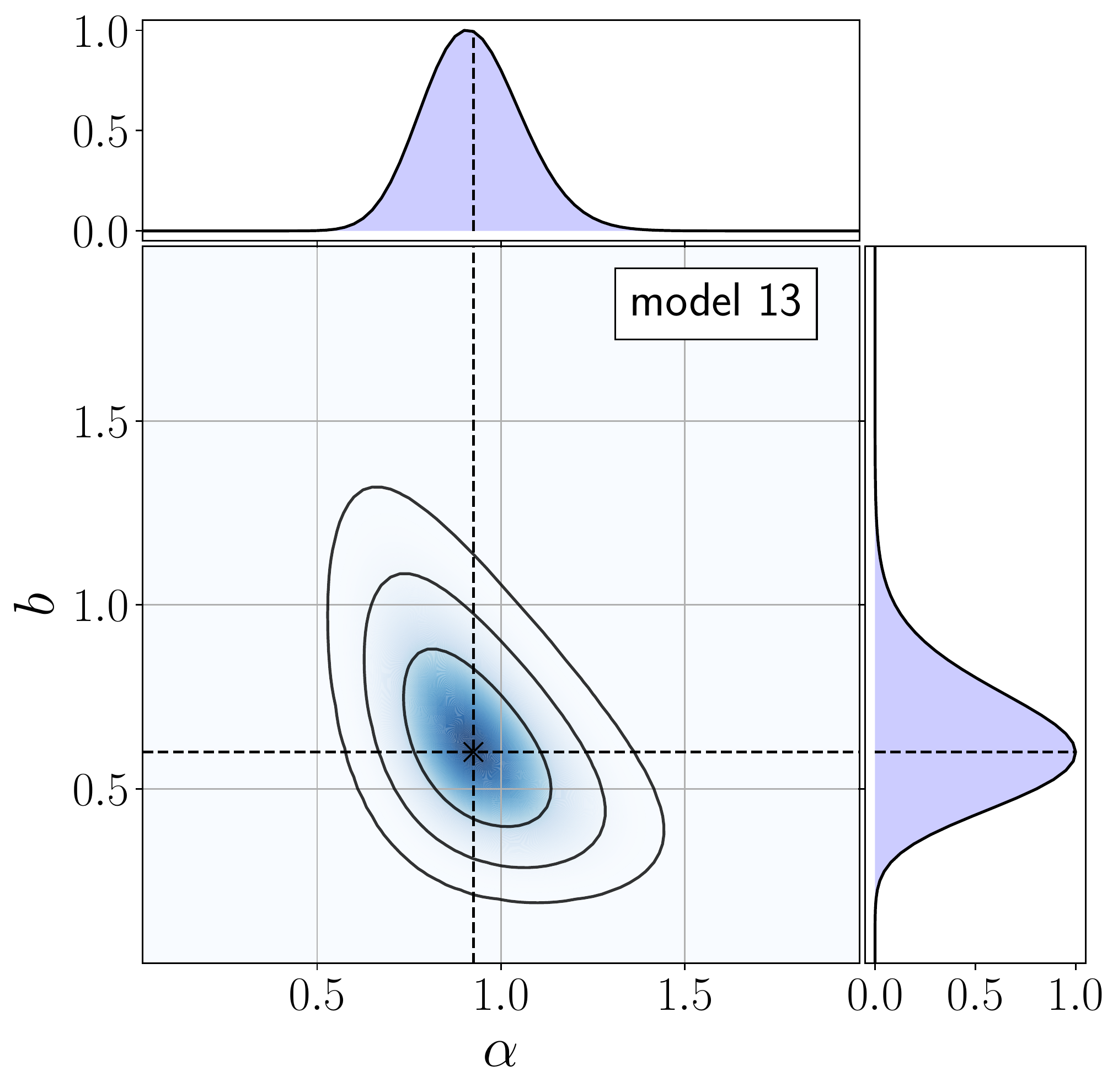}\includegraphics[width=0.33\textwidth]{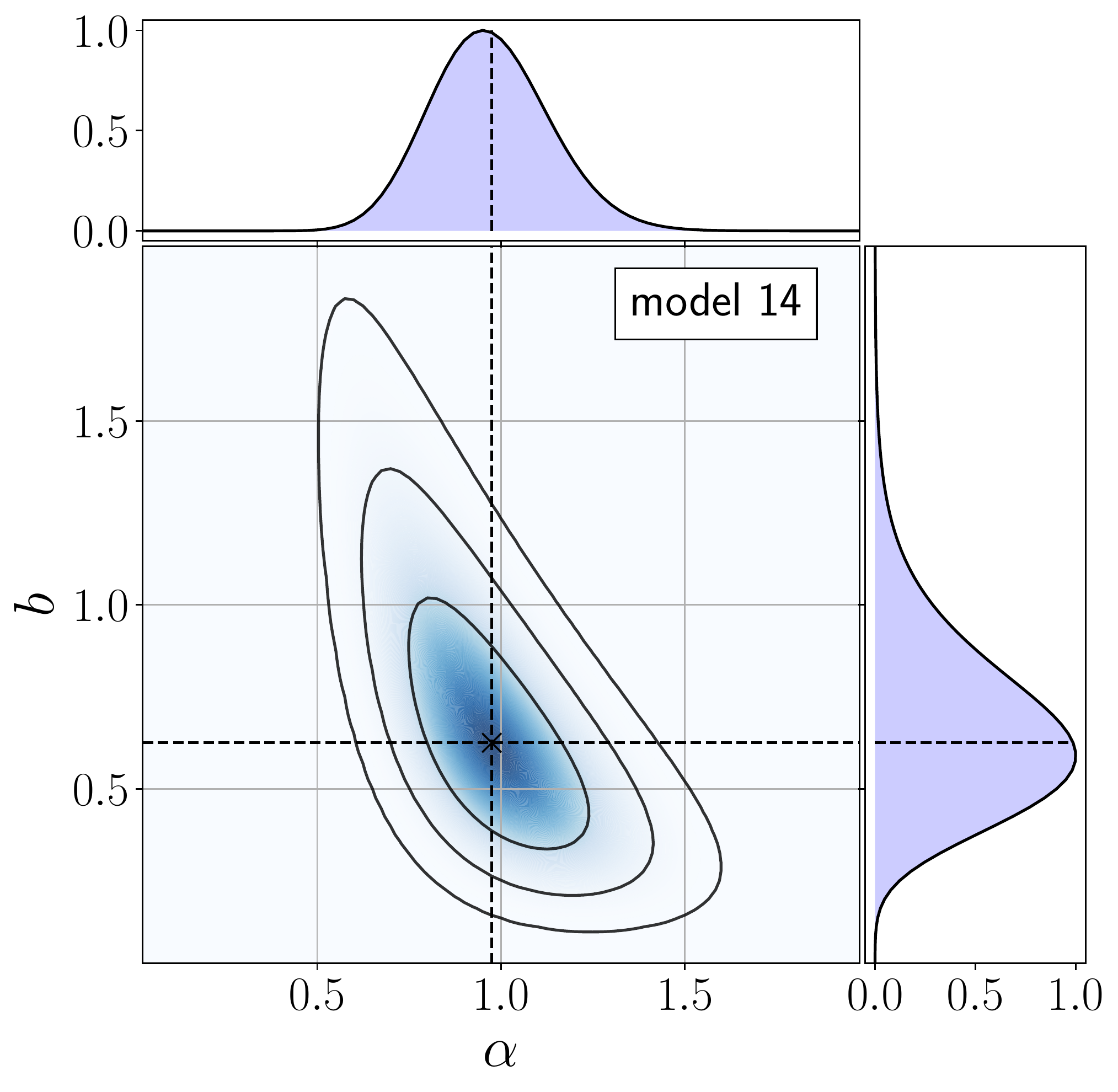}\includegraphics[width=0.33\textwidth]{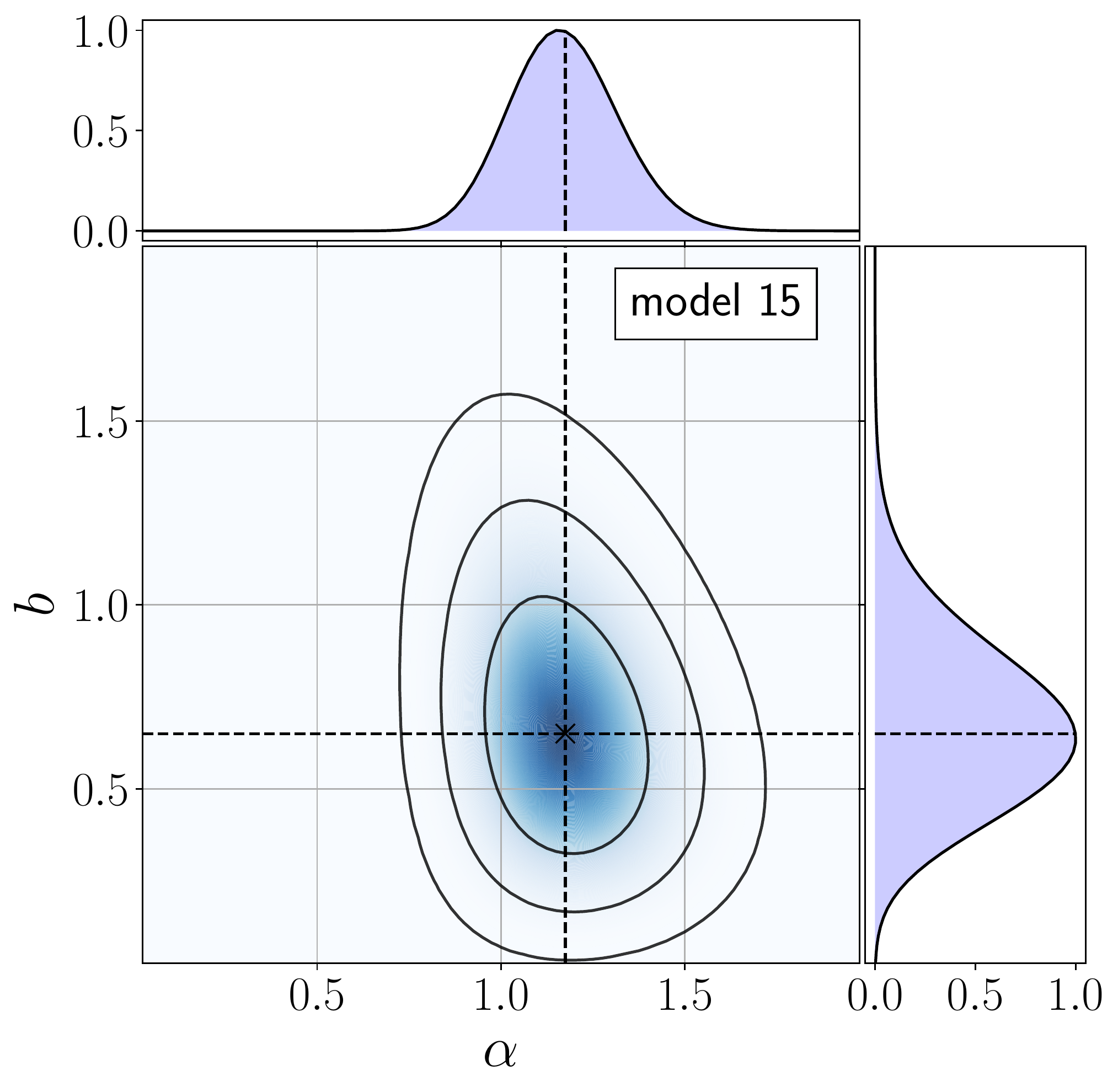}
	\includegraphics[width=0.33\textwidth]{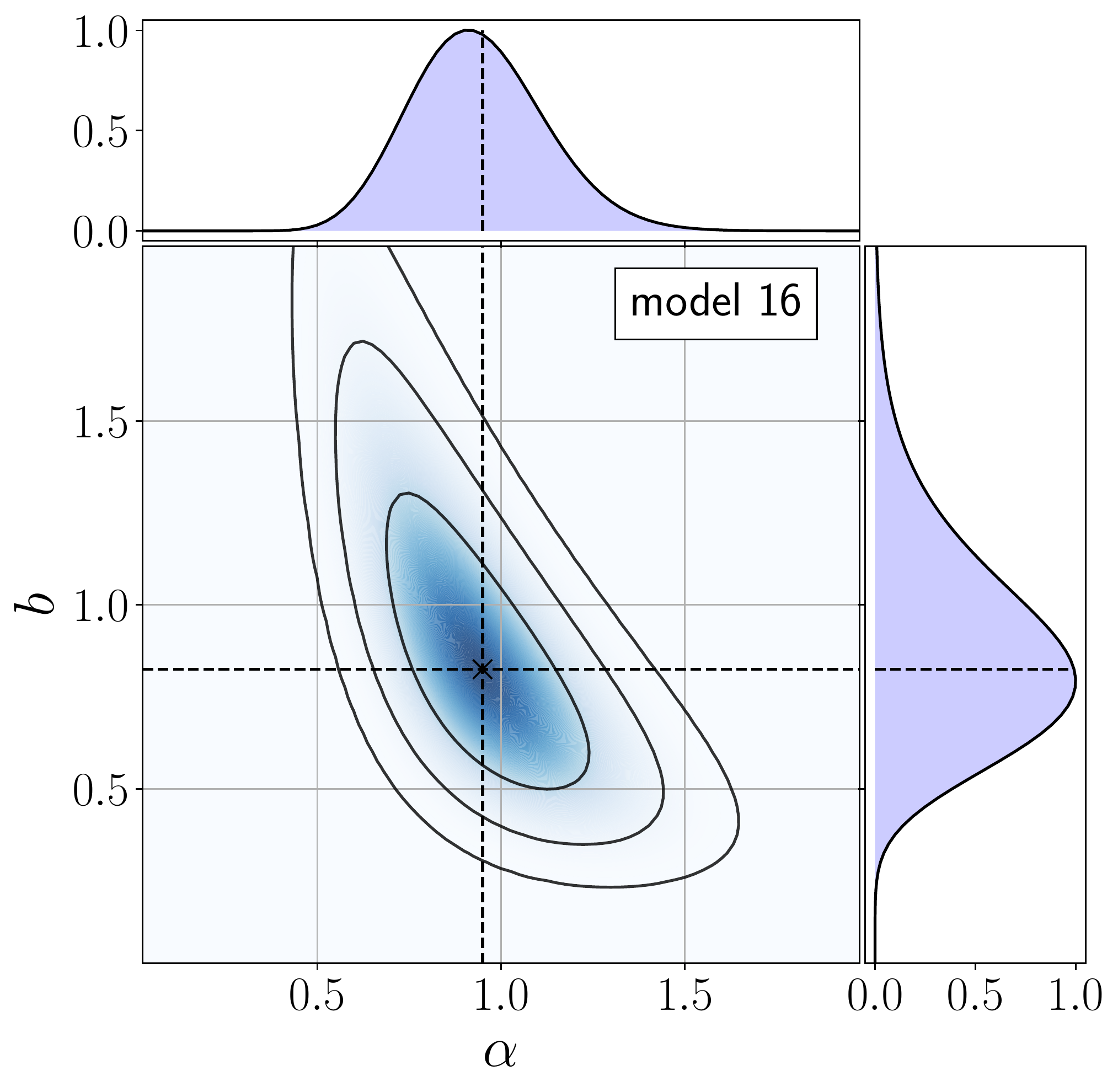}\includegraphics[width=0.33\textwidth]{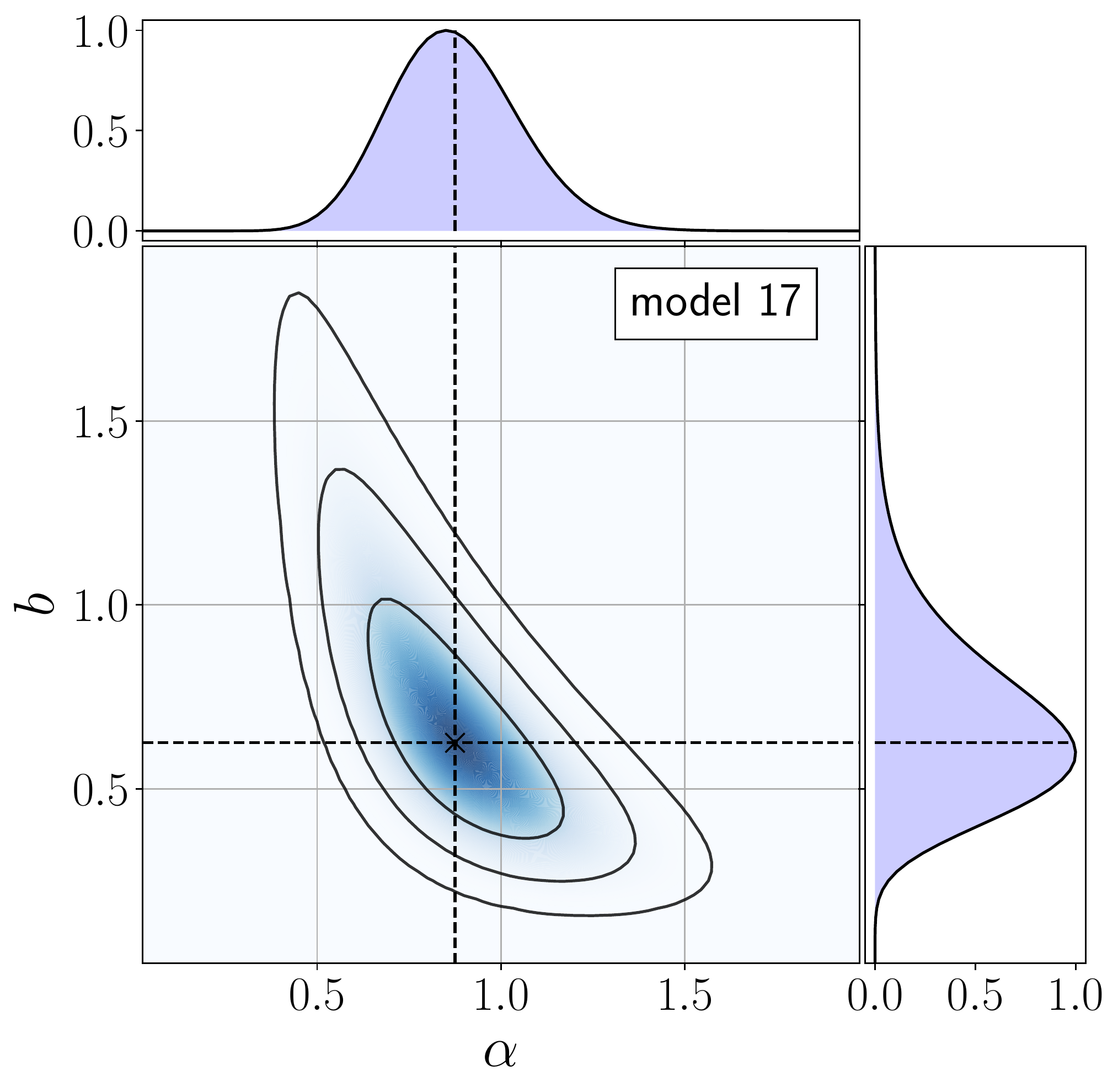}\includegraphics[width=0.33\textwidth]{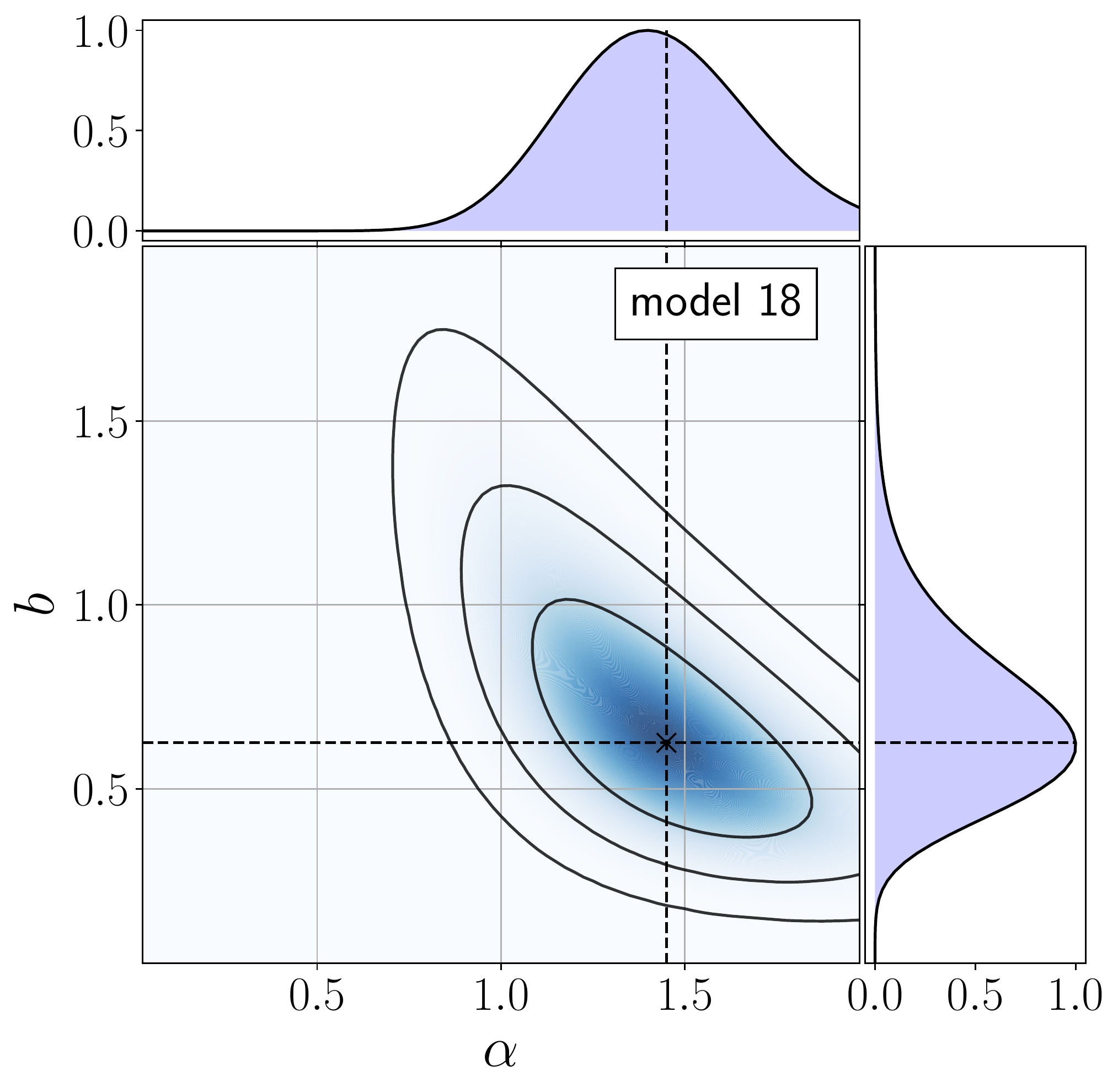}
    \caption{Same as Figure \ref{fig:chi2_1}, but for models 13-18.}
    \label{fig:chi2_3}
\end{figure*}

Figures \ref{fig:chi2_2} and \ref{fig:chi2_3} show the probability distributions for models 7-12 and 13-18 respectively.

\section{Comparison between data and models 1,2}

\begin{figure*}
	\includegraphics[width=0.5\textwidth]{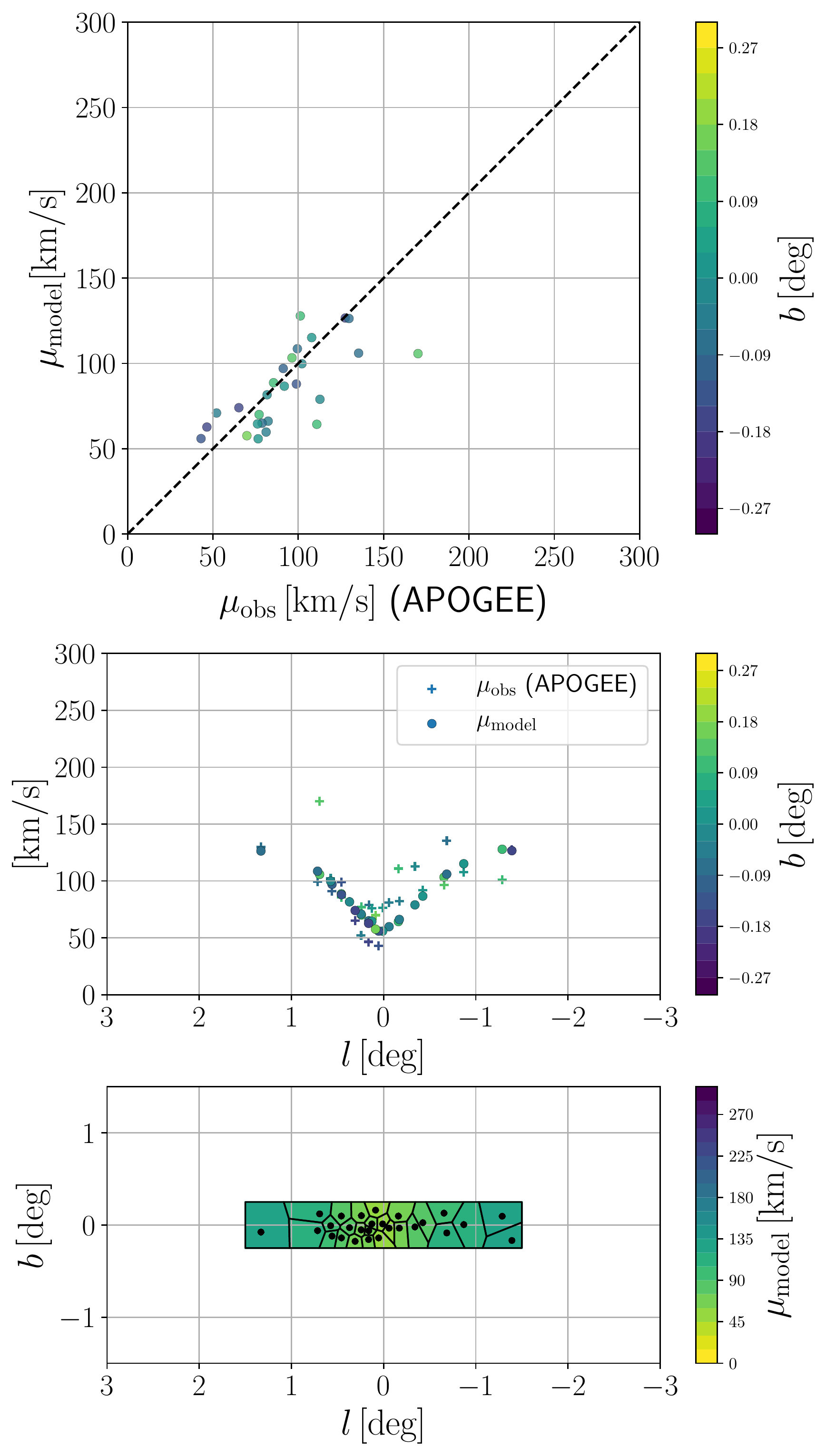}\includegraphics[width=0.5\textwidth]{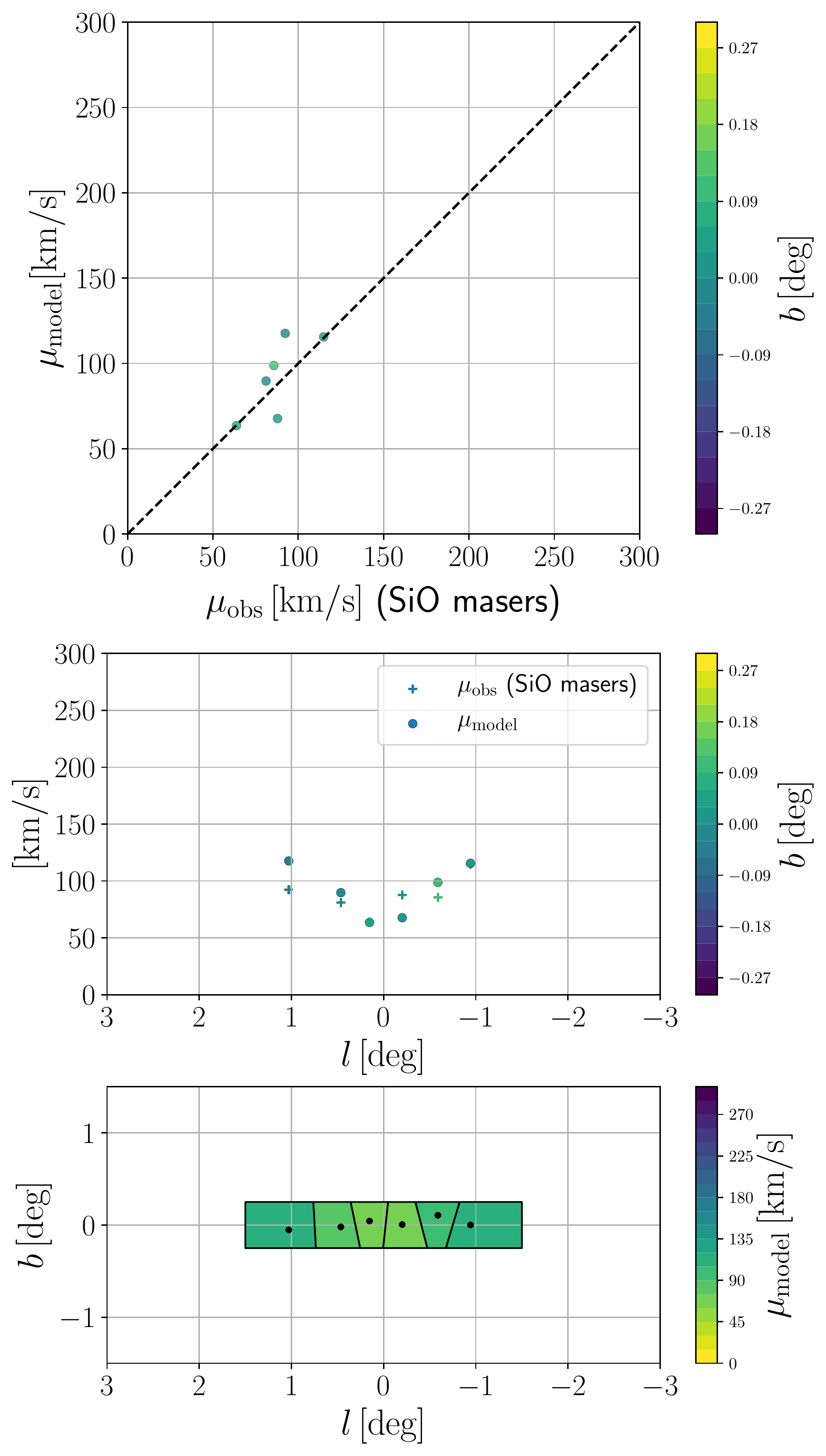}
    \caption{Same as Figure \ref{fig:bestmodel_01}, but for model 1}
    \label{fig:bestmodel_03}
\end{figure*}

\begin{figure*}
	\includegraphics[width=0.5\textwidth]{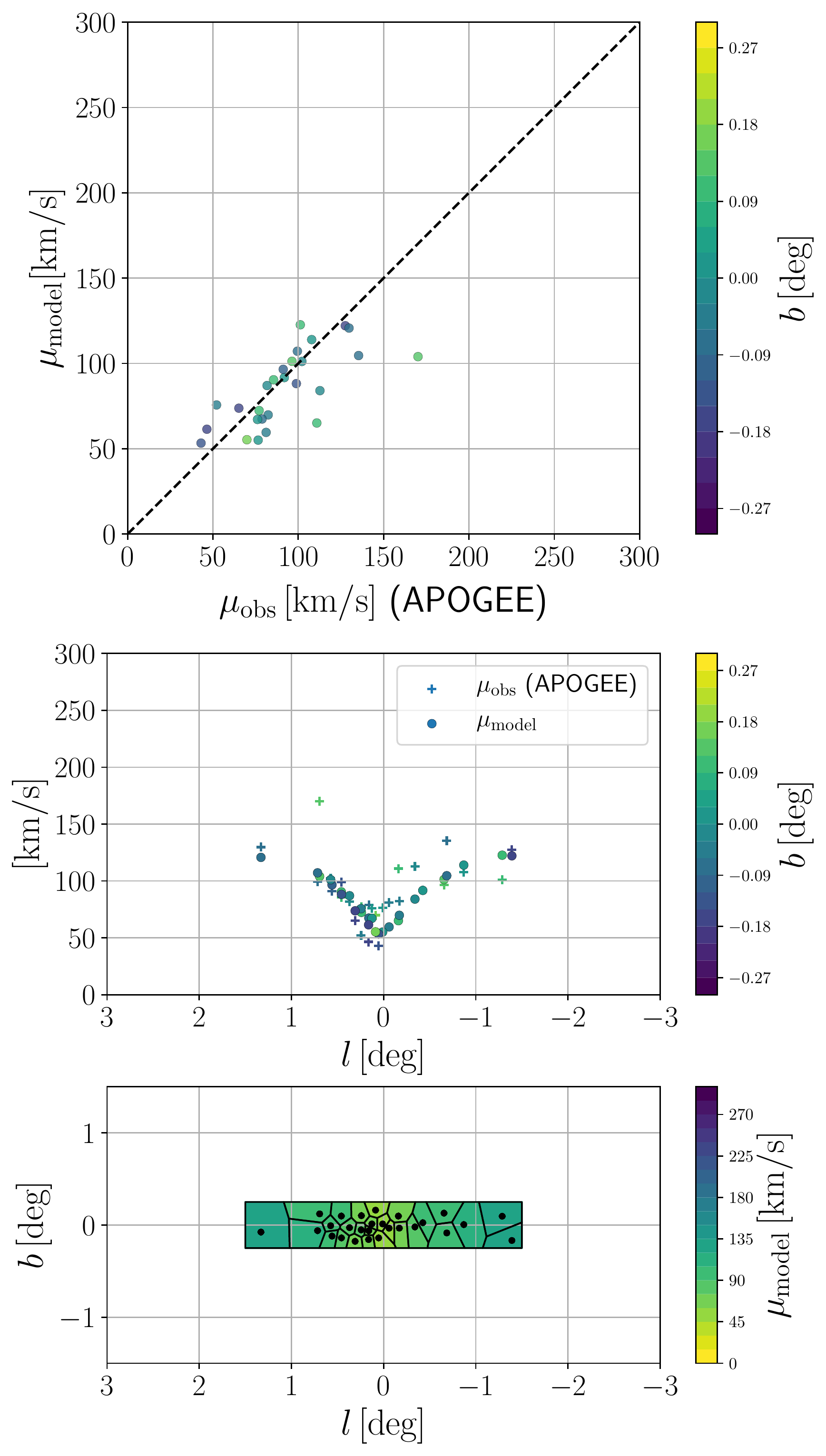}\includegraphics[width=0.5\textwidth]{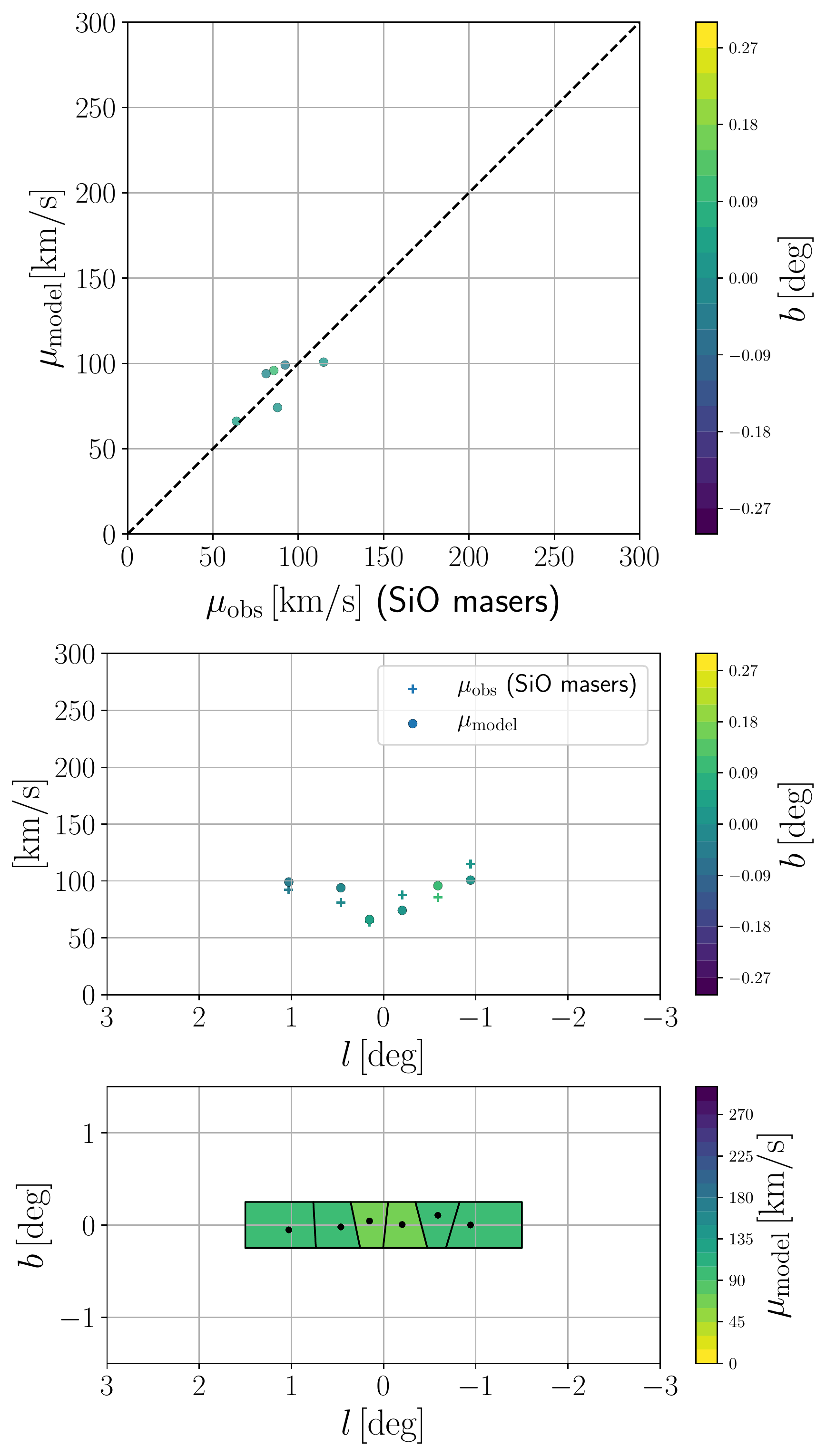}
    \caption{Same as Figure \ref{fig:bestmodel_01}, but for model 2}
    \label{fig:bestmodel_04}
\end{figure*}

Figures \ref{fig:bestmodel_03} and \ref{fig:bestmodel_04} show a comparison between the data and models 1 and 2.

\bsp	
\label{lastpage}
\end{document}